%% file: ms.tex
\documentclass[%
 reprint,
superscriptaddress,
 amsmath,amssymb,
 aps,
]{revtex4-2}

\usepackage{graphicx}% Include figure files
\usepackage{physics}
\usepackage{comment}

\usepackage{makecell}
\usepackage{dsfont}
\usepackage{booktabs}
\usepackage{tikz}
\usetikzlibrary{quantikz}
\usepackage{nameref}

\begin{document} 

\title{Demonstrating a superconducting dual-rail cavity qubit with erasure-detected logical measurements}

\newcommand{\affiliationQCI}{
\affiliation{Quantum Circuits, Inc., 25 Science Park, New Haven, CT 06511, USA}
}
\newcommand{\affiliationYale}{
\affiliation{Departments of Applied Physics and Physics, Yale University, New Haven, Connecticut 06511, USA}
\affiliation{Yale Quantum Institute, Yale University, New Haven, Connecticut 06511, USA}
}

\author{Kevin S. Chou}
\email{chou@quantumcircuits.com}
\affiliationQCI
\author{Tali Shemma}
\affiliationQCI
\author{Heather McCarrick}
\affiliationQCI
\author{Tzu-Chiao Chien}
\affiliationQCI
\author{James D. Teoh}
\affiliationYale
\author{Patrick Winkel}
\affiliationYale
\author{Amos Anderson}
\affiliationQCI
\author{Jonathan Chen}
\affiliationQCI
\author{Jacob Curtis}
\affiliationYale
\author{Stijn J. de Graaf}
\affiliationYale
\author{John W. O. Garmon}
\affiliationYale
\author{Benjamin Gudlewski}
\affiliationQCI
\author{William D. Kalfus}
\affiliationYale
\author{Trevor Keen}
\affiliationQCI
\author{Nishaad Khedkar}
\affiliationQCI
\author{Chan U Lei}
\affiliationQCI
\author{Gangqiang Liu}
\affiliationQCI
\author{Pinlei Lu}
\affiliationQCI
\author{Yao Lu}
\affiliationYale
\author{Aniket Maiti}
\affiliationYale
\author{Luke Mastalli-Kelly}
\affiliationQCI
\author{Nitish Mehta}
\affiliationQCI
\author{Shantanu O. Mundhada}
\affiliationQCI
\author{Anirudh Narla}
\affiliationQCI
\author{Taewan Noh}
\affiliationQCI
\author{Takahiro Tsunoda}
\affiliationYale
\author{Sophia H. Xue}
\affiliationYale
\author{Joseph O. Yuan}
\affiliationQCI
\author{Luigi Frunzio}
\affiliationYale
\author{Jos\'e Aumentado}
\affiliationQCI
\affiliation{National Institute of Standards and Technology, Boulder, Colorado 80305, USA}
\author{Shruti Puri}
\affiliationYale
\author{Steven M. Girvin}
\affiliationYale
\author{S. Harvey Moseley, Jr.}
\affiliationQCI
\author{Robert J. Schoelkopf}
\email{robert.schoelkopf@yale.edu}
\affiliationQCI
\affiliationYale

%TC:ignore
\begin{abstract}
A critical challenge in developing scalable error-corrected quantum systems is the accumulation of errors while performing operations and measurements.
One promising approach is to design a system where dominant errors can be detected and converted into erasures~\cite{grassl_codes_erasure_1997,wu_erasure_2022}. Such a system utilizing erasure qubits are known to have relaxed requirements for quantum error correction~\cite{grassl_codes_erasure_1997, stace_codes_loss_2009, barrett_ftqc_loss_2010}. 
A recent proposal aims to do this using a dual-rail encoding with superconducting cavities~\cite{teoh_dual-rail_2022}.
However, experimental characterization and demonstration of a dual-rail cavity qubit have not yet been realized. 
In this work, we implement such a dual-rail cavity qubit; we demonstrate a projective logical measurement with integrated erasure detection and use it to measure the dual-rail qubit idling errors.
We measure logical state preparation and measurement errors at the $0.01\%$-level and detect over $99\%$ of cavity decay events as erasures. 
We use the precision of this new measurement protocol to distinguish different types of errors in this system, finding that while decay errors occur with probability $\sim$ 0.2\% per microsecond, phase errors occur 6 times less frequently and bit flips occur at least 140 times less frequently. 
These findings represent the first confirmation of the expected error hierarchy necessary to concatenate dual-rail cavity qubits into a highly efficient erasure code.
\end{abstract}
%TC:endignore

%TC:ignore
\maketitle 
%TC:endignore

%\section*{Introduction}

Practical quantum error correction requires achieving error rates well below threshold in the physical qubits during all operations. Most platforms still require substantial improvements in fidelity of not just gates but all operations, including state preparation and measurement. 
In addition to the quest for better physical qubits with lower error rates and fewer error channels, various avenues are being investigated to address these challenges.
One approach is to implement error correction using qudits or multi-level bosonic modes to provide the necessary redundancy within a single physical element~\cite{mirrahimi_cat_qubits_2014,michael_binomial_code_2016,joshi_bosonic_review_2021,cai_bosonic_2021,ma_quantum_2021}, and is the only architecture to have achieved breakeven for the lifetime of a logical qubit~\cite{Ofek2016,sivak_real-time_2023,ni_beating_2023,hu_qec_2019}.
% Further, error correction thresholds can be less demanding if one can tailor codes~\cite{aliferis_biased_noise_2008,tuckett_tailoring_2019, guillaud_repetition_cat_qubits_2019,darmawan_practical_2021, claes_tailored_2023} to exploit a natural or engineered structure or noise bias in the qubits ~\cite{aliferis_FT_biasednoise_scqubits_2009,grimm_kerr_cat_2020,puri_bias_preserving_2020,lescanne_suppression_bit_flips_2020,berdou_long_bit_flip_2022}.
Further, error correction thresholds can be less demanding if one can tailor codes~\cite{aliferis_biased_noise_2008,tuckett_tailoring_2019, guillaud_repetition_cat_qubits_2019,darmawan_practical_2021, claes_tailored_2023} to exploit a natural or engineered noise bias in the qubits ~\cite{aliferis_FT_biasednoise_scqubits_2009,grimm_kerr_cat_2020,puri_bias_preserving_2020,lescanne_suppression_bit_flips_2020,berdou_long_bit_flip_2022}. 
While on the path to error correction, there are benefits in detecting errors as they occur in the physical qubits through the use of 
% dedicated flag qubits
additional flag qubits dedicated to error detection~\cite{chao_flag_2020,chamberland_topological_2020,ryan_anderson_quantinuum_rt_ft_qec_2021,ryan_anderson_implementing_2022,chen_calibrated_2022}.
 
A relatively new approach~\cite{wu_erasure_2022} is to directly build into the physical qubit a capability to appropriately detect~\cite{kubica_erasure_2022,kang_trapped_ions_erasure_2023,scholl_erasure_2023, ma_high-fidelity_2023} the dominant errors, thereby converting them to erasures, which are defined as detected errors that occur at a specific time and location in the physical qubits~\cite{grassl_codes_erasure_1997}.
The use of such so-called erasure qubits extends error-detection to full error correction when incorporated in a higher-level stabilizer code. 
An erasure qubit-based stabilizer code has a distinct advantage, as it is known that any stabilizer code can correct for twice as many erasure errors compared to Pauli errors~\cite{grassl_codes_erasure_1997}. Additionally, large-scale error correction codes benefit from a higher threshold for erasure errors~\cite{stace_codes_loss_2009,barrett_ftqc_loss_2010,wu_erasure_2022}.

A simple way to realize an erasure qubit is with the dual-rail encoding, defined by the codewords $\ket{0_\mathrm{L}} = \ket{01}$ and $\ket{1_\mathrm{L}} = \ket{10}$ (see Fig.~1A). 
The encoding of this qubit in two spatial modes~\cite{Chuang_1995_OG_dual_rail} or polarizations of an optical photon~\cite{knill_KLM_linear_optics_2001} is a well-known concept~\cite{kok_linear_2007} that has been widely explored in linear optics platforms, and remains an active area of research~\cite{bartolucci_fusion_based_2023}.
The recently proposed circuit quantum electrodynamics (cQED) implementation of the dual-rail cavity qubit~\cite{teoh_dual-rail_2022} has the distinct advantage of strong and controllable non-linearity via a dispersively coupled transmon ancilla, which enables on-demand arbitrary state preparation, entangling gates, and measurements enabled by efficient single-photon detection.
Other approaches to encode a qubit in a pair of superconducting microwave modes have been both proposed and experimentally demonstrated~\cite{zakka-bajjani_quantum_2011, shim_semiconductor-inspired_2016, campbell_small_gap_qubits_2020}.
More recently, the idea of building erasure qubits out of transmon qubits themselves has also been proposed~\cite{kubica_erasure_2022}.
The bosonic cQED realization uses two microwave cavities, and offers the prospect of converting not only cavity photon loss, but also dominant ancilla errors~\cite{Rosenblum2018,reinhold_error_corrected_gates_2020,ma_error_transparent_2020}, into detectable erasures. 
Crucially, this can be achieved for a full set of single and two-qubit gates, as well as state preparation and measurement.
A key design principle of the dual-rail cavity qubit is to engineer a system that exhibits a strong hierarchy of errors, with the majority of errors detectable as erasures and with the residual Pauli and leakage error rates orders of magnitude smaller.
This hierarchy results in a qubit optimized for integration into a higher-level error correction code.
Already, recent work has demonstrated single-qubit operation fidelities for a dual-rail cavity qubit in excess of 99.95\%~\cite{chapman_beamsplitter_2022,lu_beamsplitter_2023}.
However, to fully exploit the benefits of this encoding, several other aspects are required.

In this work, we implement logical state preparation and measurement (SPAM) in a dual-rail cavity qubit.
We design our logical measurement to have built-in erasure detection, rendering it insensitive to any single occurrence of the dominant hardware errors, including those arising from decoherence, initialization, and readout.
The type of logical measurement we show is an example of end-of-the-line erasure detection, which finds use when measuring qubits at the end of an algorithm or when measuring error syndromes in a stabilizer circuit. 
This is in contrast with mid-circuit erasure detection, which preserves the logical information within the codespace when an erasure is not detected. Such a mid-circuit measurement is necessary for erasure conversion or correction, and is possible~\cite{Stijn} with a different technique~\cite{teoh_dual-rail_2022} within the same architecture.
With our end-of-the-line measurement, we obtain SPAM errors that are among the lowest for any physical qubit platform, with logical errors at the $0.01\%$ level. 
Additionally, we show that over $99\%$ of decay errors can be detected as end-of-the-line erasures.
We use this logical measurement to probe idling bit-flip and dephasing error rates in a dual-rail qubit, finding they are smaller than the measured erasure rates.
Specifically, for this system we measure a cavity decay rate $\sim0.2\%$ per microsecond, with phase and bit-flip errors measured to be a factor of 6 times and at least 140 times less likely, respectively.
The development of a logical measurement with precise erasure detection is a key step for the dual-rail cavity qubit approach and enables detailed characterization of both erasures and the underlying small errors.

%TC:ignore
\begin{figure*}[h]
     \centering
     \includegraphics[width=0.7\textwidth]{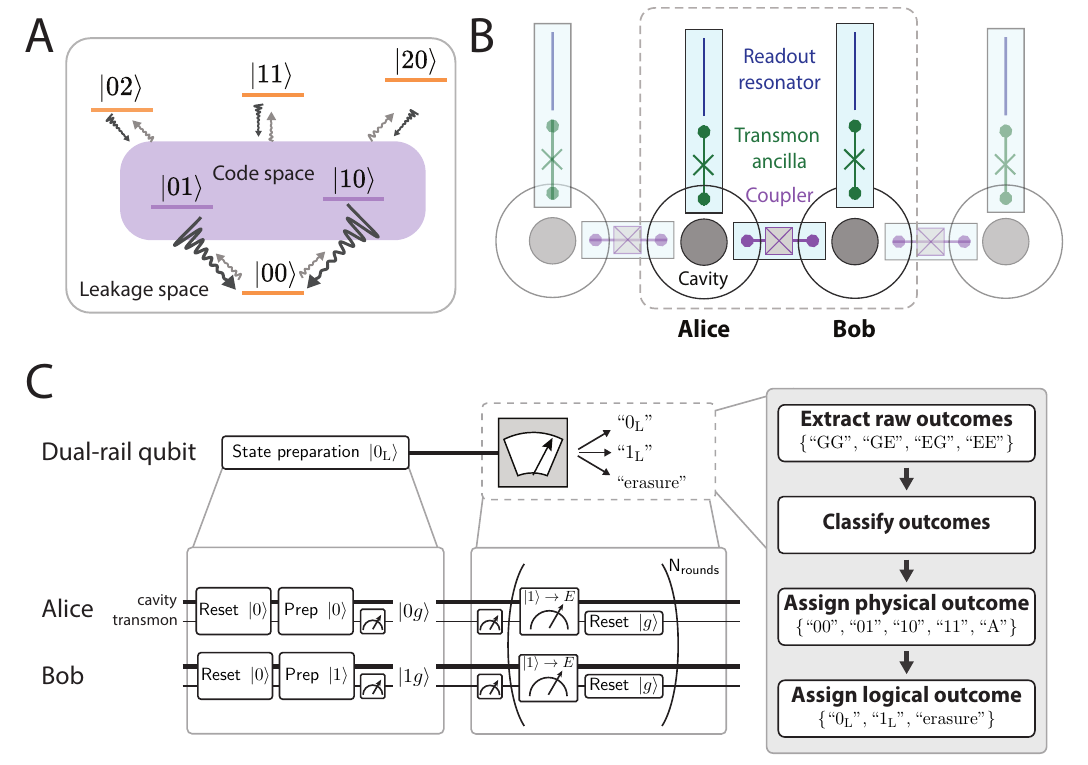}
\caption{\label{fig:fig_1} 
\textbf{Dual-rail qubit concept, implementation, and measurement.}
(A) The dual-rail codespace spans the states $\{\ket{01}, \ket{10}\}$. Transitions out of the codespace induced by either relaxation or heating events bring the system into a leakage space that can be detected with the appropriate measurement protocol.
(B) The dual-rail qubit is implemented in a superconducting cQED module consisting of several physical modes: two superconducting cavities here implemented as three-dimensional $\lambda/4$-coaxial cavities (gray), and each dispersively coupled to a transmon (green) and a resonator (blue) for control and readout. Single-qubit operations are engineered by beamsplitter interactions~\cite{chapman_beamsplitter_2022,lu_beamsplitter_2023} enabled by a nonlinear coupler, here constructed as an array of three SNAIL elements operated at the zero-bias sweet-spot~\cite{frattini_SNAIL_2017,verney_driven_JJ_2019}, resulting in single-qubit gate fidelities as shown in supplementary text 2. There are adjacent cavity systems within this module that are not used in this experiment, shown as shaded elements.
(C) 
Protocol for logical state preparation and measurement experiment.
% Logical state preparation and measurement protocols used in this work. 
The dual-rail logical measurement incorporates erasure detection, leading to a measurement with three outcomes: logical outcomes $``0_\mathrm{L}"$ and $``1_\mathrm{L}"$, and erasures. There are several steps required to decode the logical measurement, starting with assignment of the transmon outcomes, classification and assignment to one of the physical cavity outcomes (``00", ``01", ``10", or ``11") or to an ambiguous outcome (``A") in the case of multiple rounds of measurements disagreeing, before finally assigning the logical outcome.
}
\end{figure*}
%TC:endignore

%TC:ignore
\begin{figure*}[h]
     \centering
     %  Science format
     \includegraphics[width=0.6\textwidth]{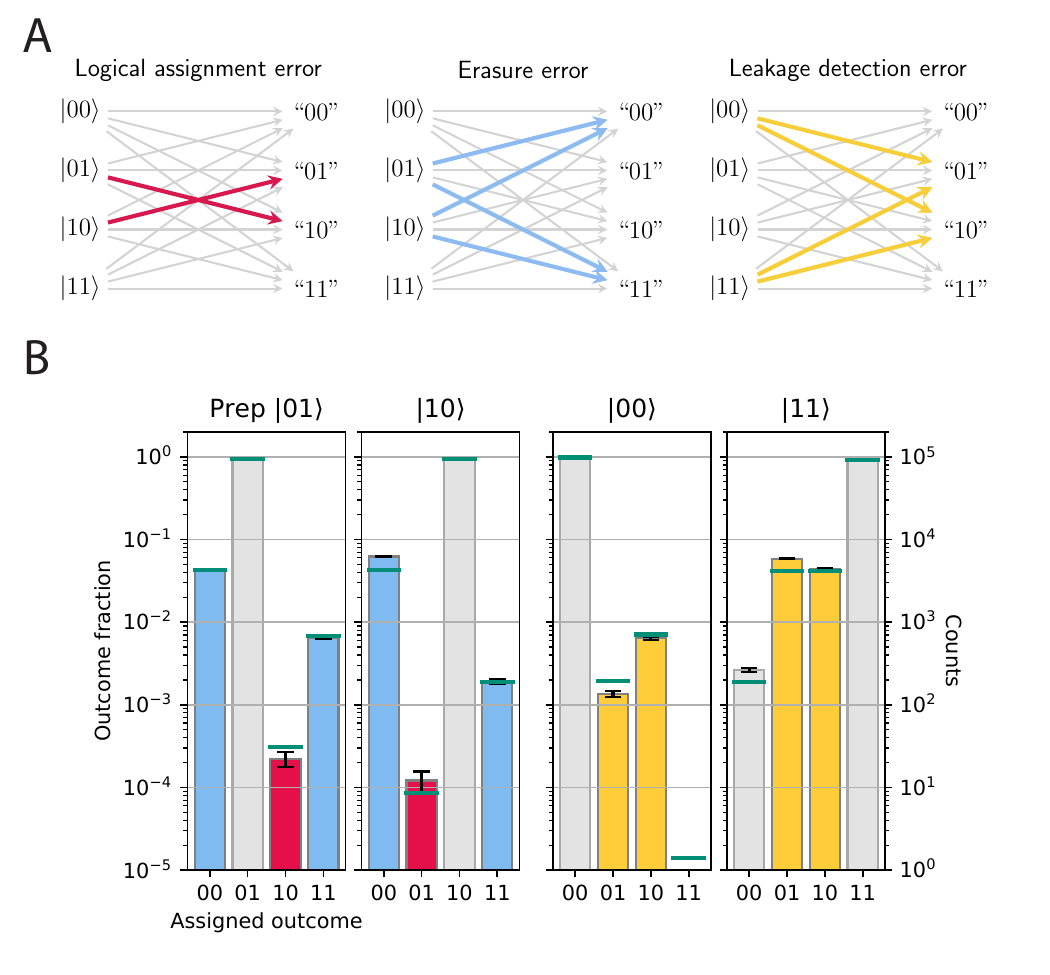}
     \caption{\label{fig:fig_2}
\textbf{State preparation and measurement of a dual-rail qubit.}
(A) The set of possible measurement outcomes can be summarized in the assignment channel diagram, color-coded to highlight several different figures of merit. Apparent bit-flip errors (pink) are a logical misassignment of $\ket{01}$ for $\ket{10}$ and vice versa. Erasure errors (blue) indicate an erasure outcome was measured despite preparation in one of the logical states. Finally, leakage detection errors  (yellow) are misassignments of a known leaked state: e.g. $\ket{00}$ or, less likely $\ket{11}$. 
(B) State assignment data using 1 round of measurements.
In our SPAM experiment, we prepare each of the four dual-rail states and perform a logical measurement. Results for each dual-rail state are shown in each panel; and each bar corresponds to a different assigned physical state. The model results are shown as blue lines for each outcome. Error bars represent the standard error, showing $\pm 1\sigma$. 
}
\end{figure*}
%TC:endignore

\section*{Results}
\subsection*{Characterizing dual-rail SPAM}
\label{sec:SPAM_data}

We first describe the principle of measurement for the dual-rail cavity qubit. 
Logical measurement of a dual-rail cavity qubit detects which cavity, if any, contains the photon.
In contrast to conventional qubit measurements, where performance is limited by misassignment and transition errors during the readout~\cite{gambetta_optimal_readout_2007,elder_msmts_2020}, the logical measurement has the additional capability of labeling most of these readout errors and also cavity leakage errors as erasures. 
The logical measurement is expected to have both low logical misassignment errors and be a highly sensitive detector of leakage due to decay errors (photon loss), while ideally incurring only a small penalty from additional erasures due to readout errors. 
These properties are essential for a good dual-rail cavity qubit.

We describe our protocols for state preparation and measurement in the dual-rail encoding (supplementary text 4).
In our hardware implementation, each cavity has an individual ancilla transmon and readout for control and measurement (Fig.~1B), which allows simultaneous measurement of both cavities.
As shown in Fig.~1C, the state preparation protocol consists of three steps: reset to $\ket{00}$, initialization in a logical state, and optional ``check" measurements to verify the state preparation, which can boost the preparation fidelity by flagging failures for postselection.  
While there are many options for the state preparation protocol, in this work we have utilized two different methods, with both expected to have similar performance (supplementary text 4).
For SPAM experiments described in this section, we use an optimal control pulse (OCP)~\cite{Heeres_2017_OCP} to initialize a cavity in $\ket{1}$ and a single transmon measurement after the OCP to check that the transmon is in $\ket{g}$, as intended.
For subsequent experiments in this work, we adopt a state preparation protocol that not only includes a transmon state check, but also repeated cavity state check measurements.
Importantly for our SPAM protocols, any errors in the state preparation not caught by the check measurements can still be detected later in the logical measurement as an erasure, which results in an increase in the erasure rate, but without necessarily impacting the fidelity of the logical outcome.

In our experiment, the logical measurement is implemented by measuring both cavities simultaneously to determine whether the dual-rail cavity qubit state is $\ket{01}$, $\ket{10}$, or a leakage state. 
For each cavity, we first perform a photon-number selective $\pi$-pulse to map the state of the cavity onto the transmon, flipping the transmon to $\ket{e}$ only if the cavity is in $\ket{1}$. This is followed by a standard dispersive readout of the transmon, with a total duration of $4.8~\mu\mathrm{s}$ for this sequence.
There are many options for implementing the logical measurement and our chosen mapping operation is optimized for detection of the dominant leakage state, $\ket{00}$ (supplementary text 3). 
In addition to leakage, any single ancilla error during the dual-rail logical measurement, including decoherence, control, or a transmon readout error, results not in a logical misassignment, but rather in an erasure assignment.
Importantly, as these photon-number-resolving measurements are non-demolition on the cavity photon number~\cite{elder_msmts_2020,curtis_single_shot_2021}, cavity measurements may be repeated to form multiple rounds of measurements to further suppress assignment errors. In this case, after each round of cavity measurements, we reset both transmons to $\ket{g}$ by applying a conditional $\pi$-pulse if either transmon was found in $\ket{e}$ and append a subsequent transmon check to confirm correct reset of the transmon. 
Finally, prior to the first logical measurement, we perform transmon readouts to verify that both are in the ground state, also assigning failures as an erasure.

Because the logical measurement protocol consists of one or more rounds of cavity measurements, each of which extracts one or more bits of information, a variety of decoding strategies can be used to assign a logical outcome: $0_\mathrm{L}$, $1_\mathrm{L}$, or erasure (supplementary text 5).
Each round of cavity measurements results in raw outcomes of the transmon readouts from which we assign one of the four possible physical outcomes: $\{``00", ``01", ``10", ``11"\}$ (see Fig.~1C). 
In the case of a single-round logical measurement, we can directly assign the logical state, $``01" \rightarrow 0_\mathrm{L}$, $``10" \rightarrow 1_\mathrm{L}$, and $\{``00", ``11"\} \rightarrow \mathrm{erasure}$. 
In the case of multiple rounds, we select a decoding strategy (supplementary text 5) to assign the most likely cavity state; in the event that a cavity state cannot be assigned, e.g. if measurement outcomes disagree, we declare an ambiguous outcome ``A" and is subsequently labeled as a logical erasure.
Finally, we discard the assigned erasures and compute the logical outcome $1_\mathrm{L}$ ($0_\mathrm{L}$), defined as $P_{1_\mathrm{L}\;(0_\mathrm{L})} = N_{1_\mathrm{L}\;(0_\mathrm{L})} / \left( N_{0_\mathrm{L}} + N_{1_\mathrm{L}} \right) $, where $N$ is the number of shots of the assigned logical state.
Given our mapping choice, each cavity measurement informs us whether or not the cavity is in $\ket{1}$. Higher leakage states in an individual cavity, such as $\ket{2}$, while less likely, are nonetheless assigned to outcome ``0''. As such, dual-rail cavity qubit leakage states, such as $\ket{02}$ or $\ket{20}$, will be assigned to ``00" and labeled as an erasure. 

We characterize our logical SPAM by quantifying logical misassignment errors and demonstrating the unique erasure detection capability of our logical measurement.
We perform our SPAM experiment by preparing logical states $\ket{0_{\mathrm{L}}}=\ket{01}$ and $\ket{1_{\mathrm{L}}}=\ket{10}$ and then performing a logical measurement. From this experiment, we extract two figures of merit. First, the logical misassignment error is defined as the fraction of counts assigned to $``10"$ ($``01"$) when preparing $\ket{01}$ ($\ket{10}$); this is represented by the assignment channel shown as pink arrows in Fig.~2A. 
From our data, shown as pink bars in Fig.~2B, we determine the logical misassignment error to be $(1.8\pm0.3)\times10^{-4}$, averaged over both state preparations.
Second, we can quantify the erasure fraction as the relative number of counts labeled $``00"$ or $``11"$ (channel shown in blue arrows in Fig.~2A), which we measure to be $(6.03\pm0.05)\times10^{-2}$, shown as blue bars in Fig.~2B. 
In addition to preparation errors, which are a logical leakage error, assigned erasures can also arise from any single error during the dual-rail measurement, including leakage in a cavity, transmon decoherence, or readout errors. 
False erasure assignments, for which the dual-rail cavity qubit is still in the codespace, are false positives and set a lower bound on the erasure fraction.

By intentionally preparing leakage states $\ket{00}$ and $\ket{11}$, we directly test the erasure detection capability of our logical measurement.
This assignment channel is shown with yellow arrows in Fig.~2A and the experimental results are shown in the right two panels in Fig.~2B. 
The critical figure of merit is what we call the leakage detection error, which is the fraction of false negatives, i.e. when we fail to detect a leakage event. 
Because undetected leakage errors are amongst the most damaging errors in a stabilizer code, it is important to keep the fraction of leaked qubits small~\cite{fowler_qubit_leakage_2013,ghosh_leakage_2013,bultink_protecting_2020,mcewen_removing_2021}.
In our system, photon loss from the codespace to $\ket{00}$ is by far the dominant leakage channel, and we measure a leakage detection error of $(7.7\pm0.3)\times 10^{-3}$.
As such, we can convert $>99\%$ of leakage errors to erasures.
The leakage detection error when preparing $\ket{11}$ is higher, at the $10^{-2}$ level but leakage to this state is much rarer, at least $1000\times$ less likely, requiring a cavity heating event.

We have developed a detailed model to simulate the state preparation and measurement protocol (supplementary text 9), finding close agreement between these simulations (plotted as green lines in Fig.~2B) and our experimental data.
Further, we have developed a simplified error model (supplementary text 10) to determine the physical error mechanisms that contribute to incorrectly assigned outcomes.
Using this model, we infer that outcomes assigned as erasures are often transmon $T_{1}$ events during readout, contributing between 30\%-50\% of the total error. Leakage detection errors are dominated by readout classification errors, contributing between 60\%-90\% of the total error. 

Finally, by performing the SPAM experiment with two rounds of measurements we observe an exceedingly low logical misassignment error of $(4\pm 2)\times 10^{-5}$ and leakage detection error of $(1.2\pm0.1)\times10^{-3}$.
We attribute this $\sim5\times$ improvement to the decoding strategy where we require agreement between both rounds of measurements (supplementary text 6).
Such a strategy reduces suppresses the effect of ancilla errors on logical misassignment and leakage detection error, instead assigning these outcomes as additional erasures.
We do observe this trade-off and report a higher erasure fraction, measured in our system to be $(17\pm0.1)\times10^{-2}$. 
Nevertheless, this experiment highlights the flexibility of our measurement protocol, with options for both multiple rounds of measurements and strategies both for reduced SPAM error or for reduced erasures.

\subsection*{Measuring bit-flip Error Rates}
\label{sec:bit_flip_results}

%TC:ignore
\begin{figure*}[h]
     \centering
     \includegraphics[width=0.6\textwidth]{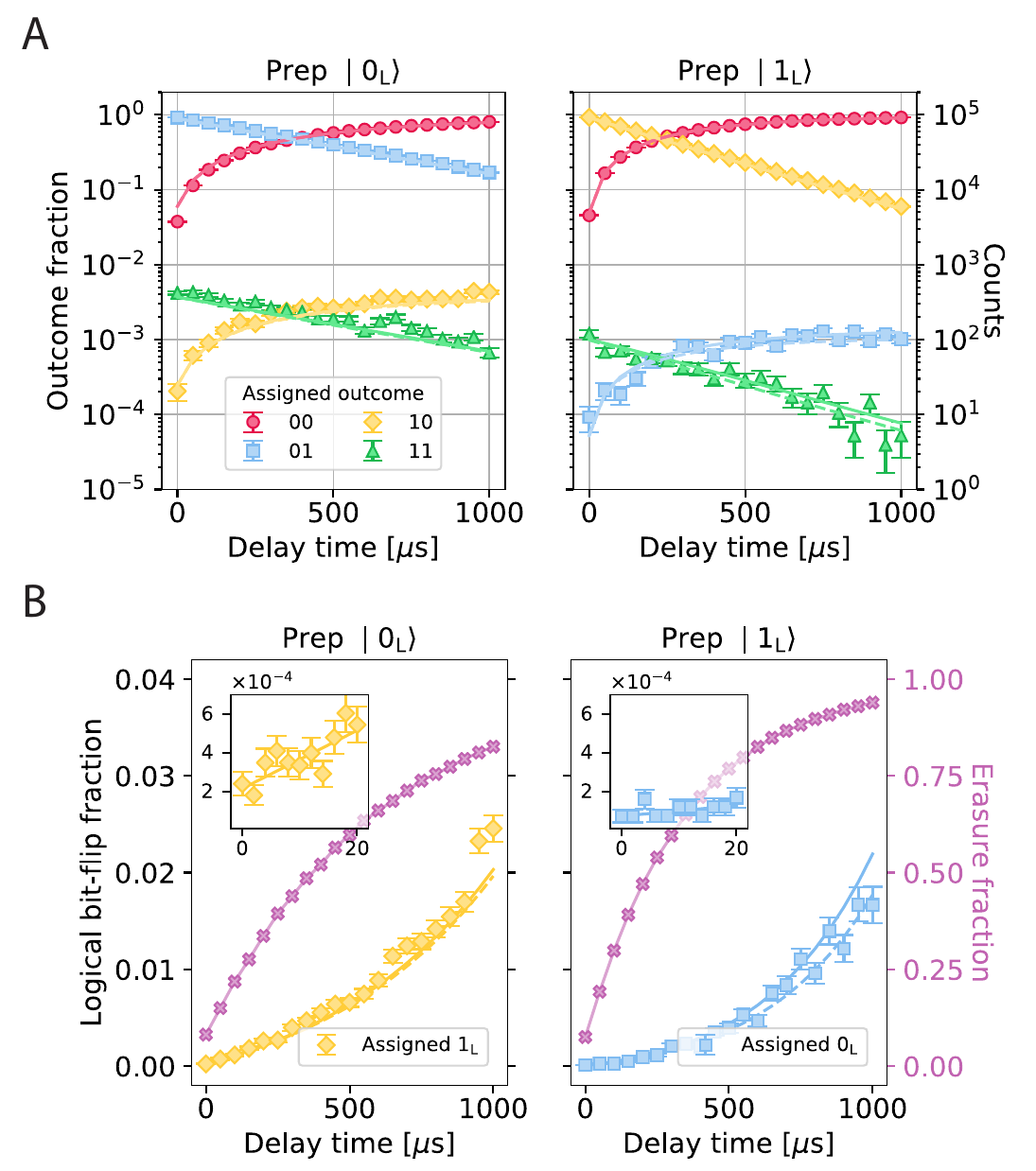}
    \caption{\label{fig:fig_3} 
    \textbf{Dual-rail bit-flip error measurement.}
    (A) Assigned physical state outcomes measured as a function of delay for the dual-rail cavity qubit prepared in $\ket{0_\mathrm{L}}$ (left) and $\ket{1_\mathrm{L}}$ (right).
    The occurrence of each measured outcome is shown both as a fraction (left axis) and as total counts (right axis). 
    The solid line corresponds to a simulation performed with measured parameters; while the dashed line is a simulation performed without any intrinsic bit-flip errors by setting the cavity heating rates to zero.
    (B) We compute the logical outcomes and plot the bit-flip error fraction when initializing in $\ket{0_\mathrm{L}}$ (left) and $\ket{1_\mathrm{L}}$ (right). The solid and dashed lines are generated using the same simulations as in Fig.~3A, showing the results with nominal and with no intrinsic bit-flip errors, respectively. The small difference between the two simulations provide evidence that intrinsic bit-flip errors are yet a small contribution to the apparent bit-flip error.
    The inset shows results for an additional experiment performed with delays out to 20~$\mu\mathrm{s}$.
    We perform a linear fit (solid line) and extract an apparent bit-flip error probability of $\left(1.5\pm0.4\right)\times 10^{-5}$ and $\left(3\pm1\right)\times10^{-6}$ in a microsecond, for $\ket{0_\mathrm{L}}$ (left) and $\ket{1_\mathrm{L}}$, respectively.
    }
\end{figure*}
%TC:endignore

Having demonstrated a high-fidelity logical measurement, we now use this tool to probe idling errors in our dual-rail qubit.
First, we study the dual-rail bit-flip error rate, defined here as the rate of transition from one logical state to the other: $\ket{01} \rightarrow \ket{10}$ and vice versa.
We expect these transitions to be exceedingly rare, caused by a double error via photon-loss in one cavity and photon-gain in the other cavity, thereby inducing a bit-flip in the logical state.
In our system, the cavities have relaxation times of $T_1 = 1/\kappa = 350~\mu \mathrm{s}$ and $592~\mu \mathrm{s}$ and thermal populations of $n_\mathrm{th} = 1.7\times10^{-4}$ and $3.5\times10^{-4}$ (supplementary text 1).
% decide if we need more discussion
In this experiment, the qubit is prepared in either of the logical states $\ket{01}$ or $\ket{10}$ and a logical measurement is performed after a variable delay; the results are shown in Fig.~3A.
At zero delay, the assigned outcomes are consistent with the $\ket{01}$ and $\ket{10}$ SPAM results, and we find the probability of assigning the bit-flip outcome to be $\sim10^{-4}$. 
As the delay increases, we observe the expected exponential increase in the $\ket{00}$ population due to decay errors out of the codespace. 
At longer delay times, the assigned outcomes tend toward the measurement outcomes when preparing $\ket{00}$ as shown in our SPAM experiment.
Our apparent bit-flip rate will have contributions both from the intrinsic bit-flip error rate, defined as the logical transition rate in the absence of logical assignment errors, and from leakage detection errors.
We compare our data with two simulations, one with measured hardware parameters and the other with no intrinsic bit-flip errors by setting the cavity thermal populations $n_{\mathrm{th}}=0$ (see Fig.~3).
We find both simulations largely agree with the measurement, suggesting that the bit-flip outcome counts are not dominated by actual state transitions in the cavities, but rather by measurement errors.

% paragraph about logical outcome
From the assigned physical outcomes we plot the logical bit-flip outcomes, shown in Fig.~3B.
Given the high-fidelity logical measurement, we also perform these experiments at short times ($20~\mu s$) in order to directly observe idling errors on the few microsecond scale, the relevant timescale for a logical entangling gate~\cite{tsunoda_error-detectable_2022}.
We perform a linear fit to the short-time data, and show that the probability of apparent logical errors grows on the order of 0.002\% in 1~$\mathrm{\mu \mathrm{s}}$, corresponding to a rate slower than 1/(66$\pm$16~ms). 
Our model suggests this rate is a lower bound for the intrinsic dual-rail bit-flip rate, as the intrinsic bit-flip contributes only $\sim0.5\%$ of the apparent bit-flip even at $20~\mathrm{\mu \mathrm{s}}$ (supplementary text 10).
Given the good agreement with the simulation, we infer that the intrinsic bit-flip rate should be non-exponential~\cite{teoh_dual-rail_2022} and extremely slow, with a probability $\sim n_{th}(\kappa t)^2$ at short times, that would be only a few parts per billion in a microsecond (supplementary text 11).

\subsection*{Measuring dephasing rates}

%TC:ignore
\begin{figure*}[h]
    \centering
     \includegraphics[width=\textwidth]{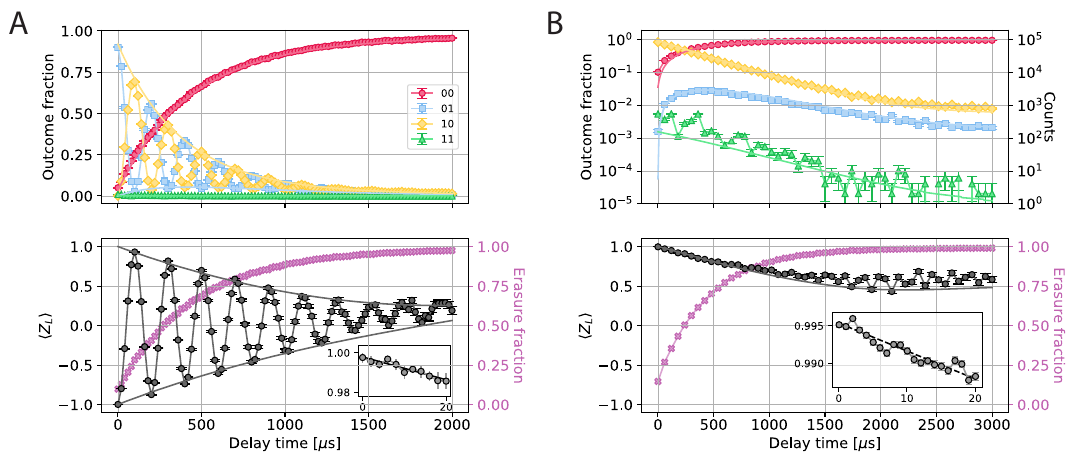}
    \caption{\label{fig:fig_4}
    \textbf{Dual-rail phase error measurement.}
    (A) Dual-rail qubit Ramsey and (B) echo experiment showing the assigned physical outcomes in the top panel; the logical outcome $\langle\hat{Z}_\mathrm{L}\rangle$ (black) and erasure fraction (purple) are shown in the bottom panel. Simulation results are shown in solid lines. The inset shows results from an additional experiment performed with shorter delays and the black line is a linear fit to the data. We extract phase-flip error probabilities of $(0.023 \pm 0.003)\%$ and $(0.019 \pm 0.001)\%$ in a microsecond for the Ramsey and echo experiments, respectively.
    }
\end{figure*}
%TC:endignore

We perform Ramsey and echo experiments in order to make a first estimate of the phase errors of the dual-rail cavity qubit.
This measurement is particularly important as the efficacy of the dual-rail encoding relies on preserving the low dephasing errors of the individual cavities when adding additional non-linear elements.
After state preparation in a logical dual-rail state, the Ramsey sequence is performed by applying a logical $\pi/2$-pulse, implemented via a parametric beamsplitter interaction~\cite{chapman_beamsplitter_2022,lu_beamsplitter_2023}, and, after a variable delay, applying a second logical $\pi/2$-pulse, before measuring the qubit. 
In the case of the echo measurement, an additional logical $\pi$-pulse is inserted mid-way during the delay. 
For the Ramsey experiment, the beamsplitter is intentionally detuned by a small (5 kHz) amount, giving rise to periodic behavior of the superposition of states. 

We present the results of the Ramsey and echo dephasing experiments in Fig.~4.
In both experiments we observe a decrease in the expectation value of $\hat{Z}_{\mathrm{L}}$ that is consistent with dephasing of the logical state. 
However, one difference that we observe in our data is an offset in $\langle \hat{Z}_\mathrm{L}\rangle$ at long times, where it is expected to decay to zero.  
This can be explained by leakage detection errors that bias the logical measurement at long times, due to the increased probability of being in the $\ket{00}$ state.
As such, we first perform coherence experiments at shorter times (up to $20~\mu\textrm{s}$), where this bias is suppressed by the smaller probability of being in $\ket{00}$. 
For these short-time Ramsey experiments, we implement an extended version of this experiment where, in addition to varying the Ramsey delay, we sweep the phase of the second $\pi/2$ pulse (supplementary text 8).
From these experiments we extract dephasing rates from the linear slope of $\Gamma_\phi^\mathrm{R} =
1/(2.2\pm0.2 \mathrm{ms})$ and $\Gamma_\phi^\mathrm{E} =1/(2.7\pm0.2 \mathrm{ms})$, respectively for Ramsey and echo.
From these quantities, we infer phase-flip error probabilities, $p_\phi = \Gamma_\phi t / 2$ for time $t$, for Ramsey and echo of $p_\phi^\mathrm{R} = (0.023 \pm 0.003)\%$ and $p_\phi^\mathrm{E} = (0.019 \pm 0.001)\%$ in microsecond, respectively (supplementary text 7).
Using these measured dual-rail dephasing rates in our simulations of the Ramsey and echo experiments, we indeed find good agreement with the long-time dephasing behavior.
The reported dual-rail dephasing rates are likely still dominated by extrinsic sources, introduced by finite heating rates and state transitions of the dispersively coupled nonlinear elements, which in our system are two measurement transmons and three Josephson coupling elements (see Fig.~1B).
This initial bound on phase-flip rate measured here is encouraging, and is approximately a factor of six slower than the erasure rate.
Further improvements are achievable, for example, by
decreasing the heating rates in these elements and detecting their residual state transitions as erasures.

\section*{Conclusion}

We have developed a logical measurement with built-in erasure detection with logical misassignment on the order of $10^{-4}$ and with efficiency for detecting leakage to $\ket{00}$ in excess of $99\%$.
Our measurements show that the logical bit-flip rate is at least two orders of magnitude smaller than the erasure rate, and that the logical phase-flip rate is a factor of 6 smaller, providing initial confirmation that dual-rail cavity qubits can have a favorable hierarchy of errors in the idling state. 
Further improvements in the measurement and characterization of dual-rail qubits should also be possible.
First, while we have already demonstrated low error rates, a more detailed analysis of the physical phenomena and the ultimate limits on dephasing in the dual-rail qubit are key questions and the topic of ongoing investigation~\cite{Winkel}. 
Second, this logical measurement can be used as a syndrome measurement for a stabilizer code and in such an application it will be important to minimize the erasure fraction. Our analysis (supplementary text 10) indicates that a significant fraction of erasure assignments are false positives and can be improved by further hardware optimizations.
Finally, asymmetric errors in erasure detection can introduce a bias in the measured outcomes, particularly for longer circuits where the erasure fraction is high, and so ways to estimate and mitigate these effects should be pursued. 

The ability to perform high-fidelity preparation and measurement is a vital tool for all further characterization of dual-rail erasure qubits, including other critical functions such as the joint parity operation~\cite{teoh_dual-rail_2022, Stijn} to perform mid-circuit erasure detection. This function is also the basis for two-qubit entangling gates~\cite{tsunoda_error-detectable_2022} and combining it with the capabilities shown in this work would complete a full toolbox of error-detected quantum operations for the dual-rail cavity architecture. 
Leakage detection in a single dual-rail qubit is an important first step on the path to correct erasures, which will require a next-level code made by concatenating many dual-rail qubits. Further exploration and confirmation of the expected error hierarchy, during not only idling but operations,  could therefore enable a new and faster path to fault-tolerant computing. 
Finally, the ability to convert errors into measurable erasures also creates near-term opportunities for error mitigation and improved fidelity in short-depth circuits. 
\emph{Note:} During completion of this manuscript, we became aware of related work on the dual-rail encoding using tunable transmon qubits~\cite{levine_dual_rail_transmons_2023}.\\

\textbf{Acknowledgments} We thank the broader mechanical and control system groups at QCI, particularly Caleb Clothier, Richard Chamberlin, Michael Maxwell, and Charles Wehr. 
\textbf{Funding:} This research was supported by the U.S. Army Research
Office (ARO) under grant W911NF-23-1-0051,
and by the U.S. Department of Energy, Office of Science,
National Quantum Information Science Research Centers,
Co-design Center for Quantum Advantage (C2QA)
under contract number DE-SC0012704. The views and
conclusions contained in this document are those of the
authors and should not be interpreted as representing
official policies, either expressed or implied, of the ARO
or the U.S. Government. The U.S. Government is
authorized to reproduce and distribute reprints for
Government purpose notwithstanding any copyright
notation herein.

\textbf{Author contributions}
Conceptualization, methodology, formal analysis, and investigation: K.C., T.S, H.M., T.C., P.L., N.M., S.O.M., A.N..
Supervision: S.P., S.M.G., S.H.M, R.J.S..
Writing:  K.C., T.S, H.M., J.D.T., P.W., R.J.S..
All authors were part of discussion, review, and editing. 
%Conceptualization
%Methodology
%Software
%Validation - don't use 
%Formal analysis 
%Investigation
%Resources - dont use
% Data Curation - don't use
%Writing - Original Draft 
%Writing - Review & Editing 
%Visualization - don't use 
%Supervision - don't use
% Project administration - don't use
% Funding acquisition - don't use

\textbf{Competing interests}
R.J.S. and L.F. are founders and shareholders of Quantum Circuits, Inc (QCI).  S.P. and S.M.G. receive consulting fees and/or are equity holders in QCI.
Authors affiliated with QCI have financial interest in the company. 
The provisional patent "Dual-rail qubits based on superconducting resonant cavities and their application for quantum error correction" has been filed (63/433,411) by R.J.S., S.M.G., S.P., J.D.T.,S.J.G, S.H.X., B.~Chapman, J.W.O.G., A. Maiti, Y.L, W.D.K., N.~Thakur, T.T., P.W.  with Yale University. 
%
% \textbf{Data and materials availability:} Data is available at Dryad. 
%TC:endignore

\textbf{Data availability:} 
The data that support the findings of this study are available from the corresponding author upon reasonable request.

\textbf{Code availability:} 
The code that supports the findings of this study are available from the corresponding author upon reasonable request.

\textbf{Additional information\\}
\textbf{Supplementary information} Supplementary Information is available for this paper.

\textbf{Correspondence and requests for materials} should be addressed to K.C. or R.J.S.

\clearpage

%TC:ignore
\bibliography{ms}
%TC:endignore

\clearpage
\widetext

%TC:ignore
\input{supplement}
%TC:endignore

\clearpage
\end{document}

%% file: supplement.tex
\section*{Supplementary Materials}

% \section*{Materials and Methods}

\section{System Properties}
\label{sec:system_params}

% https://netorg928361.sharepoint.com/:x:/s/Labteamandresults/EXQNfZuXB-pKnsocGMMgCN0B801gbGF2Ubq3KbJ2cnifxw?e=4n2iyV

\begin{center}
\begin{tabular}{| c | c | c | c || c | c | }
\hline
& Description & Parameter & Units & $\textbf{Alice}$ & $\textbf{Bob}$ \\ 
 \hline\hline
 \rule{0pt}{2.6ex}
$\textbf{System}$ & Cavity-transmon cross-kerr & $\chi_{\mathrm{cm}}/2\pi$ & MHz & -2.14 & -3.72 \\[1mm] 
 & Cavity-coupler cross-kerr & $\chi_{\mathrm{cc}}/2\pi$ & kHz & -99 & -110\\[1mm] 
\hline
  \rule{0pt}{2.6ex}
$\textbf{Cavity}$  & Relaxation time & $T_{1}^{\mathrm{c}}$ &$\mu$s& 592 & 350\\
[1mm]  
  & Dephasing time, Ramsey & $T_{\mathrm{2R}}^{\mathrm{c}}$ &$\mu$s& 768 & 543\\[1mm]
  & Dephasing time, Echo & $T_{\mathrm{2E}}^{\mathrm{c}}$ &$\mu$s& 932 & 588\\[1mm]
& Thermal population & $n_\mathrm{\mathrm{th}}^{\mathrm{c}}$ &\%& 0.035\;* & 0.017\;* \\ [1mm]
& Self-kerr & $K/2\pi$ & kHz & -5.79 & -11.02 \\ [1mm]
& Resonance frequency & $\omega_{\mathrm{c}}/2\pi$ & GHz & 7.16 & 6.76 \\[1mm]
\hline
\rule{0pt}{2.6ex}
$\textbf{Transmon}$ & Relaxation time & $T_{1}^{\mathrm{t}}$ &$\mu$s& 125 & 179\\[1mm] 
  & Dephasing time, Ramsey & $T_{\mathrm{2R}}^{\mathrm{t}}$ &$\mu$s& 92 & 206\\[1mm]
  & Dephasing time, Echo & $T_{\mathrm{2E}}^{\mathrm{t}}$ &$\mu$s& 144 & 303\\[1mm]
  & Thermal population & $n_{\mathrm{th}}^{\mathrm{t}}$ &\%& 1 & 1 \\[1mm]  
  & Anharmonicity & $\alpha/2\pi$& MHz & -184.63 & -185.71\\[1mm]
  & Resonance frequency & $\omega_{\mathrm{t}}/2\pi$ & GHz & 4.83 & 4.92\\
  & Selective $\pi$-pulse sigma & $\sigma_{\mathrm{s}}$ &ns& 300 & 300 \\[1mm]
  & Unselective $\pi$-pulse sigma & $\sigma_{\mathrm{us}}$ &ns& 16 & 16 \\[1mm]
  \hline
  \rule{0pt}{2.6ex}
$\textbf{Readout}$ & Duration & $t_{\mathrm{M}}$ &$\mu$s & 2.0 & 3.6 \\[1mm]
 & Processing time & $t_{d}$ &$\mu$s& 0.4 & 0.4 \\[1mm]
  & Transmon relaxation time during readout& $T_{1\;\mathrm{RO}}^{\mathrm{t}}$ &$\mu$s& 80\;* & 100\;*\\[1mm] 
  & Transmon Thermal population during readout & $n_{\mathrm{th\;RO}}^{\mathrm{t}}$ &\% & 2\;* & 11\;*\\[1mm]
  & Prob. of misassigning $|g\rangle$ as $|e\rangle$ & $p_{\mathrm{gE}}$ & $\times10^{-3}$ & 1.00\;* & 3.22\;*\\[1mm]
  & Prob. of misassigning $|e\rangle$ as $|g\rangle$ & $p_{\mathrm{eG}}$& $\times10^{-3}$ & 4.03\;* & 1.18\;*\\[1mm]
  \hline
\end{tabular}
\end{center}

* See subsequent sections for details of these measurements.

\subsection{Measurement of Cavity Thermal Population}
\label{sec:cavity_nth_msmt}

We bound the cavity thermal population $n^c_\mathrm{th}$ by conducting a modified bit-flip experiment. This is particularly important as cavity heating is expected to be the main mechanism for actual cavity transitions back into the codespace after a decay event. The measurement is performed as follows: the dual-rail qubit is prepared in the physical state $\ket{00}$, and after a variable time-delay the resulting state is measured to determine whether either cavity experienced a heating event and gained a photon. 

\begin{figure*}
     \centering
     \centering
     \includegraphics[width=0.5\textwidth]{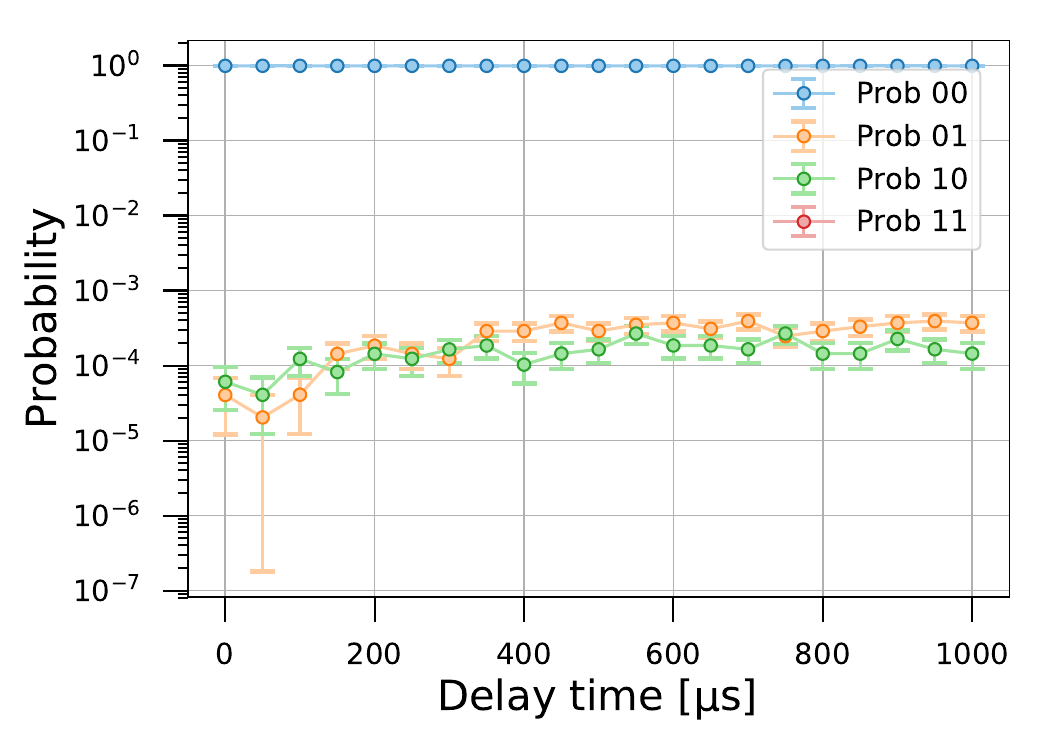}
     % \caption{Cavity $n_\mathrm{th}$ plot}
    \caption{\label{fig:nth} 
    (a) \textbf{Measurement of cavity $n_{th}$} Plot of physical state probability as a function of delay time after being prepared in the ground state $\ket{00}$. We perform this experiment to measure the cavity heating, which sets the limit of the intrinsic bit-flip rate. Given that misassignment dominates over the cavity heating, we use two end-of-line measurements to suppress the measurement error. We interpret the long-time behavior where the excited states ($\ket{01}$ and $\ket{10}$) saturate as bounding the cavity heating rate. 
}
\end{figure*}

The results of the measurement are shown in Fig.~\ref{fig:nth}. Given that we expect errors associated with the measurement (misassignment errors) to dominate over those due to cavity heating, we show results using two-rounds of measurements to suppress ancilla errors that cause logical misassignment errors. We use the same decoding protocol described in the Main text and also in supplementary text 6: both rounds of measurements must agree otherwise an erasure is assigned. With the two rounds measurements, the excited states ($\ket{01}$ and $\ket{10}$) show a behavior typical of heating, where the probability of a photon in either cavity increases and then saturates at long time. We bound the cavity $n^c_\mathrm{th}$ to be less than $4 \times 10^{-4}$ by examining the long time, saturated behavior. We use this quantity to estimate the intrinsic bit-flip errors in supplementary text 11.
% This value for the average cavity occupation suppresses the rate of actual transitions between logical states, which would then be estimated to be $<500$ ppm in one microsecond, or a remarkably low rate of 1/($\sim2000$ seconds) in the absence of other intrinsic bit-flip processes, and assuming ideal state preparation.  

%%%%% SEPARATE FIGURES FOR ARXIV

%%%%% SEPARATE FIGURES FOR ARXIV
\begin{figure*}
     \centering
     \centering
     \includegraphics[width=0.5\textwidth]{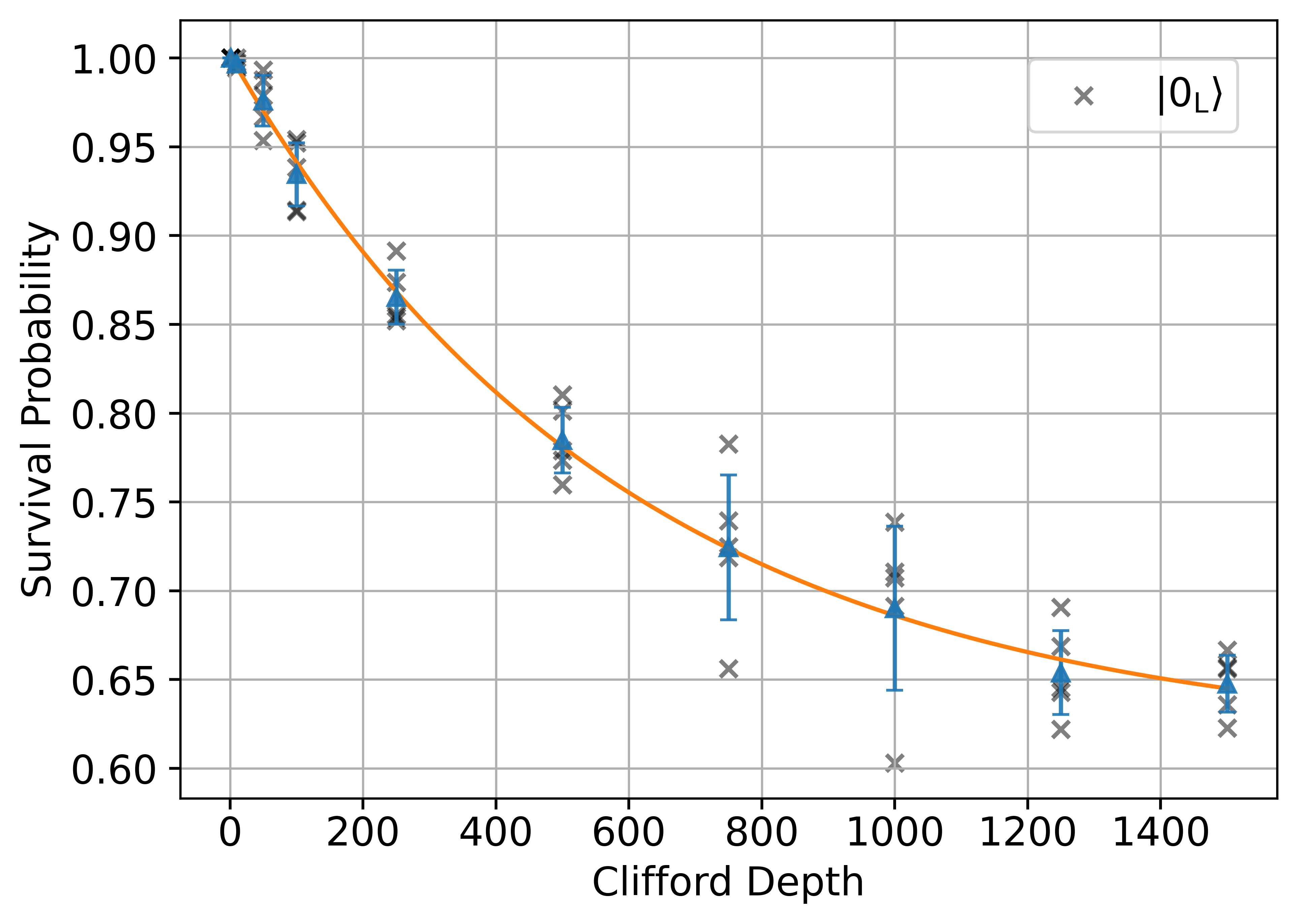}
     % \caption{}
    \caption{\label{fig:rb} 
    \textbf{Single-qubit randomized benchmarking (RB) measurement} We perform single-qubit randomized benchmarking in the dual-rail basis with five seeds (represented by x) at each Clifford Depth. The single-qubit gates are used in this paper when measuring the phase error rates. The mean (triangle) measured survival probability is fit to a exponential and from this we extract the error-per-Clifford as $8.4 \times 10^{-4}$. 
}
\end{figure*}

%%%%% COMBINED FIGURES FOR SCIENCE
% \begin{figure*}
%      \centering
%       \begin{subfigure}[b]{0.48\textwidth}
%          \centering
%          \includegraphics[width=\textwidth]{figs/figs_appendix/230530_nth.pdf}
%          \caption{Cavity $n_\mathrm{th}$ plot}
%          \label{fig:nth}
%      \end{subfigure}
%      \begin{subfigure}[b]{0.48\textwidth}
%          \centering
%          \includegraphics[width=\textwidth]{figs/figs_appendix/survival_probability.png}
%          \caption{}
%          \label{fig:rb}
%      \end{subfigure}
%     \caption{\label{fig:nth_and_rb} 
%     %
%     (a) \textbf{Measurement of cavity $n_{th}$} Plot of physical state probability as a function of delay time after being prepared in the ground state $\ket{00}$. We perform this experiment to measure the cavity heating, which sets the limit of the intrinsic bit-flip rate. Given that misassignment dominates over the cavity heating, we use two end-of-line measurements to suppress the measurement error. We interpret the long-time behavior where the excited states ($\ket{01}$ and $\ket{10}$) saturate as bounding the cavity heating rate. 
%     %
%     (b) \textbf{Single-qubit randomized benchmarking (RB) measurement} We perform single-qubit randomized benchmarking in the dual-rail basis with five seeds (represented by x) at each Clifford Depth. The single-qubit gates are used in this paper when measuring the phase error rates. The mean (triangle) measured survival probability is fit to a exponential and from this we extract the error-per-Clifford as $8.4 \times 10^{-4}$. 
% }
% \end{figure*}

\subsection{Measurement of Readout Properties}
\label{sec:RO_properties}

% Describe how we extract T1_RO and nth_RO. Include plots from t1_vs_nbar experiment?

% Describe filter for extracting p_miss

We measure the readout properties of our transmons with two experiments. 

First, we measure $T_{1}$ during readout by performing a ``$T_{1}$ vs. $\Bar{n}$" experiment on each of the ancilla transmons.
We perform this experiment with the insight that due to the phenomenological effect where the transmon lifetime gets suppressed at certain readout amplitudes, there may be an increased  probability of a transition event during readout~\cite{thorbeck2023readoutinduced}. As the measured system properties are the input for both of our models (see supplementary texts 9 and 10), our aim is to characterize all aspects of our system accurately, including any changes to coherences during readout.

In the experiment, we choose an array of values for the pulse amplitude; for each pulse amplitude we perform a $T_{1}$ experiment on the transmon (excite the transmon to $\ket{e}$ and measure the probability of it remaining in $\ket{e}$ for variable delays between preparation and measurement) while driving the readout tone at the chosen amplitude. We  fit the data at each point to a decaying exponential and extract $T_{1\;RO}^{t}$ at the nominal readout amplitudes.
%; the results of this experiment for both the Alice and Bob ancillae are presented in Fig.~\ref{fig:ancilla_T1_ro}. From these plots we read off the $T_{1\;RO}^{(t)}$ at the readout amplitudes we use.

% \begin{figure}[h]
% \centering
% \includegraphics[width=0.9\textwidth]{figs/figs_appendix/ancilla_T1_RO.pdf}
% \caption{Results of the ``$T_{1}$ vs. $\Bar{n}$" experiments, performed as $T_{1}$ experiments while driving the readout tone at varying amplitudes for both Alice (right) and Bob (left). Dashed line indicates the readout amplitude that was used in the data presented in this paper.}
% \label{fig:ancilla_T1_ro}
% \end{figure}

We also know that the transmons may suffer some residual heating during readout compared to their ambient thermal populations. In order to extract a value for the effective transmon thermal population during readout, $n_{th\;RO}^{t}$, we examine the offsets of the fits to the data as described above.

% We also know that the transmons may suffer some residual heating during readout compared to their ambient thermal populations. In order to quantify the thermal population of the transmons during readout, $n_{th\;RO}^{(t)}$, we perform an experiment where we prepare the transmon in $\ket{g}$ and then repeatedly measure it 10 times; we repeat the experiment many times. In this experiment the signature of a transition is different from the signature of a misassignment, as in the former case consecutive measurements will have the same outcome and in the latter case they will not.
% We average over the traces of all runs of the experiment, and from this extract the probability of the transmon transitioning from $\ket{g}$ to $\ket{e}$ during readout, denoted $p_{g\rightarrow e}$. 

% The functional form of the probability of this event is
% \begin{equation*}
%     p_{g\rightarrow e} \ = 1 - e^{-\frac{n_{th}\cdot t_{M}}{T_{1}}}
% \end{equation*}
% and since we have measurements of all other quantities in this expression, we solve for $n_{th\;RO}^{(t)}$ to get a close estimate of this parameter for each of Alice and Bob's ancilla transmons. \\

Second, in order to extract $p_{gE}$ and $p_{eG}$, we perform an experiment where we prepare the transmon in $\ket{g}$ and then repeatedly measure it 10 times; we repeat the experiment $10^{5}$ times. In this experiment the signature of a transition is different from the signature of a misassignment, as in the former case consecutive measurements will have the same outcome and in the latter case they will not.
By applying a filter over different measurement records, we are able to extract the probability of missassigning $\ket{g}$ as $E$ and $\ket{e}$ as $G$.
% We average over the traces of all runs of the experiment, and from the signature traces of misassignments we extract the probabilities of misassigning $\ket{g}$ as $E$, and $\ket{e}$ as $G$.  

% \subsection*{Dual-Rail State Preparation}
% \label{sec:DR_state_prep}

\section{Single-Qubit Gate Benchmarking}

In this paper, the single-qubit gates are used when measuring the phase error rates. We characterize the quality of the gates using single-qubit randomized benchmarking~\cite{Magesan_2011, Qiskit} in the dual-rail space, detecting and postselecting out erasures using a method similar to that used in~\cite{lu_beamsplitter_2023}. As shown in Fig.~\ref{fig:rb}, the resulting error-per-Clifford (EPC) from this measurement is $8.4 \times 10^{-4}$. This rate is obtained by generating five seeds across Clifford depths ranging from 0 to 1500. The physical gates are comprised of 50-50 beamsplitters ($\pi$/2) and SWAPs ($\pi$). We plot the survival probability, meaning the prepared state is returned, as a function of Clifford depth. 
%This data is fit to an exponential, from which we extract the EPC and single-qubit fidelity estimates of 0.08\% and 0.16\% for $\pi$/2 and $\pi$-gates, respectively. 

\section{cQED Toolbox for Bosonic Operations}
\label{sec:toolbox}

All operations necessary for state preparation, measurement, and reset of the dual-rail cavity qubit are implemented using the now well-established cQED toolbox of operations~\cite{Blais_2021}. This includes all the operations necessary for state preparation, measurement, and system reset of the dual-rail qubit.

In the subsequent sections we explain each element of the dual-rail qubit's toolbox in some detail.  

\subsection{Mapping}
\label{sec:mapping}

Measurement of the cavity state is realized by mapping the state to the cavity's ancilla transmon, before performing a subsequent readout of the transmon in order to infer the state of the cavity.

In the data presented in this paper, the mapping was implemented by way of a selective $\pi$-pulse on the ancilla transmon; the pulse was executed as a truncated Gaussian pulse. Specifically, this means that we perform a $\pi$-pulse on the transmon--conditioned on the cavity being in the Fock state $\ket{n}$:
% \begin{equation*}
%     X_{n}^{s} = |n\rangle\langle n|\otimes \hat{\sigma}_{x} + \sum_{m\neq n} |m\rangle\langle m|\otimes \mathds{1}
% \end{equation*}
\begin{equation*}
    X_{n}^{s} = |n\rangle\langle n|\otimes \hat{\sigma}_{x} + \sum_{m\neq n} |m\rangle\langle m|\otimes \hat{I}%\mathds{1}
\end{equation*}

The ancilla is dispersively coupled to the cavity with its resonance frequency given by $\tilde{\omega}_{q} = \omega_{q}-n\chi$, where $\omega_{q}$ is the transmon's bare resonance frequency, $n$ is the number of photons in the cavity, and $\chi$ is the cavity-transmon cross-Kerr. This interaction dictates our two main degrees of freedom in tailoring the selective-pulse mapping for our dual-rail protocol. 

First, we choose the pulse duration. The pulse, in the time domain, is parameterized as $chop\cdot\sigma$, where $\sigma$ is the standard deviation of the Gaussian pulse that is truncated at $\pm chop/2$ standard deviations of the Gaussian. In the frequency domain, the bandwidth of the pulse is given by $1/(2\pi\sigma)$, and as such given the cross-Kerrs of our subsystems, we choose a pulse sigma that is long enough such that the pulse's bandwidth is narrow enough to be sufficiently selective on one of the Fock states $\ket{n}$. We choose our pulse sigmas by numerically computing the unselectivity error of the pulse as $Err = 1-|\langle 1,g|X_{n=0}^{s}|1,g\rangle|^{2}$ taking $\chi$ and $\sigma$ into account.

Second, we can choose the frequency of the selective-pulse for the mapping. Namely, we can detune the pulse by $n\chi$, where $n$ is any integer and will lead to the transmon being rotated conditioned on the cavity being in the Fock state $\ket{n}$. In our dual-rail protocol, the dominant leakage mechanism is cavity decay events to the $\ket{00}$ state. As such, in order to minimize the misassignment of leakage states to the codespace, we detune our selective pulses by $\chi$ such that we obtain the following cavity-to-transmon mapping:
\begin{equation*}
    \begin{aligned}
        \ket{0}&\rightarrow\ket{g}\\
        \ket{1}&\rightarrow\ket{e}\\
    \end{aligned}
\end{equation*}
and the dual-rail leakage state $\ket{00}$ maps to $\ket{gg}$ rather than to the more error-prone state (for readout purposes) $\ket{ee}$.

\subsection{Transmon Readout}
\label{sec:MT_readout}

Transmon readout in the dual-rail protocol is realized by way of standard dispersive readout~\cite{Blais_2004}.

% In tailoring the transmon readout to the dual-rail protocol, we have two primary degrees of freedom: the readout pulse amplitude, and the readout pulse's duration. 
In tailoring the transmon readout to the dual-rail protocol, one primary degree of freedom that we have is the readout pulse's duration. 

% First, in order to decide on an appropriate amplitude for the readout pulses, we perform a ``T1 vs nbar" experiment. In the experiment, we choose an array of values for the pulse amplitude. For each pulse amplitude we perform a $T_{1}$ experiment on the transmon (excite the transmon to $\ket{e}$ and measure the probability of it remaining in $\ket{e}$ for variable delays between preparation and measurement) while driving the readout tone at the chosen amplitude. We perform this experiment in order to determine the effective $T_{1}$ during readout at the chosen amplitude, where we know that the transmon $T_{1}$ might be significantly suppressed at certain readout amplitudes (denoted $T_{1\;RO}^{(t)}$)~\cite{thorbeck2023readoutinduced}.
% We also check how much the chosen amplitude heats the transmon during readout (denoted $n_{th\;RO}^{(t)}$), a number which we extract by fitting the ``T1 vs nbar" experiment curves to an exponential and extracting the offset. 

% Second, having chosen a readout amplitude, we also choose the duration of the readout pulse. 
In choosing the readout duration we seek to optimize the trade-off between readout separation and transitions during readout--that are inversely parameterized by the readout duration $t_{M}$.
Specifically, the readout separation, or the distinguishability of the states in the IQ plane is characterized by the parameter $\Bar{\mathrm{I}}_{\mathrm{m}}/\sigma$ that scales as $\sqrt{t_{M}}$; as such, the separation increases, and the classification error accordingly decreases as $t_{M}$ increases. However, transmon decay and heating events scale as $t_{M}/T_{1\;RO}^{t}$ and $n_{th\;RO}^{t}\cdot t_{M}/T_{1\;RO}^{t}$, respectively, and as such the probability of a transmon transition during readout (i.e. a non-QND measurement) increases as $t_{M}$ increases. We quantify these probabilities given our system parameters and choose $t_{M}$ accordingly in order to balance this trade-off.

\subsection{Conditional Reset}
\label{sec:conditional_reset}

In our dual-rail protocols, we always immediately reset the transmon back to the ground state $\ket{g}$ subsequent to it being rotated to $\ket{e}$ by any component of the protocol. 

We perform this reset conditioned on having learned that the transmon is in $\ket{e}$, i.e. following a transmon readout, and implement it via an unselective $\pi$-pulse. An unselective pulse, in contrast with the selective pulse described in the Mapping section, is a short pulse in the time-domain or, equivalently a broad pulse in the frequency-domain that is thus intended to be unselective on the cavity photon-number.

We tailor this unselective $\pi$-pulse for the purpose of conditional reset with the same considerations in mind as when tailoring the selective pulse as described in the Mapping section. 

First, we choose the pulse duration to be short enough such that it is sufficiently unselective given the system's $\chi$, with the error computed numerically as $\epsilon = 1-|\langle 1,e|X_{n=0}^{us}|1,g\rangle|^{2}$, where $us$ denotes the unselective $\pi$-pulse. Specifically we want $1/(2\pi\sigma)$ to be large enough such that the pulse is broad enough in the frequency domain to rotate the transmon from $\ket{e}$ back to $\ket{g}$ regardless of whether the cavity is in the state $\ket{0}$ or $\ket{1}$. 

Second, similar to the case of the selective pulse, we have a choice in the detuning of the unselective pulse. 
% While ideally the pulse would be sufficiently unselective such that it would rotate the transmon regardless of its resonance condition, in practice given our system's large cross-Kerrs and power limitations on how small the pulse $\sigma $ can be -- we know that we will suffer some selectivity error in rotating the transmon when the unselective pulse if off-resonant. 
For the dual-rail protocol, we detune the unselective reset pulse by $\chi$, so that it is on resonance when the cavity is in the state $\ket{1}$. We do this, since given our mapping choice (as described in the Mapping section) the transmon is rotated to $\ket{e}$ when the cavity is in the state $\ket{1}$. Consequently, in our logical measurement protocol, when the transmon is measured to be in $\ket{e}$ after a mapping of the cavity state, the cavity is mostly likely in $\ket{1}$ and therefore the conditional reset is most likely to succeed if it is resonant when the cavity is in $\ket{1}$.

\subsection{Transmon Check}
\label{sec:MT_check}

The post-reset check is realized by way of a transmon readout, exactly as described in the Transmon Readout section, with the only difference being in the logic of the way the measurement outcome is processed. 

In the dual-rail protocol, we perform these transmon checks at points in the protocol where a measurement outcome of $\ket{e}$ will indicate that an error has occurred. As such, in the cases where the measurement outcome of a check measurement is $\ket{e}$, that shot, or run of the experiment is discarded in post-selection, with this run of the experiment contributing statistically to runs where an erasure was detected. 

\subsection{Cavity-State Initialization}
\label{sec:OCP}

% expand

In the data presented in this paper, cavities were initialized in $\ket{1}$ using an optimal control pulse (OCP)~\cite{Heeres_2017_OCP}, with a duration of $1\;\mu s$.

There are other methods of cavity-state initialization, such as sideband operations~\cite{Rosenblum2018} to load a photon from the transmon to the cavity, or measurement-based approaches~\cite{elder_msmts_2020}. Exploring other such methods and optimizing state preparation for the purpose of the dual-rail qubit will be explored in future work.

% \subsubsection*{Transmon-Cavity SWAP}
% \label{sec:TC_SWAP}

% In our dual-rail experiments we perform many runs of each experiment in order to accumulate statistics and sufficiently resolve the outcomes that we want to measure. As such, we need an efficient way to reset the system, i.e. both the dual-rail's cavities and transmons to the ground state between each run. 

% In the data presented in the paper, this reset was implemented via a sideband SWAP operation that implements the unitary $U_{SWAP} = \outerproduct{g1}{e0} + \outerproduct{e0}{g1}$, and has a duration of $1.5\;\mu s$. 

% For system reset, this operation is simultaneously implemented on both the dual-rail's subsystems, after checking that both transmon are in $\ket{g}$. In the case where the cavity is in $\ket{1}$, the excitation is swapped to the ancilla, and a subsequent measurement and  $\pi$-pulse to the transmon to reset it to the ground state result in both modes being reset to the ground state. In the case where the cavity was in $\ket{0}$ the transmon will simply remain in $\ket{g}$, indicating that both modes are in the ground state. 

\section{State Preparation and Measurement Protocols}
\label{sec:protocols}

Two fundamental components of the dual-rail qubit operations are state preparation and measurement. 
% Note that here we refer to the dual-rail measurement as a logical measurement, meaning it is a projective measurement of the logical $\hat{\sigma}_{z}$ operator. 
We construct each of these protocols using the elements in our toolbox as described in the cQED Toolbox for Bosonic Operations section.

\subsection{State Preparation}

In this section, we provide further details of our state preparation protocols. 
Initialization of a physical state in each cavity consists of first resetting each cavity and its ancilla transmon to their ground states. The dual-rail cavity qubit state preparation then requires subsequent initialization in a joint photon-number state of the two cavities by initializing each of the cavities in either $\ket{0}$ or $\ket{1}$. In the remainder of the section we outline each of these reset and initialization steps.\\

% \textbf{Reset to $\ket{00}$}.
Every experiment begins with a reset to $\ket{00}$ which we implement using a measurement-based protocol.
% Specifically, our protocol for preparing a cavity in $\ket{0}$ is as follows.
We perform a cavity measurement that consists of mapping the cavity-state to the transmon using a selective pulse followed by a readout of the ancilla transmon. As described in the Mapping section, this pulse is detuned by $\chi$ in order to map the intended outcome of $\ket{0}$ to $\ket{g}$ and mitigate assignment errors. 
If the measurement outcome is $G$, indicating that the cavity is in $\ket{0}$, the cavity measurement is repeated $N$ times, until we get a string of consecutive $G$ outcomes confirming that the cavity is indeed in $\ket{0}$. Here we require a string of six successful outcomes.
If the measurement outcome is $E$, indicating that the cavity not in $\ket{0}$, the transmon is reset and we add a short delay of $10~\mu \textrm{s}$ before restarting the protocol again. 

Represented as a circuit diagram, our $\ket{0}$ preparation protocol is:

\begin{center}
\begin{quantikz}
\lstick{Cavity}   & \ctrl{1}\gategroup[2,steps=3,style={dashed,
rounded corners,%fill=blue!10, 
inner xsep=2pt},
background,label style={label position=below,anchor=
north,yshift=-0.5cm}]{Cavity $\ket{0}$ check measurement: repeat until $G^{\otimes 6}$} &  \qw  & \qw     & \qw & \qw  \\
\lstick{Transmon} & \gate{X_{n=1}^{s}}  & \meter{G/E/...} & \gate[cwires=1]{X_{n=1}^{us}} & \qw
\end{quantikz}
\end{center}

% \textbf{$\ket{1}$ initialization}.
Initialization into the logical dual-rail qubit basis requires loading a photon into one of the two cavities while preparing the other cavity in the ground state. There are a number of protocols available in the bosonic cQED toolbox to accomplish this and we elect to use optimal control pulses (OCP)~\cite{Heeres_2017_OCP} as described in the Cavity-State Initialization section. 

% \textbf{State verification via check measurements}.
We conclude the state preparation with a series of verification measurements on the transmon or the cavity. These check measurements provide the ability to flag suspected failures in the state preparation and provide the option to postselect them out so they do not contribute to errors during the algorithm. In our experiment, we considered two types of check measurements. 
For the first type, we perform transmon check measurements after the OCP to verify that both transmons are in the $\ket{g}$ state. 
The preparation of $\ket{1}$ or $\ket{0}$ with this protocol are described by the following circuit:
\begin{center}
\begin{quantikz}
\lstick{Cavity} &   \gate[2]{\text{Reset}} & \qw\gategroup[2,steps=3,style={dashed,
rounded corners,%fill=blue!10, 
inner xsep=2pt},
background,label style={label position=below,anchor=
north,yshift=-0.5cm}]{} & \gate{\text{OCP}/\text{delay}(\tau_{ocp})} & \qw & \rstick{$|1\rangle$}\qw\\
\lstick{Transmon} &  & \qw & \qw & \gate{\checkmark} & \rstick{$|g\rangle$}\qw
\end{quantikz}
\end{center}
An OCP is implemented when preparing $\ket{1}$, and a delay equal to the duration of an OCP ($\tau_{ocp}$) is inserted instead when preparing $\ket{0}$ in order for the $\ket{0}$ and $\ket{1}$ preparation protocols to have the same total duration, as they are performed simultaneously in order to prepare a logical basis state. \\

For the second type, we also include a series of cavity measurements to directly check that the target cavity state was successfully generated. This type of check measurement is possible due to the QND nature of the cavity check measurements on photon-number cavity states, allowing us to perform multiple measurement rounds to suppress misassignment errors~\cite{elder_msmts_2020}. When using this cavity check, we label an attempt as successful when all check measurements report when both cavities are in the target state.
Namely, in this protocol the cavity-state initialization is followed by a transmon check (identically to the previous protocol described), which is then followed by a sequence of $N=5$ cavity measurements with conditional transmon reset back to $\ket{g}$ if it is measured to be in $\ket{e}$. The preparation of $\ket{1}$ or $\ket{0}$ with this protocol is described by the following circuit: 
\begin{center}
\begin{quantikz}
\lstick{Cavity} &   \gate[2]{\text{Reset}} & \qw\gategroup[2,steps=7,style={dashed,
rounded corners,%fill=blue!10, 
inner xsep=2pt, inner ysep=20pt},
background,label style={label position=below,anchor=
north,yshift=-0.5cm}]{} & \gate{\text{OCP}/\text{delay}(\tau_{ocp})} & \qw & \ctrl{1}\gategroup[2,steps=3,style={dashed,
rounded corners,%fill=blue!10, 
inner xsep=2pt, inner ysep=1pt},
background,label style={label position=below,anchor=
north,yshift=-0.2cm}]{$\times5$} & \qw & \qw & \qw & \rstick{$|1\rangle$} \qw\\
\lstick{Transmon} &  & \qw & \qw & \gate{\checkmark} & \gate{X_{n=0}^s}  & \meter{G/E/...} & \gate[cwires=1]{X_{n=0}^{us}} &\qw & \rstick{$|g\rangle$}\qw
\end{quantikz}
\end{center}

\textbf{State preparation fidelity estimates}.
We have generated a simple error model to quantify the performance of our state preparation protocols.

For the reset step of our state preparation, we model the $\ket{0}$ preparation to suffer from two main error mechanisms. First, a preparation error can occur if we obtain an unlucky sequence of $N$ measurements all with misassignment outcomes, e.g. when the cavity was in $\ket{1}$, but was miassassigned as ``0". Second, if the state was correctly prepared and measured in $\ket{0}$, but suffered an undetected heating event to $\ket{1}$ during the final measurement. We can estimate the probability of the $\ket{0}$ state reset failure in a single cavity, denoted $\epsilon_{\mathrm{reset}}$, as:
\begin{equation*}
    \epsilon_{\mathrm{reset}} = n_{\mathrm{th}}^{(c)}\cdot P\left(``0"|\ket{1}\right)^{N} + (1-n_{\mathrm{th}}^{(c)})\cdot[(1-P\left(``1"|\ket{0}\right))^{N}\cdot n_{\mathrm{th}}^{(c)}\frac{t_{M}}{T_{1}^{(c)}}]
\end{equation*}
where $N$ corresponds to the number of rounds of measurements, $P(``\alpha"|\ket{\beta})$ denotes the probability of assigning the outcome $``\alpha"$ conditioned on being in the state $\ket{\beta}$ (see supplementary text 10 for more detail). In our experiments we used $N=6$ and using this expression estimate failure probability of preparing the state $\ket{00}$ to be $\epsilon_{00}=\epsilon_{\mathrm{reset}}^A + \epsilon_{\mathrm{reset}}^B + \epsilon_{\mathrm{reset}}^A\cdot\epsilon_{\mathrm{reset}}^B = 2.8\times10^{-6}$.\\

% \textbf{Transmon-check only}

Now we turn to the subsequent state preparation following the reset, and the impact of each of the transmon check, and cavity check measurement protocols.

First, we estimate the probability of the state preparation failing when using the protocol where only a transmon check measurement is performed after the initialization. 
We can estimate the probabilities of the $\ket{0}$ or $\ket{1}$ state preparations failing with this protocol, $\epsilon_{0}^{check}$ and $\epsilon_{1}^{check}$, as: 

\begin{equation*}
\begin{aligned}
\epsilon_{0}^{check} = (1-\epsilon_{\mathrm{reset}})\cdot n_{th}^{c}\cdot\frac{\tau_{ocp}+t_{M}}{T_{1}^{c}} +\;\epsilon_{\mathrm{reset}}\cdot(1-\frac{\tau_{ocp}+t_{M}}{T_{1}^{c}})
\end{aligned}
\end{equation*}
and
\begin{equation*}
\begin{aligned}
\epsilon_{1}^{check} = \epsilon_{ocp}\cdot (1-n_{th}^{c}\cdot\frac{t_{M}}{T_{1}^{c}})
\;+ (1-\epsilon_{ocp})\cdot\frac{t_{M}}{T_{1}^{c}}
\end{aligned}
\end{equation*}

% We can estimate the probability of the $\ket{1}$ state preparation failing, denoted $\epsilon_{1}$, as either an error on the OCP (and no measurement error) or a successful OCP and decay of the cavity state during the measurement,
% \begin{equation*}
%     \epsilon_{1} = \epsilon_{ocp} + (1-\epsilon_{ocp})\cdot\frac{t_{M}}{T_{1}^{(c)}}
% \end{equation*}
where $\epsilon_{ocp}$ denotes the OCP error, or the probability that it does not prepare the state $\ket{1}$. Here, we assume that if the OCP fails the cavity is left in $\ket{0}$. In principle, there is also a probability that the cavity can be left in $\ket{2}$. While the probability of this occurring, and its effect on the logical space can be a future investigation, we assume that this probability and its effect are negligibly small for this paper. 

We monitored the fidelity of our OCP pulses for the several months while 
taking the data for this paper. Using repeated-gate quantum process tomography we extracted OCP fidelities of $(2.53\pm0.54)\%$ for Alice and $(0.67\pm0.43)\%$ for Bob, where the errors indicate the standard deviation.

% We monitored the OCP pulses of our subsystems for several months for this experiment ad performed quantum process tomography several months, extracting the X gate errors for each subsystem as $(2.53\pm0.54)\%$ for Alice and $(0.67\pm0.43)\%$ for Bob, where the errors indicate the standard deviation.

% \textbf{Cavity check}
Second, we consider the preparation protocol where additional repeated cavity check measurements are performed in order to directly confirm the state of the cavity.

The residual errors we consider for the repeated cavity-measurement part, conditioned on the outcomes of the error budget from the previous protocol, are similar to those of the repeated measurements for reset: either repeated misassignment errors or an undetected cavity transition error on the final measurement. 
We estimate the $\ket{0}$ preparation error, $\epsilon_{0}^{CM}$ (where CM stands for cavity measurements), to be:
\begin{equation*}
\begin{aligned}
\epsilon_{0}^{CM} = \epsilon_{0}^{check}\cdot P\left(``0"||1\rangle \right)^5
+\; (1-\epsilon_{0}^{check})\cdot[(1-P\left(``1"||0\rangle \right))^5\cdot n_{th}^c\cdot\frac{t_{M}}{T_{1}^{c}}]
\end{aligned}   
\end{equation*}
and the $\ket{1}$ preparation error, $\epsilon_{1}^{CM}$, to be:
\begin{equation*}
\begin{aligned}
\epsilon_{1}^{CM} = \epsilon_{1}^{check}\cdot P\left(``1"||0\rangle \right)^5 
+\; (1-\epsilon_{1}^{check})\cdot[(1-P\left(``0"||1\rangle \right))^5\cdot \frac{t_{M}}{T_{1}^{c}}]\\
\end{aligned}
\end{equation*}

Using these expressions for $\epsilon_{0}$ and $\epsilon_{1}$ we can now estimate state preparation fidelities for each protocol.
We use our system parameters (see the System Properties section), and the OCP fidelities to compute our estimates.

We estimate the error in preparing each of the cavities in $\ket{0}$ as:
\begin{center}
    \begin{tabular}{c | c || c | c }
    \textbf{Preparation Method} & \textbf{Error} & \textbf{Alice} & \textbf{Bob}\\
    \hline\hline
    \rule{0pt}{2.6ex}
    Post-initialize check only & $\epsilon_{0}^{check}$ & $3.24\times10^{-6}$& $4.07\times10^{-6}$\\[1mm]
    \hline
    \rule{0pt}{2.6ex}
    Check and repeated cavity-measurements & $\epsilon_{0}^{CM}$ & $1.02\times10^{-6}$ & $1.49\times10^{-6}$\\
    \end{tabular}
\end{center}

We estimate the error in preparing each of the cavities in $\ket{1}$ as:
\begin{center}
    \begin{tabular}{c | c || c | c }
    \textbf{Preparation Method} & \textbf{Error} & \textbf{Alice} & \textbf{Bob}\\
    \hline\hline
    \rule{0pt}{2.6ex}
    Post-initialize check only & $\epsilon_{1}^{check}$ & $2.53\times10^{-2}$ & $1.80\times10^{-2}$\\[1mm]
    \hline
    \rule{0pt}{2.6ex}
    Check and repeated cavity-measurements & $\epsilon_{1}^{CM}$ & $0.33\times10^{-2}$ & $0.97\times10^{-2}$\\
    \end{tabular}
\end{center}

In this work, we experimentally implemented the two different state preparation protocols, one with just a transmon check measurement (for our SPAM experiment), and a second with both a transmon check and repeated cavity check measurements (for our dual-rail coherence experiments). 
We develop a simple heuristic for measurement-based state prep protocols: the maximum performance for this protocol is the probability of an undetected error during the final measurement. In other words, if the fidelity of the state initialization is lower than this measurement error, then this measurement-based approach can improve the state prep fidelity. On the other hand, if the initialization fidelity is higher than the measurement error, then there is less benefit to using these measurements. As such, the system parameters that determine the preparation fidelity gain as a result of performing repeated measurements are primarily the state-initialization fidelity (OCP fidelity in this case), the measurement duration ($t_{M}$), the cavity thermal population ($n_{th}^c$), and the cavity lifetime ($T_1^c$). In our hardware, our OCP fidelities are comparable to undetected cavity decay errors, and we expect similar state preparation performance between the two methods. Namely, we can look at the ratio between the fidelity of preparing a cavity-state with the transmon-check-only protocol, and the probability of an undetected cavity transition during the last cavity measurement in the case where repeated cavity-measurements are performed:
\begin{equation*}
    \begin{aligned}
        \mathrm{Prep. \ket{0} ratio} &= \epsilon_{0}^{check}\;/\;(n_{th}^c\cdot t_{M} / T_{1}^c)\\
        \mathrm{Prep. \ket{1} ratio} &= \epsilon_{1}^{check}\;/\;( t_{M} / T_{1}^c)
    \end{aligned}
\end{equation*}

For our system, from our estimates, we find these ratios to be 
\begin{center}
    \begin{tabular}{c||c|c}
      & \textbf{Alice} & \textbf{Bob} \\
      \hline\hline
       \rule{0pt}{2.6ex}
      $\mathrm{Prep. \ket{0} ratio}$ & 3.19 & 2.73\\
      \hline
      \rule{0pt}{2.6ex}
      $\mathrm{Prep. \ket{1} ratio}$ & 7.78 & 1.85\\
    \end{tabular}
\end{center}

While we do see a performance improvement by using additional cavity-check measurements, it is important to note that this is the estimated improvement in preparing the Fock states $\ket{0}$ and $\ket{1}$ in each of the cavities. As such, a given factor of improvement in the state preparation fidelity of one of the cavities may not directly translate to an equivalent factor of improvement in the logical SPAM error. This is due to the fact that a state preparation error in one of the cavities is a single physical error, while a logical SPAM error is a second-order physical error that is the result of contributions from many error channels (see supplementary text~\ref{sec:error_budget}). Using our simple error model, we indeed find that for the factor of improvement estimated from the expression above, we would expect only a factor of $\sim 1.5$ improvement in the logical misassignment error.

We have compared the data for logical misassignment probability between state preparation methods and find similar results. With just the transmon check state preparation we find a logical SPAM error of $(1.8\pm0.3)\times10^{-4}$ while the logical SPAM error when using both transmon and cavity check measurements is $(1.6\pm0.3)\times10^{-4}$.

\subsection{Logical Measurement}

The protocol for the logical measurement of a dual-rail qubit that we have developed brings together most of the elements described in the cQED Toolbox for Bosonic Operations section in order to implement a measurement that is as robust as possible to the first-order errors, while taking into account the hierarchy of these first-order errors.

The logical measurements are first prepended by a transmon check to confirm that the logical measurement begins with the transmon in the ground state. This is especially important in our bit-flip and phase-flip experiments, where for long delay times between state preparation and measurement the transmon will most likely have thermalized and may be in the $\ket{e}$ state with around $\sim1\%$ probability.

The logical measurement then consists of mapping the cavity-state to the transmon by way of a selective-pulse, followed by a transmon readout, a reset of the transmon back to $\ket{g}$ (conditioned on the readout outcome), and, in the case of two-rounds of measurements, a post-reset transmon check to confirm that the transmon is in $\ket{g}$ after the reset.

This post-reset check is intended to suppress transmon reset failures. If the transmon state is incorrectly assigned (e.g. from a readout classification error) or if the transmon suffers an undetected transition during the readout, the conditional reset will fail to reset the transmon to $\ket{g}$. Instead, the transmon will be in $\ket{e}$ for the subsequent round of cavity measurements, propagating the reset error into a cavity error for the next cavity measurement. Thus it is critical for the transmon to be properly reset to $\ket{g}$ after each round of cavity measurements.

% This post-reset check is intended to suppress the reset decision-making infidelity, that is limited by a readout classification error ($p_{miss}$) on the post-mapping measurement. 
% Namely, in the case where we do not perform this check, if we perform an additional logical measurement it may begin with the transmon in $\ket{e}$ with probability $p_{miss}$ and lead to an incorrect outcome; in the case where we perform this post-reset check this will happen instead only with probability $p_{miss}^{2}$.

Represented as a circuit diagram, our logical measurement consists of the following protocol, implemented simultaneously on each of the dual-rail's subsystems:

\begin{center}
\begin{quantikz}
\lstick{Cavity} & \qw  & \ctrl{1}\gategroup[2,steps=4,style={dashed,
rounded corners,%fill=blue!10, 
inner xsep=2pt},
background,label style={label position=below,anchor=
north,yshift=-0.5cm}]{{}}          & \qw   & \qw   & \qw & \qw  \\
\lstick{Transmon} & \gate{\checkmark} & \gate{X_{n=1}^{s}}  & \meter{G/E/...} & \gate[cwires=1]{X_{n=1}^{us}} & \gate{\checkmark} & \qw
\end{quantikz}
\end{center}

\section{Decoding Measurement Outcomes}
\label{sec:classification}

\begin{figure}[h]
\centering
\includegraphics[width=0.9\textwidth]{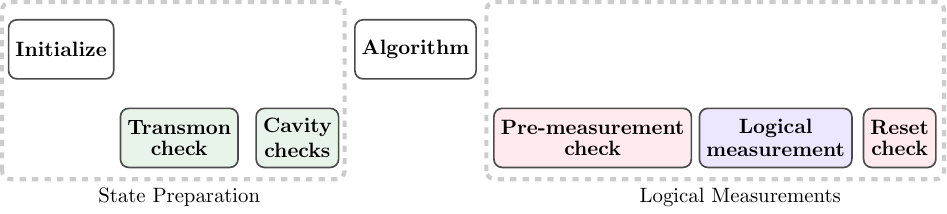}
\caption{Measurement outcomes are decoded and assigned based on the results of each of the check measurements and the logical measurement is the state preparation and measurement protocols. As described in the Protocols section, there are check measurements associated with the state preparation, and check measurements associated with the logical measurement that come after a delay or an algorithm between the state preparation and measurement. The dashed lines indicate the grouping of the check measurements to state preparation or logical measurement; the colors indicate one of the three categories that are used to assign the outcomes of each of the measurements (see Fig.~\ref{fig:buckets}).}
\label{fig:checks}
\end{figure}

In this section, we discuss our methodology for decoding measurement outcomes for our dual-rail experiments. Assignment of physical and logical outcomes for a dual-rail experiment is more involved than that of a transmon given that each dual-rail sequence can involve multiple measurements of both the cavities and the transmons.  
Our state preparation and logical measurement protocols consist of a number of ``check" measurements as well as the logical measurement, as described in the previous State Preparation and Measurement Protocols section. 
In the state preparation protocol, state initialization is followed by a transmon check and an optional sequence of cavity-check measurements. 
For the logical dual-rail measurement, we first perform transmon check and then perform the logical cavity measurement. In the case of multiple rounds, we will reset the transmons after each round of logical cavity measurement and perform an additional transmon measurement to check that the transmons are in the ground state. See Fig.~\ref{fig:checks} for the sequence of measurements.

\begin{figure}[h]
\centering
\includegraphics[width=0.6\textwidth]{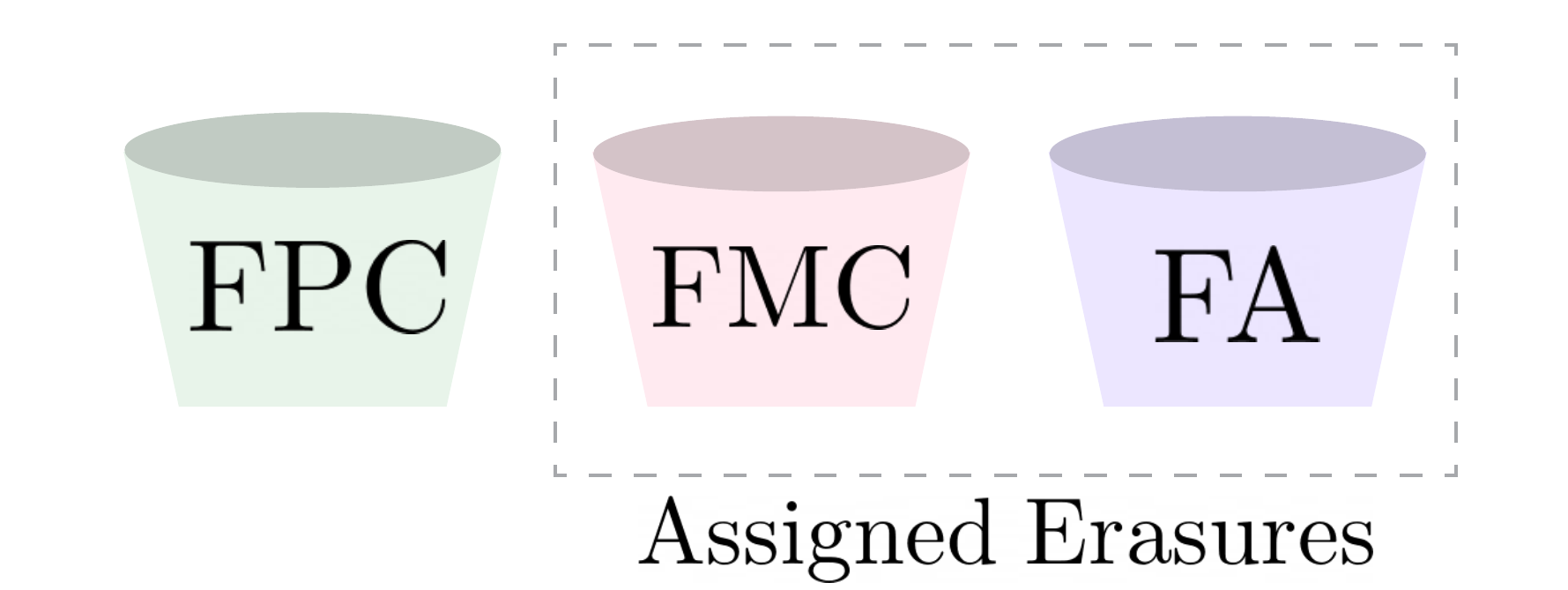}
\caption{The outcomes of each of the check or logical measurements can result in the shot being assigned either as FPC (failed state preparation check), FMC (failed measurement check), or FA (failed assignment). Each such $k$ shot is placed into an abstract ``bucket", and contributes to the total counts of the outcome, $N_{k}$. $N_{\mathrm{FPC}}$ counts are removed from the total number of counts, while $N_{\mathrm{FMC}}$ and $N_{FA}$ are two of the contributing groups of counts to the erasure fraction.}
\label{fig:buckets}
\end{figure}

We categorize the checks and logical measurements into three categories, determined by the ways in which we use the outcomes in order to classify each of the runs of the experiment: failed state preparation checks (FPC), failed measurement checks (FMC), and failed state assignments (FA). 
While FPC shots are removed from the total number of counts, FMC and FA shots contribute to the erasure counts, see Fig.~\ref{fig:buckets}. 
In this section, we describe each of the three categories and the way in which we analyze the counts assigned from each category.

\begin{enumerate}
    \item \textbf{\underline{Failed Preparation Checks (FPC)}}:\\[1mm]
    After initializing the dual-rail state, we perform measurements of the ancilla transmons in order to confirm that they are both in $\ket{g}$. Optionally, we can also perform a subsequent sequence of cavity measurements in order to confirm the correct preparation of the cavity-state. If in any run of the experiment the transmon checks indicate that either of the transmons were in $\ket{e}$, or if (in the case where cavity checks were performed) any of the cavity measurements indicate that either of the cavities were not in the intended state, then the shot is labeled as ``FPC" and removed from the total number of counts. 

    In other words, if $N_{\mathrm{All}}$ is the number of shots that were performed, the total number of counts that we consider ($N_{\mathrm{T}}$) is:
    \begin{equation*}
        N_\mathrm{T} = N_{\mathrm{All}} - N_{\mathrm{FPC}}
    \end{equation*}
    \item \textbf{\underline{Failed Measure Checks (FMC)}}:\\[1mm]
    Before performing the logical end-of-the-line measurement, we perform measurements of the ancilla transmons in order to confirm that they are both in $\ket{g}$ before the logical measurement. In the case where we perform multiple rounds of logical measurements, we also perform a transmon check before each additional logical measurement in order to confirm that the transmon reset after the previous measurement was successful, and that both transmons are indeed in $\ket{g}$ before the next logical measurement. If any of these checks indicate that either of the transmons were in $\ket{e}$, and if the shot was not already labeled as FPC, then the shot is labeled as FMC giving us counts of $N_{\mathrm{FMC}}$ and will contribute to the erasure fraction. 
    \item \textbf{\underline{Failed Assignments (FA)}}:\\[1mm]
    FA can get assigned from the logical measurements when multiple rounds of measurements are performed. Following the state assignment method as described in the Main Text, using the results of the logical measurements we can assign one of the four outcomes $\{``00",``01",``10",``11" \}$ giving us the counts for $N_{00},\;N_{01},\;N_{10},\;N_{11}$. However, if two measurements do not agree in the case where we perform two rounds of measurements, or if there is no strict majority in the case where we perform more than two measurements, the outcome is ambiguous and labeled as FA giving us the $N_{\mathrm{FA}}$ counts. 
\end{enumerate}

Having defined all the types of counts that we can assign from the ensemble of shots of an experiment, we can explicitly define each of the SPAM metrics described in the Main Text.

% \begin{enumerate}
%     \item Logical missassignment error when preparing $\ket{0_\mathrm{L}} = \ket{01}$ ($\ket{0_\mathrm{L}} = \ket{10}$) is $N_{10(01)} / \left( N_{10}+N_{01} \right)$.
%     \item Erasure fraction is $\frac{N_{\mathrm{FMC}}+N_{00}+N_{11}+N_{\mathrm{FA}}}{N_{\mathrm{T}}}$.
%     \item $\ket{00}$ Leakage detection error is $\frac{N_{01(10)}}{N_{\mathrm{T}}}$
% \end{enumerate}
The logical misassignment error when preparing $\ket{01}$ ($\ket{10}$) is defined as
\begin{equation*}
    \frac{N_{10(01)}}{N_{10}+N_{01}}.
\end{equation*}
The erasure fraction is defined as
\begin{equation*}   \frac{N_{\mathrm{FMC}}+N_{00}+N_{11}+N_{\mathrm{FA}}}{N_{\mathrm{T}}},
\end{equation*}
where $N_{\mathrm{FA}}=0$ in the case where only one logical measurement is performed. And finally, the leakage detection error to the logical state $\ket{0_{\mathrm{L}}}$($\ket{1_{\mathrm{L}}}$), measured by intentionally preparing $\ket{00}$ and $\ket{11}$ is defined as
\begin{equation*}
    \frac{N_{01(10)}}{N_{\mathrm{T}}}.
\end{equation*}

% \section*{Supplementary Text}

\section{SPAM Experiment Using 2 Rounds of Measurements}
\label{sec:2M_SPAM}

% Outline %

% Finally, by performing the SPAM experiment with two rounds of measurements we observe an exceedingly low logical misassignment error of $(4\pm 1)\times 10^{-5}$ and leakage detection error of $(1.3\pm0.1)\times10^{-3}$.
% We attribute this $\sim5\times$ improvement to the decoding strategy where we require agreement between both rounds of measurements (see Supplementary Text for more details and complete data results using two rounds).
% The trade-off with this strategy is a much higher erasure fraction at $(11.2\pm0.1)\times10^{-2}$, which is well-accounted for in our model.
% This experiment highlights the flexibility of our measurement protocol, with options for both multiple rounds of measurements and strategies both for reduced SPAM error or for reduced erasures.

We discuss results for the SPAM experiment with two rounds of measurements, $N_\textrm{rounds} = 2$. Our strategy for decoding the measurement results is to require that both raw measurement outcomes agree in order to assign a physical state outcome $\{``00", ``01", ``10", ``11"\}$; if the raw measurement outcomes disagree, for example, we observe the two-round sequence $\left(``GE", ``GG"\right)$, then we assign an ambiguous outcome, $``\mathrm{A}"$, and subsequently assign as a logical erasure.
With this measurement protocol, we measure a logical misassignment error of $\left(4\pm2\right)\times10^{-5}$, leakage detection error of $\left(1.2\pm0.1\right)\times10^{-3}$, and an erasure fraction of $\left(17\pm0.1\right)\times10^{-2}$. 
% The trade-off with this strategy is a much higher erasure fraction, which is well-accounted for in our model.
With multiple rounds of measurements, the contributions to the total erasure fraction comes from both assigned erasure states $``00"$ and $``11"$, and from the ambiguous assignments ``A".

As described in the Main Text, SPAM errors can arise either from state preparation errors, cavity transitions, or from physical ancilla errors during the two-rounds of measurements.
From our error analysis we find that cavity transitions are not the dominant source of SPAM errors, which are instead dominated by physical ancilla errors and state preparation errors (see the Error Budget for the Dual-Rail Measurement section). Physical ancilla errors such as errors during the cavity-state mapping or during the dispersive readout will result in an incorrect assignment of the underlying cavity state. By performing two rounds of measurements, the impact of these physical errors on the state assignment is suppressed, as the probability of such an error occurring on both consecutive measurements and corrupting the assignment is $p^{2}$ rather than $p$, where $p$ is the probability of a transmon ancilla error.

With this in mind, we can understand the ways in which 2 rounds of measurements affect each of the three figures of merit that we extract from the SPAM experiment. 
First, we expect the logical assignment error to decrease. A logical misassignment is a second-order logical error resulting from an error in both cavities. 
Suppressing each of the cavity misassignment probabilities to $p^{2}$ means that the logical misassignment due to ancilla errors becomes a highly unlikely fourth-order event that can result only from the \emph{consecutive} misassignment of \emph{both} of the cavities.  
On the other hand, errors due to state preparation or cavity transitions between the state preparation and logical measurement will not be suppressed by multiple measurements as the logical measurement is QND and as such any number of measurements should result in outcomes that agree and as such cannot be ``caught". 
% The state itself is changed from the intended preparation
For instance, from our simplified error model (see the Error Budget for the Dual-Rail Measurement section), we find that the dominant error channel contributing to logical assignment errors with 2 rounds of measurements is the case where on one of dual-rail's cavities we misassign both the post cavity-state mapping readout and the post-reset transmon check, and suffer a state preparation error on the other cavity.

Second, we expect the leakage detection error to decrease slightly. On the one hand, the probability of actual leakage during two measurements increases by a factor of 2, i.e. the probability of a cavity state transition increases from $\sim t_{RM}/T_{1}$ to $\sim 2t_{RM}/T_{1}$, where $t_{RM}$ is the duration of each round of measurement. However, as subset of the leakage detection errors are the result of misassignemnt due to ancilla errors, these errors will be suppressed by a power of 2 by the second measurement, offsetting the increase in actual leakage errors. 

Finally, we expect the erasure fraction to increase due to two primary contributions compared to the erasure fraction from one round of measurement. First, with our decoding strategy, any 2-round outcomes that disagree with one another are decoded as ``A" and contribute to the erasure fraction. In such cases, if the outcome of the first round of measurement is ``00" or ``11", it would of have contributed to the erasure fraction even in the case of one round of measurement and thus is not an additional contribution. However, cases where the result of the first round is assigned to be in the code space but the outcome of the second round is assigned as a leakage state or the opposite logical state will get decoded as residual ``A" counts compared to one round of measurement. For our system, we both estimate and measure this contribution to be $\sim4\%$.
Second, as described in the Protocols section, after each round of measurement the ancilla transmon is reset to $\ket{g}$ if it is measured to be in $\ket{e}$ after the cavity measurement by way of an unselective $\pi$-pulse. This conditional reset is followed by a transmon check in order to confirm that it was correctly reset to ground state. In cases where this check fails, the shot is treated as an erasure and is an additional contribution to the ``A" fraction, as this check is not performed in the case of a single round of measurement.

\begin{figure}[h]
\centering
\includegraphics[width=0.7\textwidth]{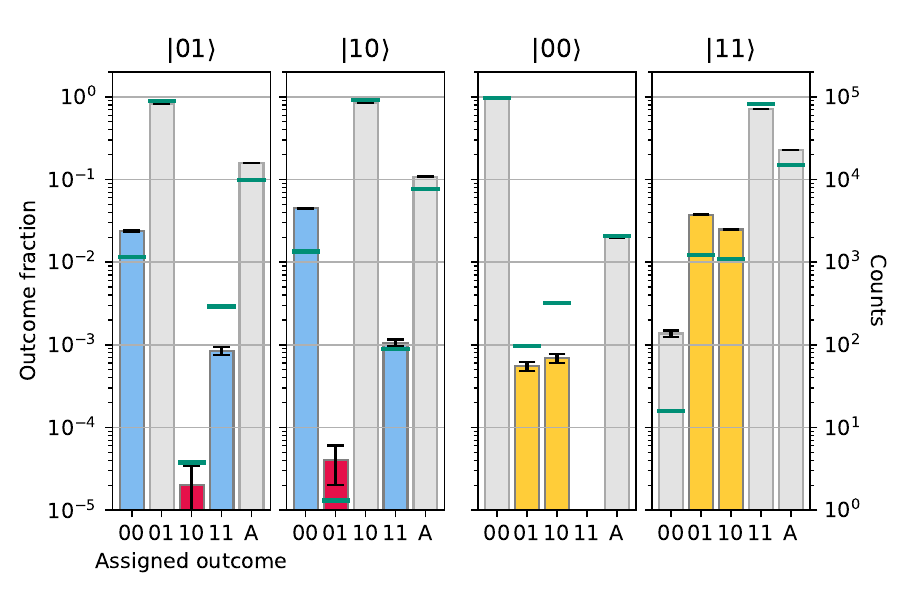}
\caption{State assignment data using 2 rounds of measurements. As in the case of the SPAM experiment with 1 round of measurements, we prepare each of the four dual-rail basis states. Using the results of the 2 rounds 0f measurements and our decoding strategy, we assign one of the four states $\{``00", ``01", ``10", ``11"\}$, or the ambiguous outcome $``\mathrm{A}"$ in the case where the two measurement outcomes disagree. Results of preparing the two logical state $\ket{01}$ and $\ket{10}$ are shown in the left two panels; results of intentionally preparing the leakage states $\ket{00}$ and $\ket{11}$ are shown in the right two panels.}
\label{fig:2M_SPAM}
\end{figure}

% With the assignment strategy, we measure exceptionally low logical misassignment and leakage detection errors, that we attribute to the strictness of the assignment strategy. By requiring both measurements to agree, misassignment errors are suppressed~\cite{elder_msmts_2020}.

% Decoding strategy has two consequences: requiring two measurements to agree decrease misassignment errors, see Sal
% we see that this results in a modest decrease in the SPAM error and in the leakage detection error
% what about the erasure fraction? its much higher

% From this data, we observe that ambiguous ``A" outcomes will contribute a majority of the erasure assignments, which we attribute to this strict assignment strategy.
% The ambiguous outcomes that disagree arise either because of cavity transitions between the two rounds of measurement, or because of physical ancilla errors during either of the logical measurements.
% From modeling we see that its mostly physical ancilla errors and not demolition of the logical dual-rail cavity state induced by the measurement.

% Comment on the ways we expect 2 rounds of measurements to effect each of the three FOMs we identified in the text:
% (1) erasure errors (blue)
% (2) logical assignemnt errors (red)
% (3) leakage detection errors (yellow)

\section{Extracting Dephasing Probabilities}

In this work, we extract dual-rail qubit dephasing errors by linearizing the dephasing error at short times. This method is used to minimize the impact of state transitions of coupled nonlinear elements that are known to cause photon-shot noise dephasing. Here we provide more details how we extract a dephasing error rate by considering a simple qubit model that suffers a $\hat{\sigma_z}$ error with probability $p_\phi$. This can be described with the following Kraus operators: $\hat{K}_0 = \sqrt{1-p_\phi}\; \hat{\mathds{1}}$ and $\hat{K}_1 = \sqrt{p_\phi}\;\hat{\sigma_z}$. Given the initial state $\ket{+} = \left(\ket{0} + \ket{1}\right) / \sqrt{2}$, this noise channel causes the state to evolve as 
\begin{equation*}
    \hat{\rho} = \ketbra{+} \rightarrow \frac{1}{2}
        \begin{pmatrix}
        1 & 1 - 2p_\phi \\
        1 - 2p_\phi & 1
        \end{pmatrix}
\end{equation*}
or in the continuous limit where we set $2p_\phi = \Gamma_\phi \Delta t \ll 1$,
\begin{equation*}
    \hat{\rho}(t) = 
    \begin{pmatrix}
        1 & e^{-\Gamma_\phi t} \\
        e^{-\Gamma_\phi t} & 1
    \end{pmatrix}.
\end{equation*}
The coherence of this state is given by $\expval{X} = e^{-\Gamma_\phi t}$, and this is the quantity measured in experiment (labeled $\expval{Z_L}$ in experiment because the second $\pi/2$ pulse implements a transformation converting X-information into Z-information).
At short times, we expect a linear decay with slope $m = -\Gamma_\phi$. Then we can extract the probability of a phase-flip error in a time $t$ to be $p_\phi = \Gamma_\phi t / 2$.

\section{Short-time Ramsey experiment}
\begin{figure}
    \centering
    \includegraphics[width=0.7\textwidth]{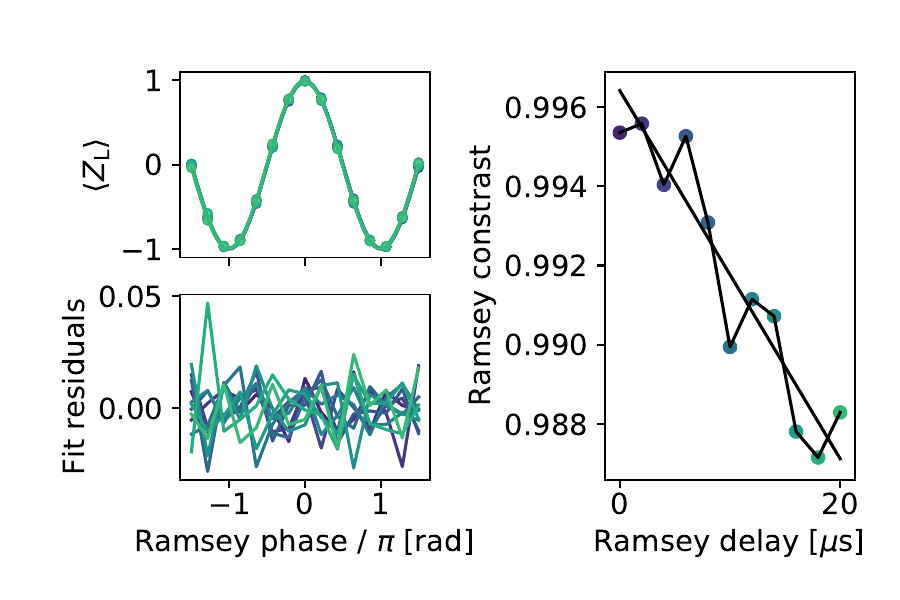}
    \caption{Short-time Ramsey experiment analysis. Top left: Logical results showing Ramsey oscillations for each delay out to $20~\mu \mathrm{s}$. The markers correspond to experimental results while the lines correspond to the fit. Bottom left: Fit residuals. Right: Extracted Ramsey contrast (valued between 0 and 1) as a function of Ramsey delay. Each marker corresponds to a different delay, color-coded to the data in the left column. Straight black line is a linear fit to the data.}
    \label{fig:ramsey_phase_expt}
\end{figure}

As described in the Main text, we used an extended protocol for performing short-time Ramsey experiment to better resolve the dephasing at these shorter delays, mitigating effects arising from small detuning errors. In this version, we vary both the Ramsey delay as well as the phase of the second $\pi/2$ pulse. 
% We present results by plotting $\expval{Z_\mathrm{L}}$ for this 2D Ramsey experiment and results of another example experiment measured out to longer delays to show the decrease in Ramsey contrast in Figure XX. 
For each delay, as a function of the phase of the second $\pi/2$-pulse, we extract a characteristic Ramsey oscillation in the expectation value and the coherence of the qubit is encoded in the amplitude of the oscillation. We fit and extract the contrast of the oscillation for each delay. Finally we fit the constrast as a function of delay to extract the dephasing error rate. These results are shown in Fig~\ref{fig:ramsey_phase_expt}.

\section{Dual-Rail Measurement Numerical Simulations}
\label{sec:model}

In order to numerically simulate the expected performance of the dual-rail cavity qubit and the expected outcomes of the characterization experiments, we developed a model for numerical simulation.

With this model we run QuTiP-based simulations~\cite{QuTiP_1,QuTiP_2}, that take the full set of independently-measured hardware parameters as input, and from which we extract the expected measurement outcome of the dual-rail qubit as output.
Specifically, we perform a time-domain simulation of the dual-rail processes (delays, mapping, reset etc.) using QuTiP's master equation solver, and integrate this with our projective measurement model. Similar to the way in which we treat the measured dual-rail outcomes, the simulated readout outcomes are then correlated and histogrammed; this allows us to simulate and analyze any number of rounds of measurements, as well as simulating and extracting the underlying quantum state in the hypothetical case of no measurement errors. \\

The main feature of the model is the way in which we model the transmon readout. 
In the dual-rail measurement protocol, after the the cavity state is mapped to the transmon, a dispersive readout is performed on the transmon.

Dispersive readout is a continuous process in which a probe tone is sent through the readout resonator, and the discrete state of the transmon is effectively mapped to the transmitted microwaves due to the transmon state-dependent frequency shift of the resonator~\cite{Blais_2004}. 
In a transmon readout, it is nominally the $\sigma_{z}$ operator, or observable that is measured, and as such the readout ``projects" the transmon state to one of the $\sigma_{z}$ eigenstates i.e. one of the computational basis states.
In the dual-rail measurement protocol, the cavity state is always mapped to one of the transmon's computational basis states (i.e. $\ket{g}$ or $\ket{e}$, not a superposition state). This means that in the case where the measurement is a high-fidelity, single-shot QND (quantum non-demolition) readout, the measurement outcome should correspond to the physical state of the transmon that was measured -- and the transmon state should be left unchanged, as it was in an eigenstate of the measured operator.

However, in practice, transmon readout is an error-prone process. Non-idealities of the measurement are primarily introduced by two sources: 
\begin{enumerate}
    \item State transitions of the transmon during the finite-duration readout. Such transitions will lead to a recorded outcome that is incorrect w.r.t to either the transmon's true initial or final underlying state. If the transmon undergoes a state transition event during the measurement, the assigned state w.r.t the transmon's initial state has some uncertainty. If the transition happens at an early point in the readout, then the transmon will be in the post-transition state for the majority of the readout time, and as such the assigned state will most likely reflect the transmon's post-readout state, and not its pre-readout state. Conversely, if the transition occurs closer to the end of the readout then the pre-transition state would already mostly have been mapped and the assigned state will most likely reflect the transmon's pre-readout state, but not its final post-readout state. Readout errors due to transmon state transitions are somewhat enhanced due to the increase in the transmon's thermal population during readout, and the decrease in its $T_{1}$ during readout~\cite{thorbeck2023readoutinduced}.
    \item Classification errors due to insufficient separation of the readout peaks in the IQ plane.
\end{enumerate}

These two distinct sources of error motivate the way in which we model and implement the transmon readout in our model, and we handle these two sources of error separately. We do this since a number of components of the dual-rail protocol are conditioned on the transmon readout outcome, and we want to be able to separate the assigned measurement outcome (G/E) from the underlying state ($\ket{g}$/$\ket{e}$) that undergoes subsequent evolution in the protocol. In particular we want to capture the cases where the assigned state does not agree with the underlying state, and the ways in which this propagates and limits the performance of the overall protocol.

\begin{figure}[h]
\centering
\includegraphics[width=0.6\textwidth]{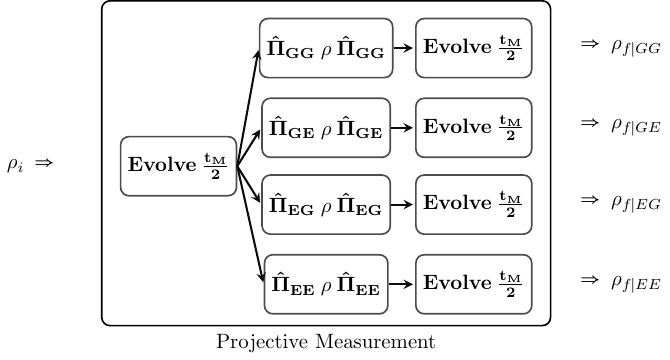}
\caption{}
\label{fig:projmsmt}
\end{figure}

Specifically, in our modelling of the measurement we handle these two sources as follows. First we freely evolve the density matrix of the initial state $\rho_{i}$ for half the measurement duration time, $t_{M}$, using a generalized amplitude damping Kraus map to account for state transitions~\cite{siddhu2021queuechannel}, we then apply ``imperfect" measurement projectors (that account for the probability of classification errors) for each of the $j$ possible measurement outcomes, and then apply the Kraus map for then second half of $t_{M}$ to each of the projected states to account for the remaining transition probabilities and give us the final states $\rho_{f|j}$. $\rho_{f|j}$ is the density matrix of the final state, given that outcome $j$ was measured, where $\sum_{j}$ equals the number of possible measurement outcomes. For a more conceptual description see Fig.~\ref{fig:projmsmt}. We apply the measurement operators to the states mid-way through the measurement time (as opposed to at the beginning or the end) as a heuristic for the average probabilistic effect on the projected outcome, given the fact that a transmon transition can occur at any point during the measurement (as described in point 1 above).\\

The generalized amplitude damping Kraus map that we use was chosen in order to account for both heating and decay events in each of the dual-rail's modes during the readout time, and consists of the following operators:

\begin{equation*}
    \begin{aligned}
        K_{0}(t) &= \sqrt{1-n_{th}}\;(\;|0\rangle\langle0| + \sqrt{1-p_{\downarrow}(t)}\;|1\rangle\langle1|\;)\\[1mm]
        K_{1}(t) &= \sqrt{p_{\downarrow}(t)\;(1-n_{th})}\;|0\rangle\langle1|\\[1mm]
        K_{2}(t) &= \sqrt{n_{th}}\;(\;(1-p_{\downarrow}(t))\;|0\rangle\langle0|+|1\rangle\langle1|\;)\\[1mm]
        K_{3}(t) &= \sqrt{p_{\downarrow}(t)\;n_{th}}\;|1\rangle\langle0|
    \end{aligned}
\end{equation*}
where $n_{th}$ is the mode's thermal population, $p_{\downarrow}=1-e^{-t/T_{1}}$ is the probability of decay of the mode's excited state, and $\ket{0}$ ($\ket{1}$) denotes the ground (excited) state of the mode (sub $\ket{g}$ and $\ket{e}$ in the case where the mode is one of the transmons).
The map is then constructed and applied as
\begin{equation*}
    \mathcal{A}(\rho)(t) = \sum_{i=0}^{3}\;K_{i}(t)\rho K_{i}^{\dagger}(t)
\end{equation*}
In order to account for transitions in each of the four dual-rail's modes (the two cavities and two transmons) during the readout, we tensor four such Kraus maps where the cavities' maps take the measured $n_{th}^{(c)}$ and $T_{1}^{(c)}$, and the transmons' maps take the $n_{th\;RO}^{(t)}$ and $T_{1\;RO}^{(t)}$ values measured during readout. \\ 

The ``imperfect" measurement projectors we use are constructed such as to form positive operator-valued measurements (POVM) elements that take classification errors into account as follows. In general, given the probabilities of misassigning $|g\rangle$ as $E$ (denoted as $p_{gE}$) or $|e\rangle$ as $G$ (denoted as $p_{eG}$), the assignment probability matrix is:
\begin{equation*}
    M =\begin{pmatrix}
    1-p_{gE} & p_{eG}\\
    p_{gE} & 1-p_{eG}
    \end{pmatrix}
\end{equation*}

In order to construct the corresponding set of POVMs, since each measurement will yield some outcome, we want to construct a complete set of $m$ projectors $\hat{\Pi}_{m}$, such that $\sum_{m}\;\hat{\Pi}_{m}^{\dagger}\hat{\Pi}_{m}=\mathds{1}$~\cite{nielsen_chuang_2010}. The ideal projectors would be 
\begin{equation*}
    \hat{\Pi}_{G}=\begin{pmatrix}
        1&0\\
        0&0
    \end{pmatrix},\;\;
    \hat{\Pi}_{E}=\begin{pmatrix}
        0&0\\
        0&1
    \end{pmatrix}
\end{equation*}

As such, from the measurement outcome probabilities we can see that we can construct ``imperfect" projectors that satisfy the completeness requirement, and are consistent with the zero-error limit as:
\begin{equation*}
    \hat{\Pi}_{G}=\begin{pmatrix}
        \sqrt{1-p_{gE}} & 0\\
        0 & \sqrt{p_{eG}}
    \end{pmatrix},\;\;
    \hat{\Pi}_{E}=\begin{pmatrix}
        \sqrt{p_{gE}} & 0 \\
        0 & \sqrt{1-p_{eG}}
    \end{pmatrix}
\end{equation*}
or concisely, in general as $\hat{\Pi}_{m}=\sum_{i=0}^{d-1}\sqrt{M_{mi}}\;|i\rangle\langle i|$. 

The projectors for the two-transmon dual-rail readout are constructed from these as:
\begin{equation*}
    \begin{aligned}
        \hat{\Pi}_{GG} &= \mathds{1}\otimes\hat{\Pi}_{G}\otimes\mathds{1}\otimes\hat{\Pi}_{G} \\[1mm]
        \hat{\Pi}_{GE} &= \mathds{1}\otimes\hat{\Pi}_{G}\otimes\mathds{1}\otimes\hat{\Pi}_{E} \\[1mm]
        \hat{\Pi}_{EG} &= \mathds{1}\otimes\hat{\Pi}_{E}\otimes\mathds{1}\otimes\hat{\Pi}_{G} \\[1mm]
        \hat{\Pi}_{EE} &= \mathds{1}\otimes\hat{\Pi}_{E}\otimes\mathds{1}\otimes\hat{\Pi}_{E}
    \end{aligned}
\end{equation*}
where we tensor the transmon POVMs with the identity matrix on each of the cavities, and use independently-measured values of $p_{gE}$ and $p_{eG}$ for each of the two transmons projectors.\\

Conditioned on the projected outcome, we simulate additional components of the protocol. In the case of a transmon check, we only propagate forward the state $\rho_{f|GG}$, as this corresponds to the density matrix with the probabilistic distributions of the case where both transmons would get assigned as $G$ and ``pass" the check. In the case of a conditional transmon reset, for any of the projected outcomes $\rho_{f|GE},\;\rho_{f|EG},\;\rho_{f|EE}$ we apply an unselective pi-pulse to the transmon mode that was projected to $E$.\\

In order to perform such unselective pulses, as well as selective pulses (for the cavity-state mapping), and logical dual-rail gates for the purpose of simulating the bit-flip and phase-flip experiments, we use functions native to the QuTiP wrapper package Sequencing~\cite{sequencing}.

\section{Error Budget for the Dual-Rail Measurement}
\label{sec:error_budget}

% \subsubsection*{}

% In tandem with establishing the experimental toolbox for the dual-rail architecture, we have developed an error budget for analyzing the dual-rail error channels and estimating its expected performance given the hardware parameters. The motivation for developing this model can be explained by the fact that the dual-rail encoding introduces a number of novel considerations that must be taken into account when seeking to understand and build intuition for its performance.  
In tandem with the numerical simulations, we have also developed an error budget ``model" that provides us with a tool to perform analysis on contributing and limiting error channels in a different way. Namely, the error budget model is constructed in such a way that it takes all first-order error mechanisms of the dual-rail SPAM protocol into account, and provides us with the functional form of the individual error channels that contribute to the experimentally-measured SPAM outcomes. This provides us with a simple tool to: understand the scaling of the individual errors as a function of our particular hardware parameters, to quantify the relative contribution of various errors w.r.t each other, and to optimize the implementation of our protocol within the parameter space available to us. 

The dual-rail qubit is encoded using multiple physical modes (namely two coupled bosonic modes), but is logically processed in a way that is different to a standard repetition code that might similarly be encoded in multiple physical modes. In the dual-rail encoding, the standard logical errors that we are interested in (Pauli errors, leakage etc.) now acquire contributions from multiple physical modes, and as such, the ways in which the underlying physical error mechanisms propagate and manifest in the space of the error-detected logical qubit are quite novel.\\

A convenient metric for quantifying the dual-rail performance is the SPAM matrix:  

\newcommand{\ProbAssign}[2]{$p\left(``#1" \mid \ket{\tilde{#2}}\right)$}

\begin{equation}
\label{eqn:M_matrix}
\hat{M}=\begin{pmatrix}
\begin{tabular}{c c c c}
\ProbAssign{00}{00}&\ProbAssign{00}{10}&\ProbAssign{00}{10}&\ProbAssign{00}{11}\\[1mm]
\ProbAssign{01}{00}&\ProbAssign{01}{01}&\ProbAssign{01}{10}&\ProbAssign{01}{11}\\[1mm]
\ProbAssign{10}{00}&\ProbAssign{10}{01}&\ProbAssign{10}{10}&\ProbAssign{10}{11}\\[1mm]
\ProbAssign{11}{00}&\ProbAssign{11}{01}&\ProbAssign{11}{10}&\ProbAssign{11}{11}\\
\end{tabular}
\end{pmatrix}
\end{equation}

which is the matrix form of the sixteen measurement mapping outcomes presented in Fig.~\ref{fig:fig_2}, and comprehensively quantifies the SPAM errors in the dual-rail space. 

As it is an assignment matrix, if our SPAM were perfect $\hat{M}$ would be the identity matrix. As such, in order to understand our errors, the method of the error budget model is to decompose each of the twelve off-diagonal matrix element probabilities (the error elements) into its first-order contributing physical mechanisms. To do this, we decompose each outcome into its subsystems, e.g.:
$$ p(``00"|\;|\Tilde{10}\rangle) = p(``0"|\;|\Tilde{1}\rangle)_{\text{B}}\;\otimes\;p(``0"|\;|\Tilde{0}\rangle)_{\text{A}}$$

Under this decomposition of each of the matrix elements, it is clear that all of the matrix elements can in fact be computed from all the possible combinations (for each of the two subsystems) of just two error budgets: 
\begin{enumerate}
    \item $p(``0"|\;|\Tilde{1}\rangle)_{i}$: Probability of assigning subsystem $i$'s cavity state to be $``0"$, given that we tried to prepare it in $|1\rangle$.
    \item $p(``1"|\;|\Tilde{0}\rangle)_{i}$: Probability of assigning subsystem $i$'s cavity state to be $``1"$, given that we tried to prepare it in $|0\rangle$.
\end{enumerate}

In some cases of the error budget decomposition we will be interested in the probability of assigning a certain outcome conditioned on already being in a certain state. This is equivalent to ideal state preparation (i.e. the case where we are initially in a given state with unit probability), and we denote this case by removing the tilde on the prepared or initial state as $p(``0"|\;|1\rangle)_{i}$ and $p(``1"|\;|0\rangle)_{i}$. Using this method, in the model we can effectively ``decouple" the state preparation errors from the measurement errors.\\
% and construct a measurement matrix:
% \newcommand{\ProbAssignPerfect}[2]{$p\left(``#1" \mid \ket{#2}\right)$}

% \begin{equation}
% \label{eqn:M_matrix}
% \hat{M}=\begin{pmatrix}
% \begin{tabular}{c c c c}
% \ProbAssignPerfect{00}{00}&\ProbAssignPerfect{00}{10}&\ProbAssignPerfect{00}{10}&\ProbAssignPerfect{00}{11}\\[1mm]
% \ProbAssignPerfect{01}{00}&\ProbAssignPerfect{01}{01}&\ProbAssignPerfect{01}{10}&\ProbAssignPerfect{01}{11}\\[1mm]
% \ProbAssignPerfect{10}{00}&\ProbAssignPerfect{10}{01}&\ProbAssignPerfect{10}{10}&\ProbAssignPerfect{10}{11}\\[1mm]
% \ProbAssignPerfect{11}{00}&\ProbAssignPerfect{11}{01}&\ProbAssignPerfect{11}{10}&\ProbAssignPerfect{11}{11}\\
% \end{tabular}
% \end{pmatrix}
% \end{equation}

We develop this as a first-order model, by which we mean two things. First, we do not treat the cases of $p(``0"|\;|\Tilde{0}\rangle)$ and $p(``1"|\;|\Tilde{1}\rangle)$ -- as there is no first-order (single-error) way for such outcomes to occur. Second, the error expressions for each step in the protocol are constructed such that they would lead to the outcome we are computing -- if it were true that no errors occurred in any of the previous steps. It is important to note that we are considering only first-order errors in \emph{each of the subsystems}, therefore when we consider the logical bit-flip case this is of course second order in the dual-rail space, but still analyzed considering only first order error contributions from each of the subsystems.

In addition to the system parameters defined in the System Properties section, we define the following parameters:

\begin{center}
\begin{tabular}{| c | c |}
\hline
 \textbf{Parameter} & \textbf{Description} \\ [0.5ex]
 \hline\hline
 % $\epsilon_{0}$ & \makecell[l]{Probability of an error in preparing the cavity in $|0\rangle$.\\[1mm]
 % Given our state preparation protocol, the functional form of this error is\\
 % $\epsilon_{0} = n_{th}^{c}\cdot P(``0"|\ket{1})+(1-n_{th}^{c})[(1-P(``1"|\ket{0}))\cdot\frac{n_{th}^{c}t_{M}}{T_{1}^{c}}]$\\
 % i.e. limited by the cases where the cavity is excited but misassigned as $``0"$, \\ or where the cavity is in the ground state but heats during the transmon\\ check measurement.}
 % \\[1mm]
 % \hline
 % $\epsilon_{1}$ & \makecell[l]{Probability of an error in preparing the cavity in $|1\rangle$.\\[1mm]
  % Given our state preparation protocol, the functional form of this error is\\
 % $\epsilon_{1} = \epsilon_{ocp}+(1-\epsilon_{ocp})\cdot\frac{t_{M}}{T_{1}^{c}}$\\
 % i.e. limited by the cases where the OCP pulse initializing $\ket{1}$ failed ($\epsilon_{ocp}$),\\ or where the initialization succeeded but the cavity relaxed during \\the transmon check measurement.
 % }
 
 % \\[1mm]
 % \hline
 $t_{map}$ & \makecell[l]{Duration of cavity-state mapping.\\ Implemented with a truncated Gaussian selective $\pi$-pulse, therefore equals $4\sigma_{s}$} \\[1mm]
 \hline
 $t_{reset}$ & \makecell[l]{Duration of transmon reset.\\ Implemented with a truncated Gaussian unselective $\pi$- pulse, therefore equals $4\sigma_{us}$} \\[1mm]
 \hline
 $p_{us}$ & \makecell[l]{Probability of an unselectivity error on the selective pulse. \\Taking $\chi$ and $\sigma_{s}$ into account, it is computed numerically\\ as $\mathrm{Err} = 1-|\langle 1,g|X_{n=0}^{s}|1,g\rangle|^{2}$.}\\[1mm]
 \hline
 $p_{s}$ & \makecell[l]{Probability of a selectivity error on the unselective pulse. \\Taking $\chi$ and $\sigma_{us}$ into account, it is computed numerically\\ as $\mathrm{Err} = 1-|\langle 1,e|X_{n=0}^{us}|1,g\rangle|^{2}$} \\[1mm]
 \hline
\end{tabular}
\end{center}

\vspace{1cm}

The particular protocol considered is the following protocol for state preparation and measurement (as described in the State Preparation and Measurement Protocols section), where the circuit describes the protocol on one of the dual-rail's subsystems. State preparation followed by $M$ logical measurements with a transmon check before the logical measurements:

\begin{center}
% \hspace*{-2.5cm}   
\begin{quantikz}
\lstick{Cavity}   & \gate[2]{\mathrm{State\;Prep.}}  & \qw & \ctrl{1}\gategroup[2,steps=4,style={dashed,
rounded corners,%fill=blue!10, 
inner xsep=2pt},
background,label style={label position=below,anchor=
north,yshift=-0.5cm}]{{\sc M times}}          & \qw   & \qw   & \qw & \qw  \\
\lstick{Transmon} & \qw   & \gate{\checkmark} & \gate{X_{n=1}^{s}}  & \meter{G/E/...} & \gate[cwires=1]{X_{n=1}^{us}} & \gate{\checkmark} & \qw
\end{quantikz}
\end{center}

In this paper we concentrate on state preparation and single-measurement. Only processes that occur until the end of the first post-mapping transmon measurement are reflected in the outcome of this first logical measurement. Therefore, we consider only the following section of the protocol for the state preparation and single-measurment error budget: 

\begin{center}
\hspace*{-1.5cm}   
\begin{quantikz}
\lstick{Cavity}   & \gate[2]{\mathrm{State\;Prep.}}  & \qw & \ctrl{1}\gategroup[2,steps=2,style={dashed,
rounded corners, inner xsep=2pt},
background,label style={label position=below ,anchor=
north,yshift=-0.5cm}]{Logical Measurement}  & \qw   & \qw  \\
\lstick{Transmon} & \qw   & \gate{\checkmark} & \gate{X_{n=1}^{s}}  & \meter{G/E/...}\slice{} & \qw
\end{quantikz}
\end{center}

The error budgets (for the case where we also account for the state preparation i.e. SPAM) can then be constructed as follows:

\begin{enumerate}
    \item \underline{$p(``0"|\;|\tilde{1}\rangle)$ Error Budget}:
\begin{center}
\hspace*{-2.5cm}
    \begin{tabular}{ c | c | c }
     &\textbf{Error Mechanism} & \textbf{Parameterization} \\ [0.5ex]
     \hline\hline
     (1) & Prep 1 failure & $\epsilon_{1}$\\ [0.5ex]
     \hline
      (2) & \makecell{Cavity decay during transmon check \\ or first half of cavity-state mapping} & $\frac{2\;t_{M}+t_{map}}{2\;T_{1}^{c}}$\\ [0.5ex]
     \hline
      (3) & $T_{2}$ event on transmon during cavity-state mapping & $\frac{t_{map}}{T_{2R}^{t}}$\\ [0.5ex]
     \hline
      (4) & Transmon decay during first half of readout & $\frac{t_{M}}{2\;T_{1\;RO}^{t}}$\\ [0.5ex]
     \hline
      (5) & Readout classification error & $p_{miss}$\\ [0.5ex]
     \hline
    \end{tabular}
\end{center}
    \item \underline{$p(``1"|\;|\tilde{0}\rangle)$ Error Budget}:
\begin{center}
\hspace*{-2.5cm}
    \begin{tabular}{ c | c | c }
     &\textbf{Error Mechanism} & \textbf{Parameterization} \\ [0.5ex]
     \hline\hline
     (1) & Prep 0 failure & $\epsilon_{0}$\\ [0.5ex]
     \hline
     (2) & \makecell{Cavity heating during transmon check, \\ or first half of cavity-state mapping} & $\frac{n_{th}^{c}(2\;t_{M}+t_{map})}{2\;T_{1}^{c}}$\\ [0.5ex]
     \hline
      (3) & \makecell{Transmon heating during second half of cavity-state mapping} & $\frac{n_{th}^{t}\;t_{map}}{2\;T_{1}^{t}}$\\ [0.5ex]
     \hline
      (4) & \makecell{Transmon heating during first half of readout} & $\frac{n_{th\;RO}^{t}\;t_{M}}{2\;T_{1\;RO}^{t}}$\\ [0.5ex]
     \hline
      (5) & Unselectivity of selective pulse erroneously rotating the transmon & $p_{us}$\\ [0.5ex]
     \hline
      (6) & Readout classification error & $p_{miss}$\\ [0.5ex]
     \hline
    \end{tabular}
\end{center}
\end{enumerate}

Since we choose to account only for first-order events in this model, we make the assumption that there are no correlations between the errors--as any such correlation would be second-order--and therefore compute the probability of the expected outcome by simply summing the terms in the error budget. 

For instance, for each of the three error events that we care about: \emph{erasure} assignments, \emph{leakage detection error} assignments, and \emph{logical SPAM error} assignments, we  construct the error budget as:
\begin{itemize}
    \item \underline{Erasure:} 
    \begin{equation*}
    \begin{aligned}
        p(``00"|\;|\tilde{01}\rangle) &=
        p(``0"|\;|\Tilde{0}\rangle)_{B}\cdot p(``0"|\;|\Tilde{1}\rangle)_{A}\\
        &=  (1-\sum_{i=1}^{6}[\;p(``1"|\;|\tilde{0}\rangle)\;\text{Budget}\;]_{B}^{(i)} )\cdot(\sum_{i=1}^{5}[\;p(``0"|\;|\tilde{1}\rangle)\;\text{Budget}\;]_{A}^{(i)})
    \end{aligned}   
    \end{equation*}
    \item \underline{Leakage-Detection Error:}
    \begin{equation*}
        \begin{aligned}
            p(``10"|\;|\Tilde{00}\rangle) &= p(``1"|\;|\Tilde{0}\rangle)_{B}\cdot  p(``0"|\;|\Tilde{0}\rangle)_{A} \\
            &= (\sum_{i=1}^{6}[\;p(``1"|\;|\Tilde{0}\rangle)\;\text{Budget}\;]_{B}^{(i)})\cdot(1-\sum_{i=1}^{6}[\;p(``1"|\;|\tilde{0}\rangle)\;\text{Budget}\;]_{A}^{(i)})
        \end{aligned}
    \end{equation*}
    \item \underline{Logical SPAM Error:}
    \begin{equation*}
        \begin{aligned}
            p(``10"|\;|\tilde{01}\rangle) &= p(``1"|\;|\Tilde{0}\rangle)_{B}\cdot p(``0"|\;|\Tilde{1}\rangle)_{A} \\
            &= (\sum_{i=1}^{6}[\;p(``1"|\;|\tilde{0}\rangle)\;\text{Budget}\;]_{B}^{(i)})\cdot(\sum_{i=1}^{5}[\;p(``0"|\;|\tilde{1}\rangle)\;\text{Budget}\;]_{A}^{(i)})
        \end{aligned}
    \end{equation*}
\end{itemize}

Similarly, we can construct the error budget for all other dual-rail SPAM error outcomes. In the next two subsections we show two ways in which we can use the error budget formalism to analyze the individual contributions of various error mechanism to our measured dual-rail performance. 

\subsection{Breakdown of SPAM Error}

One immediate application of the error budget is to breakdown the each of the SPAM error-outcomes to its contributing factors. Namely, we can focus on the SPAM experiment described in the Characterizing dual-rail SPAM section where we prepare all four joint cavity states in turn and then measure the probabilities of assigning each of the four possible joint cavity states. In this experiment there are twelve error outcomes, or the twelve off-diagonal elements of the corresponding SPAM matrix (see eqn.~\ref{eqn:M_matrix}). 

As established in this paper, our measured dual-rail outcomes have contributions both from the intrinsic hardware coherences as well as from errors introduced by the protocols of the state preparation and measurement. With an eye towards future work on hardware and protocol optimization in order to improve the dual-rail cavity qubit's performance, we show the error breakdown of the six most important SPAM error terms, in three groups, as an initial guide to where such work would ideally first be focused. 

Specifically, the three error groups that we care about are erasure assignments, where we prepare a logical state and it gets assigned to $``00"$ (see Table~\ref{tab:erasure_assignments}); leakage detection error assignments, where the leakage state $\ket{00}$ gets assigned to one of the logical states $``01"$ or $``10"$ (see Table~\ref{tab:erasure_conversion_assignment_error}); and logical SPAM errors, where one of the logical states gets assigned to the other logical state (see Table~\ref{tab:logical_SPAM_erros}). For the bit-flip SPAM outcome we quote only the top five error contributions.

From our error analysis we notice two things. First, we see that the particular values of the hardware parameters of each of the dual-rail's subsystems give rise to very different relative error contributions. In other words, for example, a ``00" assignment having prepared $|\tilde{01}\rangle$ as opposed to $|\tilde{10}\rangle$ will have a different error budget. This is important to understand, particularly for the case of logical Pauli errors, as these are second-order error that will therefore occur due to error contributions from both subsystems. 

Second, our analysis quantifies and emphasizes that as expected, actual cavity transitions are not the dominant error source. Instead, our SPAM errors are dominated by ancillae transmon errors. \emph{Erasure} assignments (Table~\ref{tab:erasure_assignments}) where a prepared logical state gets assigned as an erasure, both get very significant contributions ($30\%-50\%$) from transmon decay events during the readout; \emph{leakage detection error} assignments (Table~\ref{tab:erasure_conversion_assignment_error}) are dominated by transmon readout classification errors (60\%-90\%); and a dominant contribution to \emph{logical SPAM error} assignments (Table~\ref{tab:logical_SPAM_erros}) is the combination of these errors on each subsystem ($\sim30\%$).

\begin{table}[h]
    \centering
    \caption{\textbf{\emph{Erasure} Assignments}}
\begin{tabular}{ c | c | c | c}
% \caption{The caption}
\centering
 \textbf{Error Outcome} & \textbf{Error Description} & \textbf{\makecell{Raw \\ Contribution}} & \textbf{\makecell{Relative \\ Contribution}} \\
 \hline\hline
 \rule{0pt}{2.6ex}
 $p(``00"||\Tilde{01}\rangle)$ & Transmon $T_{1}$ during readout & 2.77e-2 & 47.75\%\\ \cmidrule{2-4}
  &  Transmon $T_{2}$ during mapping & 1.24e-2 & 21.36 \%\\\cmidrule{2-4}
  & State ($\ket{1}$) preparation failure & 9.26e-3 & 15.99 \% \\\cmidrule{2-4}
  & Cavity $T_{1}$ during check or mapping & 4.6e-3 &7.94 \% \\\cmidrule{2-4}
  & Readout classification error & 4.03e-3 & 6.96 \% \\
 \hline\hline
 \rule{0pt}{2.6ex}
 $p(``00"||\Tilde{10}\rangle)$ & State ($\ket{1}$) preparation failure & 2.84e-2 & 40.73 \% \\\cmidrule{2-4}
  & Transmon $T_{1}$ during readout & 2.18e-2 & 30.40 \% \\\cmidrule{2-4}
  & Cavity $T_{1}$ during check or mapping & 1.21e-2 & 17.37 \% \\\cmidrule{2-4}
  & Transmon $T_{2}$ during mapping & 6.82e-3 & 9.80 \% \\\cmidrule{2-4}
  & Readout classification error & 1.18e-3 & 1.69 \% \\
 \hline\hline
\end{tabular}
    \label{tab:erasure_assignments}
\end{table}

\begin{table}[h]
    \centering
    \caption{\textbf{\emph{Erasure conversion error} assignments}}
\begin{tabular}{ c | c | c | c}
 \textbf{Error Outcome} & \textbf{Error Description} & \textbf{\makecell{Raw \\ Contribution}} & \textbf{\makecell{Relative \\ Contribution}} \\
 \hline\hline
 \rule{0pt}{2.6ex}
 $p(``01"||\Tilde{00}\rangle)$ & Readout classification error & 3.14e-3 & 89.05 \%\\\cmidrule{2-4}
  & Transmon heating during readout & 2.771e-4 & 7.86 \%\\\cmidrule{2-4}
  & State ($\ket{0}$) preparation failure & 5.48e-5 & 1.56 \%\\\cmidrule{2-4}
  & Transmon heating during mapping & 4.60e-5 & 1.30 \%\\\cmidrule{2-4}
  & Cavity heating during check or mapping & 4.60e-6 & 0.13 \%\\\cmidrule{2-4}
  & Unselectivity of selective mapping pulse & 3.75e-6 & 0.10\%\\
 \hline\hline
 \rule{0pt}{2.6ex}
 $p(``10"||\Tilde{00}\rangle)$ & Readout classification error & 3.22e-3 & 58.90 \%\\\cmidrule{2-4}
  & Transmon heating during readout & 2.12e-3 & 38.74 \%\\\cmidrule{2-4}
  & State ($\ket{0}$) preparation failure & 6.39e-3 & 1.17 \%\\\cmidrule{2-4}
  & Transmon heating during mapping & 3.54e-5 & 0.65 \%\\\cmidrule{2-4}
  & Unselectivity of selective mapping pulse & 1.75e-5 & 0.32 \%\\\cmidrule{2-4}
  & Cavity heating during check or mapping & 1.21e-5 & 0.22 \%\\
 \hline\hline
\end{tabular}
\label{tab:erasure_conversion_assignment_error}
\end{table}

\begin{table}[]
    \centering
    \caption{\textbf{\emph{Logical SPAM error} assignments}}
\begin{tabular}{ c | c | c | c}
 \textbf{Error Outcome} & \textbf{Error Description} & \textbf{\makecell{Raw \\ Contribution}} & \textbf{\makecell{Relative \\ Contribution}} \\
 \hline\hline
 $p(``10"||\Tilde{01}\rangle)$ & \makecell{Readout classification error (B) and\\
  Transmon $T_{1}$ during readout (A)} & 8.91e-5 & 28.13 \%\\\cmidrule{2-4}
  & \makecell{Transmon heating during readout (B) and\\
  Transmon decay during readout (A)} & 5.86e-5 & 18.50 \%\\\cmidrule{2-4}
  & \makecell{Readout classification error (B) and\\
  Transmon $T_{2}$ during mapping (A)} & 3.98e-5 & 12.58 \%\\\cmidrule{2-4}
  & \makecell{Transmon heating during readout (B) and\\
  Transmon $T_{2}$ during mapping (A)} & 2.62e-5 & 8.27 \%\\\cmidrule{2-4}
  & \makecell{Transmon heating during readout (B) and\\
  State ($\ket{1}$) preparation failure (A)} & 1.96e-5 & 6.19 \%\\
 \hline\hline
 $p(``01"||\Tilde{10}\rangle)$ & \makecell{State ($\ket{1}$) preparation failure (B) and \\
 Readout classification error (A)} & 8.90e-5 & 36.27 \%\\\cmidrule{2-4}
  & \makecell{Transmon $T_{1}$ during readout (B) and \\
  Readout classification error} & 6.64e-5 & 27.07 \%\\\cmidrule{2-4}
  & \makecell{Cavity $T_{1}$ during check or mapping (B) and \\
  Readout classification error (A)} & 3.79e-5 & 15.47 \%\\\cmidrule{2-4}
  & \makecell{Transmon $T_{2}$ during mapping (B) and \\
  Readout classification error (A)} & 2.14e-5 & 8.73 \%\\\cmidrule{2-4}
  & \makecell{State ($\ket{1}$) preparation failure (B) and \\
  Transmon heating during readout (A)} & 7.85e-6 & 3.20 \%\\\cmidrule{2-4}
  & \makecell{Transmon $T_{1}$ during readout (B) and \\
  Transmon heating during readout} & 5.86e-6 & 2.39 \%\\
 \hline\hline
\end{tabular}
    \label{tab:logical_SPAM_erros}
\end{table}

\subsection{Breakdown of Bit-Flip Error}
 
The logical lifetime of the dual-rail cavity qubit is characterized using the bit-flip experiment as described in the Measuring Bit-Flip Errors section. 
We know that the intrinsic physical mechanism of the logical bit flip is dominated by the rare case of a double transition event: a decay in the cavity prepared in $\ket{1}$ \emph{and} a heating event in the cavity prepared in $\ket{0}$. However, in the process of measuring the dual-rail's state in order to characterize its bit-flip time -- many more errors are introduced. These state-mapping and measurement errors result in an ``effective", or ``measured" bit-flip rate that is much faster than the estimated intrinsic bit-flip rate that is solely due to double cavity transitions.  

In order to better understand and support this distinction between the ``measured" bit-flip and the ``intrinsic" bit-flip, we can use our models to construct an error budget of the relative contributions to each of the bit-flip rates.\\

The probabilities of measuring a logical bit-flip, when preparing $0_{\mathrm{L}}$ or $1_{\mathrm{L}}$ are defined as, respectively
\begin{equation*}
\label{eqn:logical_bit_flip_prob_expr}
\begin{aligned}
   P_{1_\mathrm{L}}(t) &= \frac{P(m_{10}(t)|\;\text{prep }01)}{P(m_{10}(t)|\;\text{prep }01)+P(m_{01}(t)|\;\text{prep }01)} \\[2mm]
   P_{0_\mathrm{L}}(t) &= \frac{P(m_{01}(t)|\;\text{prep }10)}{P(m_{10}(t)|\;\text{prep }10)+P(m_{01}(t)|\;\text{prep }10)}
\end{aligned}
\end{equation*}
where $m_{01}$ and $m_{10}$ are the raw probabilities of assigning the states $``01"$ and $``10"$ (i.e before re-normalization of the logical subspace), and are given by the sum of the probabilities of assigning the outcome $``01"$ or $``10"$ (denoted $\mathcal{O}$ in general) conditioned on being in any one of the dual-rail states $\ket{k}$:
\begin{equation}
\label{eqn:m_O}
    m_{\mathcal{O}}(t) = \sum_{k \in \{\ket{00}, \ket{01}, \ket{10}, \ket{11}\}} P_{\ket{k}}(t)\;\cdot\;P \left( \mathcal{O} \mid \ket{k} \right) 
\end{equation}

By solving a simple system of ODEs we can obtain each of the $P_{\ket{k}}(t)$ terms of the \textbf{dual-rail state evolutions} in eqn.~\ref{eqn:m_O}, and using our model we can compute each of the $P \left( \mathcal{O} \mid \ket{k} \right)$ terms of the \textbf{assignment probabilities} in eqn.~\ref{eqn:m_O}. In this case when we compute the assignment probabilities we do not include state preparation errors, since we are conditioning the outcomes on already being in each of the dual-rail states.

\textbf{Dual-Rail State Evolutions}\hspace{1mm} 
As we are interested in the long-time behavior of the logical bit-flip, we solve the system of ODEs for the two-level cavity system in order to extract the exact expressions of the evolution of the ground and excited populations of the cavity states, and construct the expressions for the dual-rail state evolutions, $P_{\ket{k}}(t)$ from these. 

Treating each cavity as a two-level system, the evolution of the ground and excited state populations can be expressed as the following system of ODEs:
\begin{equation}
\label{eqn:ODE}
    \begin{pmatrix}
        \Dot{P}_{0}\\\Dot{P}_{1}
    \end{pmatrix} = 
    \begin{pmatrix}
        -\Gamma_{\uparrow} & \Gamma_{\downarrow}\\
        \Gamma_{\uparrow} & -\Gamma_{\downarrow}\\
    \end{pmatrix}
    \begin{pmatrix}
        P_{0}\\P_{1}
    \end{pmatrix}
\end{equation}
where $\Gamma_{\downarrow}=1/T_{1}$ and $\Gamma_{\uparrow}=n_{th}/T_{1}$. 

Using the error budget we can compute the $\ket{0}$ and $\ket{1}$ state preparation errors: $\epsilon_{0}$ and $\epsilon_{1}$, with which we define the initial conditions for the ODEs. For the case where $\ket{\psi_{0}}=|\tilde{0}\rangle$: $\{P_{0}(0)=1-\epsilon_{0},\; P_{1}(0)=\epsilon_{0}\}$, and for the case where $\ket{\psi_{0}}=|\tilde{1}\rangle$: $\{P_{0}(0)=\epsilon_{1},\; P_{1}(0)=1-\epsilon_{1}\}$.
Solving eqn.~\ref{eqn:ODE} with these initial conditions we find for the case where $\ket{\psi_{0}}=|\tilde{0}\rangle$:
\begin{equation*}
    \begin{aligned}
        P_{(0|\text{init}\;\ket{0})}(t) &= \frac{\Gamma_{\downarrow}}{\Gamma_{\uparrow}+\Gamma_{\downarrow}}+e^{-(\Gamma_{\uparrow}+\Gamma_{\downarrow})t}\;(\frac{\Gamma_{\uparrow}}{\Gamma_{\uparrow}+\Gamma_{\downarrow}}-\epsilon_{0})\\
        P_{(1|\text{init}\;\ket{0})}(t) &= \frac{\Gamma_{\uparrow}}{\Gamma_{\uparrow}+\Gamma_{\downarrow}}\;(1-e^{-(\Gamma_{\uparrow}+\Gamma_{\downarrow})t})+\epsilon_{0}\;e^{-(\Gamma_{\uparrow}+\Gamma_{\downarrow})t}
    \end{aligned}
\end{equation*}
and for the case where $\ket{\psi_{0}}=|\tilde{1}\rangle$:
\begin{equation*}
    \begin{aligned}
        P_{(0|\text{init}\;\ket{1})}(t) &= \frac{\Gamma_{\downarrow}}{\Gamma_{\uparrow}+\Gamma_{\downarrow}}\;(1-e^{-(\Gamma_{\uparrow}+\Gamma_{\downarrow})t})+\epsilon_{1}\;e^{-(\Gamma_{\uparrow}+\Gamma_{\downarrow})t}\\
        P_{(1|\text{init}\;\ket{1})}(t) &= \frac{\Gamma_{\uparrow}}{\Gamma_{\uparrow}+\Gamma_{\downarrow}} - e^{-(\Gamma_{\uparrow}+\Gamma_{\downarrow})t}(\epsilon_{1}-\frac{\Gamma_{\downarrow}}{\Gamma_{\uparrow}+\Gamma_{\downarrow}})
    \end{aligned}
\end{equation*}

\begin{figure}[h]
    \centering
    \includegraphics[width=0.6\textwidth]{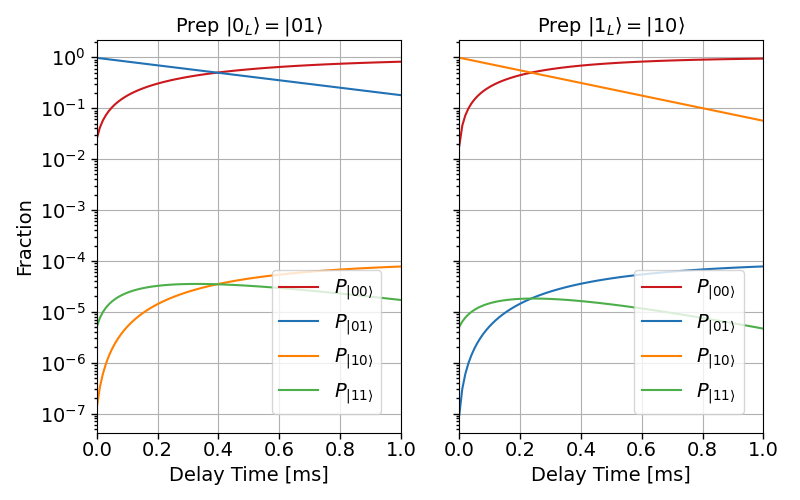}
    \caption{Time evolution of the four dual-rail basis states over 1 ms, in the case where $|0_{\mathrm{L}}\rangle$ is prepared (left), and in the case where $|1_{\mathrm{L}}\rangle$ is prepared (right).}
    \label{fig:ODEs}
    \end{figure}

With these solutions we can then construct the functional form of the evolution of the dual-rail cavity states as $P_{(\ket{00}|\text{init }\ket{01})} = P_{(0|\text{init }\ket{0})}^{B}(t)\cdot P_{(0|\text{init }\ket{1})}^{A}(t)$ etc., where $\Gamma_{\uparrow}$ and $\Gamma_{\downarrow}$ for each cavity's solution are parameterized by the cavity's measured $T_{1}$ and $n_{th}$. The result using our system's parameters are plotted in Fig.~\ref{fig:ODEs}.\\

\begin{figure}[h]
    \centering
    \includegraphics[width=0.7\textwidth]{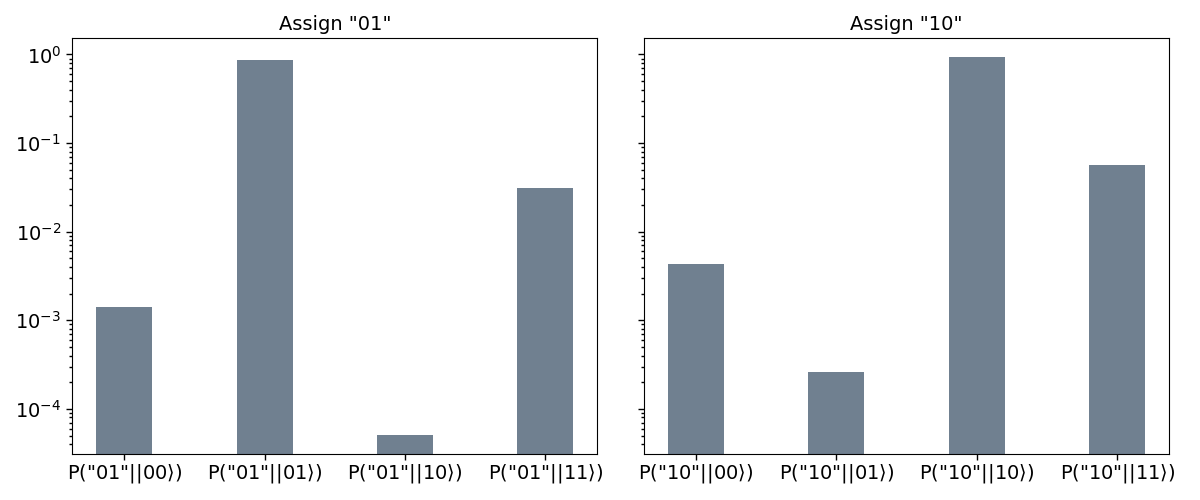}
    \caption{Probability of assigning the logical basis outcomes--conditioned on being in each of the dual-rail basis states; computed using the error budget. The probability of assigning ``01" from each of the basis states in plotted on the left, and the probability of assigning ``10" from each of the basis states is plotted on the right.}
    \label{fig:logical_assignments}
    \end{figure}

\textbf{Assignment Probabilities}\hspace{1mm}
Using our model, we can also compute the assignment probabilities conditioned on being in any one of the dual-rail states, $P \left( \mathcal{O} \mid \ket{k} \right)$. The results are plotted in Fig.~\ref{fig:logical_assignments}.\\

These assignment probabilities are similar to the left two panels in Fig.\ref{fig:fig_2}. However, as noted, in Fig.\ref{fig:fig_2} the numbers reflect the SPAM outcomes (including state preparation) whereas Fig.~\ref{fig:logical_assignments} reflects only the measurement errors, assuming perfect state preparation.

In general, we are most interested in using the model to quantify two things. We know that the measured bit-flip rate gets some contribution from the intrinsic bit-flip (a heating event on one cavity and a decay event on the other cavity), and some contribution from measurement errors introduced when trying to measure the bit-flip rate. We are interested then in:
\begin{enumerate}
    \item Understanding what fraction of the measured bit-flip rate is due to the true intrinsic bit-flip.
    \item For the measured bit-flip rate, which measurement assignment channel $P_{\ket{k}}\cdot P \left( \mathcal{O} \mid \ket{k} \right) $ dominates the measured outcome.
\end{enumerate}

To answer these questions we single out the three cases of interest:
\begin{enumerate}
    \item \textbf{Only intrinsic bit-flip}. The case where finite cavity $T_{1}$ and $n_{th}$ lead to real logical bit-flips at long delay times; we assume perfect measurements in order to isolate this mechanism.

    In the case where there is no measurement error, a logical state will only be measured if it is actually the underlying physical state, i.e. the measurement outcomes are:
\begin{equation*}
    \begin{aligned}
        m_{01} &= P_{\ket{01}}\\
        m_{10} &= P_{\ket{10}}\\
    \end{aligned}
\end{equation*}

    For delay times on the order of the bare cavity $T_{1}$s, we can approximately express the intrinsic logical bit-flip probabilities as
    \begin{equation*}
        \begin{aligned}
            P(\text{flip}\;|\;\text{initial }\ket{0_{L}}) &= P_{1_{L}} =  \frac{(1-e^{-n_{th}t/T_{1}^{(B)}})(1-e^{-t/T_{1}^{(A)}})}{(e^{-n_{th}t/T_{1}^{(B)}})(e^{-t/T_{1}^{(A)}})+(1-e^{-n_{th}t/T_{1}^{(B)}})(1-e^{-t/T_{1}^{(A)}})}
        \end{aligned}
    \end{equation*}
    and
    \begin{equation*}
        \begin{aligned}
            P(\text{flip}\;|\;\text{initial }\ket{1_{L}}) &= P_{0_{L}} =  \frac{(1-e^{-t/T_{1}^{(B)}})(1-e^{-n_{th}t/T_{1}^{(A)}})}{(1-e^{-t/T_{1}^{(B)}})(1-e^{-n_{th}t/T_{1}^{(A)}})+(e^{-t/T_{1}^{(B)}})(e^{-n_{th}t/T_{1}^{(A)}})}
        \end{aligned}
    \end{equation*}

    At short times, from first-order Taylor expansion we can extract the expression:
    \begin{equation*}P(\text{flip}\;|\;\text{initial }\ket{0_{L}})
    = n_{th}^{(B)}\frac{t^{2}}{T_{1}^{(A)}\;T_{1}^{(B)}} = n_{th}^{(B)}\;\kappa_{A}\;\kappa_{B}\;t^{2}
    \end{equation*}
    and
    \begin{equation*}P(\text{flip}\;|\;\text{initial }\ket{1_{L}})
     = n_{th}^{(A)}\frac{t^{2}}{T_{1}^{(A)}\;T_{1}^{(B)}} = n_{th}^{(A)}\;\kappa_{A}\;\kappa_{B}\;t^{2}
    \end{equation*}
    showing that at short times the intrinsic bit-flip is quadratic in time. 

    At long times compared to the bare cavity $T_{1}$ we use the exact expressions for the population probabilities, which are reflected in the plots below. 

    \item \textbf{Only measurement error bit-flip}. In this case we set the cavity $n_{th}$ to zero, such that there can by no real intrinsic bit-flip, and the probability of assigning a bit-flip is only due to the measurement.

    This means that for the case where $\ket{\psi_{0}}=\ket{1_{L}}$:
\begin{equation*}
\begin{aligned}
m_{01} &= P_{\ket{00}}\cdot P(``01"|\;\ket{00})+P_{\ket{10}}\cdot P(``01"|\;\ket{10})\\[3mm]
m_{10} &= P_{\ket{00}}\cdot P(``10"|\;\ket{00})+P_{\ket{10}}\cdot P(``10"|\;\ket{10})
\end{aligned}    
\end{equation*}
and for the case where $\ket{\psi_{0}}=\ket{0_{L}}$:
\begin{equation*}
    \begin{aligned}
m_{01} &= P_{\ket{00}}\cdot P(``01"|\;\ket{00})+P_{\ket{01}}\cdot P(``01"|\;\ket{01})\\[3mm]
m_{10} &= P_{\ket{00}}\cdot P(``10"|\;\ket{00})+P_{\ket{01}}\cdot P(``10"|\;\ket{01})
    \end{aligned}
\end{equation*}

    from which, using the ODE solutions and the assignment probabilities, we can construct the functional form of the logical bit-flip in each case. 
    
    \item \textbf{Intrinsic bit-flip and measurement errors}. The realistic case where we have both finite cavity coherences and finite measurement errors.

    In the case where we allow for an intrinsic bit-flip, i.e. $n_{th}\neq0$, as well as measurement errors, there is a probability of being in, or measuring any one of the four states. Therefore, we must account for all terms in eqn.~\ref{eqn:m_O}.
    Again, substituting the appropriate expression for $P_{\ket{k}}$ and the assignment probabilities for each of the cases we then construct the functional form of the logical bit-flips. 

    \end{enumerate}

\begin{figure}[h]
    \centering
    \includegraphics[width=0.6\textwidth]{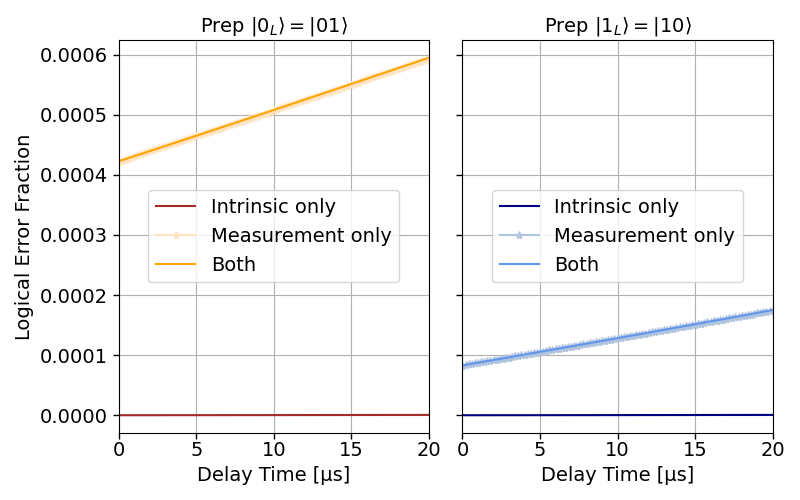}
    \caption{Probability of a logical bit-flip from different error channels, given $|0_{\mathrm{L}}\rangle$ preparation (left) and $|1_{\mathrm{L}}\rangle$ preparation (right). \emph{Intrinsic only} represents the case where measurements are perfect and therefore a bit-flip can only occur as a result of double-transitions; \emph{measurement only} represents the case where the cavity $n_{th}=0$, and therefore bit-flips are only apparent bit-flips due to measurement errors; \emph{both} represents the case where both of these error channels are finite.}
    \label{fig:bit_flip_contributions_20us}
    \end{figure}

\begin{figure}[h]
    \centering
    \includegraphics[width=0.6\textwidth]{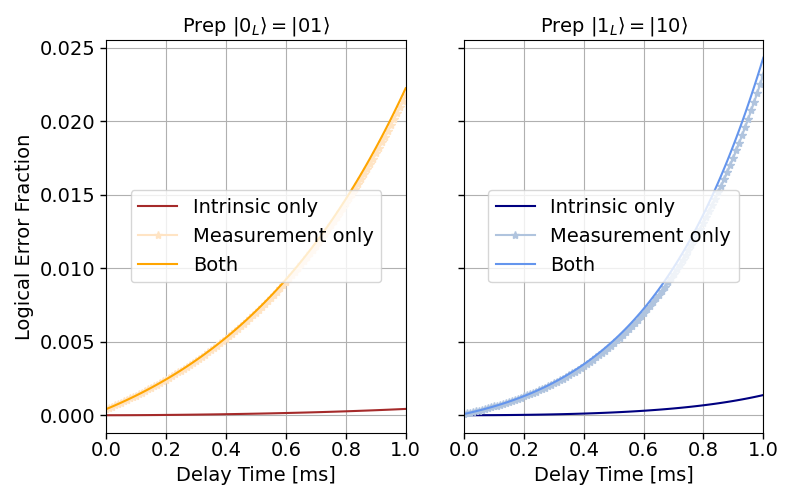}
    \caption{Same as Fig.~\ref{fig:bit_flip_contributions_20us}, but delayed out to 1 ms.}
    \label{fig:bit_flip_contributions_1ms}
    \end{figure}

\begin{figure}[h]
    \centering
    \includegraphics[width=0.6\textwidth]{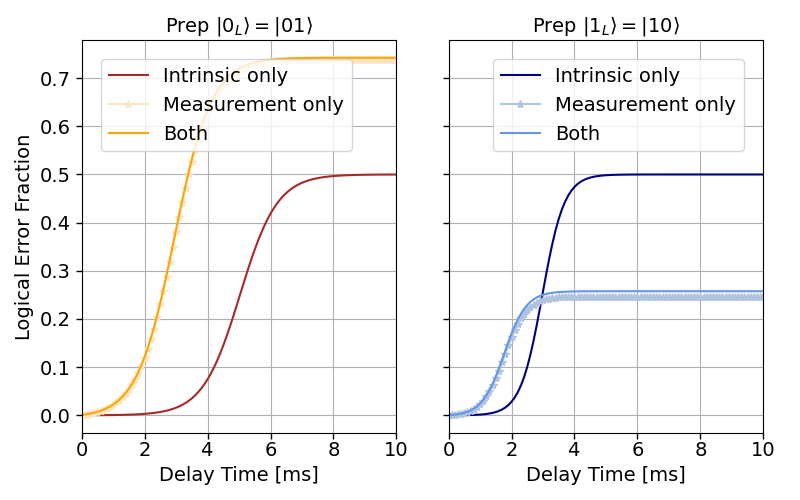}
    \caption{Same as Fig.~\ref{fig:bit_flip_contributions_20us}, but delayed out to 10 ms.}
    \label{fig:bit_flip_contributions_10ms}
    \end{figure}

    We can plot each of these three cases for the delay times of interest discussed in this paper: 20 $\mu$s and 1 ms. The results are presented in Fig.~\ref{fig:bit_flip_contributions_20us}, and Fig.~\ref{fig:bit_flip_contributions_1ms}, respectively. 
    In addition, we can plot the three cases for a delay time of 10 ms, longer than we measure, in order to observe the expected saturation of the logical bit-flip probability; the results are presented in Fig.~\ref{fig:bit_flip_contributions_10ms}. 

    The plots in Figures \ref{fig:bit_flip_contributions_20us} and \ref{fig:bit_flip_contributions_1ms} show the extent to which the measured bit-flip probability is dominated by errors of the measurement itself, and not by the true intrinsic bit-flip. As described in the Measuring Bit-Flip Errors section, this shows why via measurement we can only bound the bit-flip rate to 1/($\sim$80 ms), while given the cavity coherences we estimate the true intrinsic bit-flip rate is on the order of 1/($\sim$2.5\;s).

    Specifically, we can now further use these computed state evolutions and assignment probabilities in order to answer our two initial questions. 

\begin{figure}[h]
    \centering
    \includegraphics[width=0.6\textwidth]{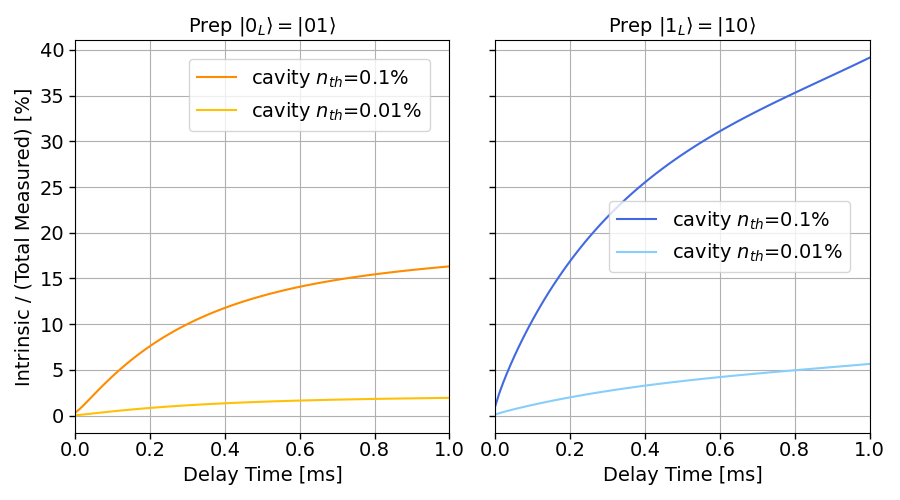}
    \caption{Fraction that the intrinsic bit-flip contributes to the total measured bit-flip, for delay times up to 1 ms, given prep $|0_{\mathrm{L}}\rangle$ (left) and $|1_{\mathrm{L}}\rangle$ (right). Plotted both for cavity $n_{\mathrm{th}}$ of $0.1\%$ and $0.01\%$.}
    \label{fig:reltive_fraction_bit_flip}
    \end{figure}

    First, the fraction that the intrinsic bit-flip constitutes of the total measured bit-flip probability is simply the ``Measurement only" outcome plotted in Fig.~\ref{fig:bit_flip_contributions_1ms}, divided by the total measured bit-flip (``Both" in Fig.~\ref{fig:bit_flip_contributions_1ms}). This ratio, as a function of the delay time, is plotted in Fig.~\ref{fig:reltive_fraction_bit_flip}. As described in section~\ref{sec:cavity_nth_msmt}, we bound the cavity thermal population to be between 1e-3 and 1e-4. Therefore, in Fig.~\ref{fig:reltive_fraction_bit_flip} we plot this ratio for both values of cavity thermal population, in order to establish a the range that we expect this contribution to be in our case. 
    For cavity thermal populations of $10^{-4}$, which is closer to what we measure, we learn from the model that the intrinsic bit-flip contributes $(0.04-0.13)\%$ at $1\;\mathrm{\mu s}$, and $(1.93-5.64)\%$ $1 \;\mathrm{ms}$, where the range is given by the two state preparations. 

    % comment on long-time saturation.
    Using the model we can also confirm our expectations of the contributions of the respective bit-flip channels at very long times compared to the cavity lifetimes. At these long times, the cavities will have thermalized, and we expect the intrinsic bit-flip for both initial state preparations to approximately (assuming similar thermal populations in both cavities) saturate to 
    \begin{equation*}
        \frac{n_{th}\cdot(1-n_{th})}{n_{th}\cdot(1-n_{th})+(1-n_{th})\cdot n_{th}} = \frac{1}{2}
    \end{equation*}
    Similarly, at long times when the cavities have thermalized, we expect the contribution of the measurement error-induced bit-flip to be given by the ratio of the probabilities of misassigning $\ket{00}$ to each of the bit-flip outcomes:
    \begin{equation*}
    \begin{aligned}
        P(\text{flip}|\text{initial\;}\ket{0_{\mathrm{L}}}) &= \frac{P(``10"|\ket{00})}{P(``01"|\ket{00})+P(``10"|\ket{00})}\\[2mm]
        P(\text{flip}|\text{initial\;}\ket{1_{\mathrm{L}}}) &= \frac{P(``01"|\ket{00})}{P(``01"|\ket{00})+P(``10"|\ket{00})}
    \end{aligned}
    \end{equation*}
    From the logical state assignments that we compute using the model (see Fig.~\ref{fig:logical_assignments}) we can extract these numbers and see that $P(``10"|\ket{00}) = 4.26\times10^{-3}$, and $P(``01"|\ket{00}) = 1.42\times10^{-3}$. Therefore we expect the logical bit-flip probability at long times to saturate to 0.75 for prep $\ket{0_{\mathrm{L}}}$, and 0.25 for prep $\ket{1_{\mathrm{L}}}$.
    
    Both of these expectations are confirmed by the model, see Fig.~\ref{fig:bit_flip_contributions_10ms}.\\

    Second, for the measured bit-flip rate, we can decompose the measured number into the contributions it gets from assigning each of the dual-rail states to the bit-flip outcome. We do this to answer our question about which of the assignment channels dominates the logical bit-flip, and to what extent.

    Specifically, we decompose it as follows. As described in eqn.~\ref{eqn:logical_bit_flip_prob_expr}, the probability of a logical bit-flip outcome for the case where the initial state is $\ket{0_{L}}$ is given by 
    \begin{equation*}
        P_{1_L}(t) = \frac{P(m_{10}(t)|\;\text{prep }01)}{P(m_{10}(t)|\;\text{prep }01)+P(m_{01}(t)|\;\text{prep }01)}
    \end{equation*}
    We can write the numerator out explicitly, in which case
    \begin{equation*}
        P_{1_L}(t) =  \frac{P_{|00\rangle}\cdot P(``10"||00\rangle)+P_{|01\rangle}\cdot P(``10"||01\rangle)+P_{|10\rangle}\cdot P(``10"||10\rangle)+P_{|11\rangle}\cdot P(``10"||11\rangle)}{P(m_{10}(t)|\;\text{prep}\;01)+P(m_{01}(t)|\;\text{prep}\;01)}
    \end{equation*}   
    
    % \begin{equation*}
    %     P_{1_L}(t) = \frac{P_{|00\rangle}\cdot P(``10"||00\rangle)}{P(m_{10}(t)|\;\text{prep}\;01)+P(m_{01}(t)|\;\text{prep}\;01)} + \frac{P_{|01\rangle}\cdot P(``10"||01\rangle)}{P(m_{10}(t)|\;\text{prep}\;01)+P(m_{01}(t)|\;\text{prep}\;01)} + \frac{P_{|10\rangle}\cdot P(``10"||10\rangle)}{P(m_{10}(t)|\;\text{prep}\;01)+P(m_{01}(t)|\;\text{prep}\;01)} + \frac{P_{|11\rangle}\cdot P(``10"||11\rangle)}{P(m_{10}(t)|\;\text{prep}\;01)+P(m_{01}(t)|\;\text{prep}\;01)}
    % \end{equation*}

    Each term in this last expression can be interpreted as the contribution to the total measured bit-flip from  the probability of assigning each of dual-rail states to the bit-flip outcome. 

\begin{figure}[h]
    \centering
    \includegraphics[width=0.7\textwidth]{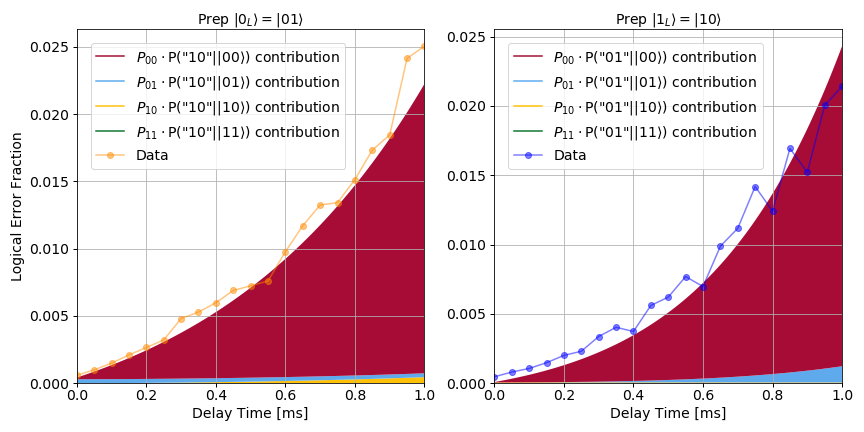}
    \caption{The measured bit-flip probability gets contributions from each of the dual-rail basis states getting assigned to the bit-flip outcome. Here we plot each of these contributions in order to show the relative contributions of each assignment channel to the measured bit flip for prep $|0_{\mathrm{L}}\rangle$ (left) and prep $|1_{\mathrm{L}}\rangle$ (right). The sum of the contributions should correspond to the data.} 
    \label{fig:decomposed_bit_flip}
    \end{figure}

    In Fig.~\ref{fig:decomposed_bit_flip} we plot the contribution of each of these bit-flip assignment channels, for each of the logical states. The sum of all these channels for each logical state corresponds to the bit-flip probability that we would measure; in Fig.~\ref{fig:decomposed_bit_flip} the data is overlaid with the model and we do indeed see good agreement. 

    From this plot we can see the extent to which the measured bit-flip outcome is dominated by misassignment of the leakage state $\ket{00}$ to the bit-flip outcome in the case of both logical states, where the measured probability can in fact be attributed almost entirely to this assignment channel.
    At $1\;\mathrm{\mu s}$, the ``$\ket{00}$ to bit-flip outcome" assignment channel contributes $\sim34\%$ to the measured bit-flip probability, and at $1\;\mathrm{ms}$ it contributes $\sim95\%$. 

\section{Intrinsic Bit-Flip Estimate}
\label{sec:intrinsic_bit_flip_estimate}

An important metric of the coherence of the logical qubit is the rate at which the logical basis states ``flip" to the other logical state, resulting in an effective bit-flip. As described in the main text, we identify two primary contributions to this rate: ``intrinsic" processes, and measurement-introduced processes. The combination of these processes result in the ``apparent" bit-flip that we measure (see the Measuring Bit-Flip Errors section).

While we can place a measured, SPAM-limited bound on the bit-flip rate, we are also interested in extracting the intrinsic bit-flip rate as a figure of merit for our dual-rail qubit. The intrinsic bit-flip rate is the true idling bit-flip rate, and thus is the true rate applicable during any operation of the qubit in a circuit. 

Specifically, there are two intrinsic processes that contribute to effective transitions between the logical states. First, there is a finite probability of actual double-transitions, where both photon-loss in one cavity and photon-gain in the other cavity occur and induce a flip in the logical state. Second, a difference in the cavity decay rates will cause evolution in the no-jump trajectory \cite{michael_binomial_code_2016} toward the cavity with the longer lifetime, with the resulting backaction flipping the logical state. 

Measurement errors contribute to the bit-flip rate when performing an experiment in order to measure the bit-flip rate. In this case non-bit-flip outcomes will get assigned to the bit-flip outcome as a result of measurement errors, and thus increase the apparent rate of the logical bit flips. 

By way of measurement we have of course have no way of decoupling these measurement errors from the intrinsic errors in order to quantify each separately. However, in theory we know exactly what the contributions to the intrinsic processes are and as such, using our measured system parameters (see the System Properties section), we can compute the intrinsic bit-flip rate of our qubit. \\

\textbf{Leakage-seepage contribution\\}

From the functional form of the time evolution of the dual-rail states (see eqn.~\ref{eqn:ODE}) given finite fidelity state-preparation, we can extract the probabilities of a leakage-seepage event at 1$\mu$s for each of the logical state preparation: $P_{(\ket{01}|\ket{10})}(t=1\mu s)$ and $P_{(\ket{10}|\ket{10})}(t=1\mu s)$, and extract a time constant from this as 
\begin{equation*}
    T_{flip} = \frac{1\mu s}{P_{\text{leakage-seepage}}(t=1\mu s)}
\end{equation*}

For our system's parameters we find that for prep $\ket{0_{\mathrm{L}}}$ and $\ket{1_{\mathrm{L}}}$ the qubit lifetime due to this bit-flip channel is $T_{\mathrm{flip}\;\ket{0_{\mathrm{L}}}\rightarrow\ket{1_{\mathrm{L}}}} = 6.74\;s$ and $T_{\mathrm{flip}\;\ket{1_{\mathrm{L}}}\rightarrow\ket{0_{\mathrm{L}}}} = 9.06\;s$.\\

\textbf{No-jump backaction contribution\\}

In the case where the lifetimes of each of the dual-rail's cavity are not the same ($T_{1}^{A}\neq T_{1}^{B}$), a dual-rail state with finite population in both of the logical states will be susceptible to the effect of the no-jump backaction~\cite{michael_binomial_code_2016,teoh_dual-rail_2022}. Under the no-jump evolution, conditioned on the case where no jump (decay event) occurs, the quantum amplitude of the logical state defined by an excitation in the longer-lived cavity will increase at the expense of the quantum amplitude of other logical state.  

In the context of the bit-flip experiment where we try to initialize in a basis state, if the state preparation was perfect the no-jump backaction would have no effect as there would be no probability of being in the other logical state. However, in practice the fidelity of state-preparation will always be finite, and as such there will always be some small finite probability of being in an undesired ensemble state--vulnerable to the no-jump backation. 

While no population transfer occurs due to this effect, and in that sense it is not comparable to a $\hat{\sigma}_{x}$ type bit-flip, under this deterministic evolution the relative amplitudes of the logical states will polarize towards the bit-flip-outcome, and then be measured as such what projecting the z-axis of the logical qubit. Therefore this is an effective ``bit-flip" channel for one of the logical states (the one defined by an excitation in the shorter-lived cavity).

We quantify this as follows. For each state preparation the populations of the logical state can be expressed as:
\begin{center}
\begin{tabular}{ l | l }
 Prep $\ket{0_{\mathrm{L}}}$ & Prep $\ket{1_{\mathrm{L}}}$  \\ 
 \hline\hline
 $P_{01} = (1-\epsilon_{0})(1-\epsilon_{1})$ & $P_{01} = \epsilon_{1}\cdot\epsilon_{0}$  \\  
 $P_{10} = \epsilon_{0}\cdot\epsilon_{1}$ & $P_{10} = (1-\epsilon_{1})(1-\epsilon_{0})$    
\end{tabular}
\end{center}
where $\epsilon_{0}$, $\epsilon_{1}$ are the probabilities of an error in preparing $\ket{0}$ and $\ket{1}$, respectively.

The no-jump backation affects the relative ratio between the population within the logical subspace, and as such we renormalize these population probabilities and define:
\begin{equation*}
    P_{01}^{\prime} = \frac{P_{01}}{P_{01}+P_{10}},\;P_{10}^{\prime} = \frac{P_{10}}{P_{01}+P_{10}}
\end{equation*}
for each of the state preparations. 

The functional form of the time evolution of expectation value of the $\hat{\sigma}_{z}$ is defined in~\cite{teoh_dual-rail_2022}, and cast in terms of the notation conventions in this paper takes the form:
\begin{equation*}
    \langle Z(t)\rangle = \frac{P_{10}^{\prime}-P_{10}^{\prime}e^{-(\kappa_{A}-\kappa_{B})t}}{1-P_{01}^{\prime}(1-e^{-(\kappa_{A}-\kappa_{B})t})}
\end{equation*}
where $\kappa_{A(B)} = 1/T_{1}^{A(B)}$

Under this convention $\langle Z(t)\rangle=+1$ corresponds to $\ket{10}$ and $\langle Z(t)\rangle=-1$ corresponds to $\ket{01}$. With this expression we can extract a time constant corresponding to the lifetime of the qubit to the no-jump evolution for prep $\ket{0_{\mathrm{L}}}$ and $\ket{1_{\mathrm{L}}}$ as:
\begin{equation*}
    T_{\mathrm{flip}} = \frac{1\mu s}{p(\langle Z(t=1\mu s)\rangle =+1\;|\;\text{prep}\;01)},\; T_{\mathrm{flip}} = \frac{1\mu s}{p(\langle Z(t=1\mu s)\rangle =-1\;|\;\text{prep}\;10)}
\end{equation*}

For our system's parameters we find that for prep $\ket{0_{\mathrm{L}}}$ and $\ket{1_{\mathrm{L}}}$ the qubit lifetime due to this bit-flip channel is $T_{\mathrm{flip}\;\ket{0_{\mathrm{L}}}\rightarrow\ket{1_{\mathrm{L}}}} = 3.28\;s$ and $T_{\mathrm{flip}\;\ket{1_{\mathrm{L}}}\rightarrow\ket{0_{\mathrm{L}}}} = 4.43\;s$.\\

\textbf{Total intrinsic bit-flip\\}

In order to extract an effective, relevant time-constant ($\text{T}_{\text{flip}}$) representing the intrinsic lifetime of the dual-rail qubit, we look at the error in the nominal duration of a logical gate: 1 $\mu$s. We then define the time-constant as:
\begin{equation*}
    \text{T}_{\text{flip}} = \frac{1\;\mu\text{s}}{\text{probability of an intrinsic bit-flip at }1\;\mu\text{s}}
\end{equation*}
where this probability is the sum of the two intrinsic contributions.

As described above, we find that both of these intrinsic bit-flip channels contribute on the order of $\sim10^{-7}$ at 1 microsecond to the bit-flip probability. As such, for our system's parameters we find that for prep $\ket{0_{\mathrm{L}}}$ and $\ket{1_{\mathrm{L}}}$ the intrinsic qubit lifetime is $\mathbf{T_{\mathrm{flip}\;\ket{0_{\mathrm{L}}}\rightarrow\ket{1_{\mathrm{L}}}} = 2.21\;s}$ and $\mathbf{T_{\mathrm{flip}\;\ket{1_{\mathrm{L}}}\rightarrow\ket{0_{\mathrm{L}}}} = 2.98\;s}$.

% \subsection*{Dual-rail phase-flip estimate}
% \label{sec:intrinsic_phase_flip_estimate}

%% file: ms.bbl
\providecommand{\noopsort}[1]{}\providecommand{\singleletter}[1]{#1}%
\begin{thebibliography}{69}%
\makeatletter
\providecommand \@ifxundefined [1]{%
 \@ifx{#1\undefined}
}%
\providecommand \@ifnum [1]{%
 \ifnum #1\expandafter \@firstoftwo
 \else \expandafter \@secondoftwo
 \fi
}%
\providecommand \@ifx [1]{%
 \ifx #1\expandafter \@firstoftwo
 \else \expandafter \@secondoftwo
 \fi
}%
\providecommand \natexlab [1]{#1}%
\providecommand \enquote  [1]{``#1''}%
\providecommand \bibnamefont  [1]{#1}%
\providecommand \bibfnamefont [1]{#1}%
\providecommand \citenamefont [1]{#1}%
\providecommand \href@noop [0]{\@secondoftwo}%
\providecommand \href [0]{\begingroup \@sanitize@url \@href}%
\providecommand \@href[1]{\@@startlink{#1}\@@href}%
\providecommand \@@href[1]{\endgroup#1\@@endlink}%
\providecommand \@sanitize@url [0]{\catcode `\\12\catcode `\$12\catcode
  `\&12\catcode `\#12\catcode `\^12\catcode `\_12\catcode `\%12\relax}%
\providecommand \@@startlink[1]{}%
\providecommand \@@endlink[0]{}%
\providecommand \url  [0]{\begingroup\@sanitize@url \@url }%
\providecommand \@url [1]{\endgroup\@href {#1}{\urlprefix }}%
\providecommand \urlprefix  [0]{URL }%
\providecommand \Eprint [0]{\href }%
\providecommand \doibase [0]{https://doi.org/}%
\providecommand \selectlanguage [0]{\@gobble}%
\providecommand \bibinfo  [0]{\@secondoftwo}%
\providecommand \bibfield  [0]{\@secondoftwo}%
\providecommand \translation [1]{[#1]}%
\providecommand \BibitemOpen [0]{}%
\providecommand \bibitemStop [0]{}%
\providecommand \bibitemNoStop [0]{.\EOS\space}%
\providecommand \EOS [0]{\spacefactor3000\relax}%
\providecommand \BibitemShut  [1]{\csname bibitem#1\endcsname}%
\let\auto@bib@innerbib\@empty
%</preamble>
\bibitem [{\citenamefont {Grassl}\ \emph {et~al.}(1997)\citenamefont {Grassl},
  \citenamefont {Beth},\ and\ \citenamefont
  {Pellizzari}}]{grassl_codes_erasure_1997}%
  \BibitemOpen
  \bibfield  {author} {\bibinfo {author} {\bibfnamefont {M.}~\bibnamefont
  {Grassl}}, \bibinfo {author} {\bibfnamefont {T.}~\bibnamefont {Beth}},\ and\
  \bibinfo {author} {\bibfnamefont {T.}~\bibnamefont {Pellizzari}},\ }\bibfield
   {title} {\bibinfo {title} {Codes for the {Q}uantum {E}rasure {C}hannel},\
  }\href {https://doi.org/10.1103/PhysRevA.56.33} {\bibfield  {journal}
  {\bibinfo  {journal} {Phys. Rev. A}\ }\textbf {\bibinfo {volume} {56}},\
  \bibinfo {pages} {33} (\bibinfo {year} {1997})}\BibitemShut {NoStop}%
\bibitem [{\citenamefont {Wu}\ \emph {et~al.}(2022)\citenamefont {Wu},
  \citenamefont {Kolkowitz}, \citenamefont {Puri},\ and\ \citenamefont
  {Thompson}}]{wu_erasure_2022}%
  \BibitemOpen
  \bibfield  {author} {\bibinfo {author} {\bibfnamefont {Y.}~\bibnamefont
  {Wu}}, \bibinfo {author} {\bibfnamefont {S.}~\bibnamefont {Kolkowitz}},
  \bibinfo {author} {\bibfnamefont {S.}~\bibnamefont {Puri}},\ and\ \bibinfo
  {author} {\bibfnamefont {J.~D.}\ \bibnamefont {Thompson}},\ }\bibfield
  {title} {\bibinfo {title} {Erasure {C}onversion for {F}ault-{T}olerant
  {Q}uantum {C}omputing in {A}lkaline {E}arth {Rydberg} {A}tom {A}rrays},\
  }\href {https://doi.org/10.1038/s41467-022-32094-6} {\bibfield  {journal}
  {\bibinfo  {journal} {Nature Communications}\ }\textbf {\bibinfo {volume}
  {13}},\ \bibinfo {pages} {4657} (\bibinfo {year} {2022})}\BibitemShut
  {NoStop}%
\bibitem [{\citenamefont {Stace}\ \emph {et~al.}(2009)\citenamefont {Stace},
  \citenamefont {Barrett},\ and\ \citenamefont
  {Doherty}}]{stace_codes_loss_2009}%
  \BibitemOpen
  \bibfield  {author} {\bibinfo {author} {\bibfnamefont {T.~M.}\ \bibnamefont
  {Stace}}, \bibinfo {author} {\bibfnamefont {S.~D.}\ \bibnamefont {Barrett}},\
  and\ \bibinfo {author} {\bibfnamefont {A.~C.}\ \bibnamefont {Doherty}},\
  }\href {https://doi.org/10.1103/PhysRevLett.102.200501} {\bibinfo {title}
  {Thresholds for {T}opological {C}odes in the {P}resence of {L}oss}} (\bibinfo
  {year} {2009})\BibitemShut {NoStop}%
\bibitem [{\citenamefont {Barrett}\ and\ \citenamefont
  {Stace}(2010)}]{barrett_ftqc_loss_2010}%
  \BibitemOpen
  \bibfield  {author} {\bibinfo {author} {\bibfnamefont {S.~D.}\ \bibnamefont
  {Barrett}}\ and\ \bibinfo {author} {\bibfnamefont {T.~M.}\ \bibnamefont
  {Stace}},\ }\bibfield  {title} {\bibinfo {title} {Fault {T}olerant {Q}uantum
  {C}omputation with {V}ery {H}igh {T}hreshold for {L}oss {E}rrors},\ }\href
  {https://doi.org/10.1103/PhysRevLett.105.200502} {\bibfield  {journal}
  {\bibinfo  {journal} {Phys. Rev. Lett.}\ }\textbf {\bibinfo {volume} {105}},\
  \bibinfo {pages} {200502} (\bibinfo {year} {2010})}\BibitemShut {NoStop}%
\bibitem [{\citenamefont {Teoh}\ \emph {et~al.}(2022)\citenamefont {Teoh},
  \citenamefont {Winkel}, \citenamefont {Babla}, \citenamefont {Chapman},
  \citenamefont {Claes}, \citenamefont {de~Graaf}, \citenamefont {Garmon},
  \citenamefont {Kalfus}, \citenamefont {Lu}, \citenamefont {Maiti},
  \citenamefont {Sahay}, \citenamefont {Thakur}, \citenamefont {Tsunoda},
  \citenamefont {Xue}, \citenamefont {Frunzio}, \citenamefont {Girvin},
  \citenamefont {Puri},\ and\ \citenamefont
  {Schoelkopf}}]{teoh_dual-rail_2022}%
  \BibitemOpen
  \bibfield  {author} {\bibinfo {author} {\bibfnamefont {J.~D.}\ \bibnamefont
  {Teoh}}, \bibinfo {author} {\bibfnamefont {P.}~\bibnamefont {Winkel}},
  \bibinfo {author} {\bibfnamefont {H.~K.}\ \bibnamefont {Babla}}, \bibinfo
  {author} {\bibfnamefont {B.~J.}\ \bibnamefont {Chapman}}, \bibinfo {author}
  {\bibfnamefont {J.}~\bibnamefont {Claes}}, \bibinfo {author} {\bibfnamefont
  {S.~J.}\ \bibnamefont {de~Graaf}}, \bibinfo {author} {\bibfnamefont
  {J.~W.~O.}\ \bibnamefont {Garmon}}, \bibinfo {author} {\bibfnamefont {W.~D.}\
  \bibnamefont {Kalfus}}, \bibinfo {author} {\bibfnamefont {Y.}~\bibnamefont
  {Lu}}, \bibinfo {author} {\bibfnamefont {A.}~\bibnamefont {Maiti}}, \bibinfo
  {author} {\bibfnamefont {K.}~\bibnamefont {Sahay}}, \bibinfo {author}
  {\bibfnamefont {N.}~\bibnamefont {Thakur}}, \bibinfo {author} {\bibfnamefont
  {T.}~\bibnamefont {Tsunoda}}, \bibinfo {author} {\bibfnamefont {S.~H.}\
  \bibnamefont {Xue}}, \bibinfo {author} {\bibfnamefont {L.}~\bibnamefont
  {Frunzio}}, \bibinfo {author} {\bibfnamefont {S.~M.}\ \bibnamefont {Girvin}},
  \bibinfo {author} {\bibfnamefont {S.}~\bibnamefont {Puri}},\ and\ \bibinfo
  {author} {\bibfnamefont {R.~J.}\ \bibnamefont {Schoelkopf}},\ }\href@noop {}
  {\bibinfo {title} {Dual-rail encoding with superconducting cavities}}
  (\bibinfo {year} {2022}),\ \Eprint {https://arxiv.org/abs/2212.12077}
  {arXiv:2212.12077} \BibitemShut {NoStop}%
\bibitem [{\citenamefont {Mirrahimi}\ \emph {et~al.}(2014)\citenamefont
  {Mirrahimi}, \citenamefont {Leghtas}, \citenamefont {Albert}, \citenamefont
  {Touzard}, \citenamefont {Schoelkopf}, \citenamefont {Jiang},\ and\
  \citenamefont {Devoret}}]{mirrahimi_cat_qubits_2014}%
  \BibitemOpen
  \bibfield  {author} {\bibinfo {author} {\bibfnamefont {M.}~\bibnamefont
  {Mirrahimi}}, \bibinfo {author} {\bibfnamefont {Z.}~\bibnamefont {Leghtas}},
  \bibinfo {author} {\bibfnamefont {V.~V.}\ \bibnamefont {Albert}}, \bibinfo
  {author} {\bibfnamefont {S.}~\bibnamefont {Touzard}}, \bibinfo {author}
  {\bibfnamefont {R.~J.}\ \bibnamefont {Schoelkopf}}, \bibinfo {author}
  {\bibfnamefont {L.}~\bibnamefont {Jiang}},\ and\ \bibinfo {author}
  {\bibfnamefont {M.~H.}\ \bibnamefont {Devoret}},\ }\bibfield  {title}
  {\bibinfo {title} {Dynamically {P}rotected {C}at-{Q}ubits: {A} {N}ew
  {P}aradigm for {U}niversal {Q}uantum {C}omputation},\ }\href
  {https://doi.org/10.1088/1367-2630/16/4/045014} {\bibfield  {journal}
  {\bibinfo  {journal} {New Journal of Physics}\ }\textbf {\bibinfo {volume}
  {16}},\ \bibinfo {pages} {045014} (\bibinfo {year} {2014})}\BibitemShut
  {NoStop}%
\bibitem [{\citenamefont {Michael}\ \emph {et~al.}(2016)\citenamefont
  {Michael}, \citenamefont {Silveri}, \citenamefont {Brierley}, \citenamefont
  {Albert}, \citenamefont {Salmilehto}, \citenamefont {Jiang},\ and\
  \citenamefont {Girvin}}]{michael_binomial_code_2016}%
  \BibitemOpen
  \bibfield  {author} {\bibinfo {author} {\bibfnamefont {M.~H.}\ \bibnamefont
  {Michael}}, \bibinfo {author} {\bibfnamefont {M.}~\bibnamefont {Silveri}},
  \bibinfo {author} {\bibfnamefont {R.~T.}\ \bibnamefont {Brierley}}, \bibinfo
  {author} {\bibfnamefont {V.~V.}\ \bibnamefont {Albert}}, \bibinfo {author}
  {\bibfnamefont {J.}~\bibnamefont {Salmilehto}}, \bibinfo {author}
  {\bibfnamefont {L.}~\bibnamefont {Jiang}},\ and\ \bibinfo {author}
  {\bibfnamefont {S.~M.}\ \bibnamefont {Girvin}},\ }\bibfield  {title}
  {\bibinfo {title} {New {C}lass of {Q}uantum {E}rror-{C}orrecting {C}odes for
  a {B}osonic {M}ode},\ }\href {https://doi.org/10.1103/PhysRevX.6.031006}
  {\bibfield  {journal} {\bibinfo  {journal} {Phys. Rev. X}\ }\textbf {\bibinfo
  {volume} {6}},\ \bibinfo {pages} {031006} (\bibinfo {year}
  {2016})}\BibitemShut {NoStop}%
\bibitem [{\citenamefont {Joshi}\ \emph {et~al.}(2021)\citenamefont {Joshi},
  \citenamefont {Noh},\ and\ \citenamefont {Gao}}]{joshi_bosonic_review_2021}%
  \BibitemOpen
  \bibfield  {author} {\bibinfo {author} {\bibfnamefont {A.}~\bibnamefont
  {Joshi}}, \bibinfo {author} {\bibfnamefont {K.}~\bibnamefont {Noh}},\ and\
  \bibinfo {author} {\bibfnamefont {Y.~Y.}\ \bibnamefont {Gao}},\ }\bibfield
  {title} {\bibinfo {title} {{Q}uantum {I}nformation {P}rocessing with
  {B}osonic {Q}ubits in {C}ircuit {QED}},\ }\href
  {https://doi.org/10.1088/2058-9565/abe989} {\bibfield  {journal} {\bibinfo
  {journal} {Quantum Science and Technology}\ }\textbf {\bibinfo {volume}
  {6}},\ \bibinfo {pages} {033001} (\bibinfo {year} {2021})}\BibitemShut
  {NoStop}%
\bibitem [{\citenamefont {Cai}\ \emph {et~al.}(2021)\citenamefont {Cai},
  \citenamefont {Ma}, \citenamefont {Wang}, \citenamefont {Zou},\ and\
  \citenamefont {Sun}}]{cai_bosonic_2021}%
  \BibitemOpen
  \bibfield  {author} {\bibinfo {author} {\bibfnamefont {W.}~\bibnamefont
  {Cai}}, \bibinfo {author} {\bibfnamefont {Y.}~\bibnamefont {Ma}}, \bibinfo
  {author} {\bibfnamefont {W.}~\bibnamefont {Wang}}, \bibinfo {author}
  {\bibfnamefont {C.-L.}\ \bibnamefont {Zou}},\ and\ \bibinfo {author}
  {\bibfnamefont {L.}~\bibnamefont {Sun}},\ }\bibfield  {title} {\bibinfo
  {title} {{B}osonic {Q}uantum {E}rror {C}orrection {C}odes in
  {S}uperconducting {Q}uantum {C}ircuits},\ }\href
  {https://doi.org/10.1016/j.fmre.2020.12.006} {\bibfield  {journal} {\bibinfo
  {journal} {Fundamental Research}\ }\textbf {\bibinfo {volume} {1}},\ \bibinfo
  {pages} {50} (\bibinfo {year} {2021})}\BibitemShut {NoStop}%
\bibitem [{\citenamefont {Ma}\ \emph {et~al.}(2021)\citenamefont {Ma},
  \citenamefont {Puri}, \citenamefont {Schoelkopf}, \citenamefont {Devoret},
  \citenamefont {Girvin},\ and\ \citenamefont {Jiang}}]{ma_quantum_2021}%
  \BibitemOpen
  \bibfield  {author} {\bibinfo {author} {\bibfnamefont {W.-L.}\ \bibnamefont
  {Ma}}, \bibinfo {author} {\bibfnamefont {S.}~\bibnamefont {Puri}}, \bibinfo
  {author} {\bibfnamefont {R.~J.}\ \bibnamefont {Schoelkopf}}, \bibinfo
  {author} {\bibfnamefont {M.~H.}\ \bibnamefont {Devoret}}, \bibinfo {author}
  {\bibfnamefont {S.~M.}\ \bibnamefont {Girvin}},\ and\ \bibinfo {author}
  {\bibfnamefont {L.}~\bibnamefont {Jiang}},\ }\bibfield  {title} {\bibinfo
  {title} {{Q}uantum {C}ontrol of {B}osonic {M}odes with {S}uperconducting
  {C}ircuits},\ }\href {https://doi.org/10.1016/j.scib.2021.05.024} {\bibfield
  {journal} {\bibinfo  {journal} {Science Bulletin}\ }\textbf {\bibinfo
  {volume} {66}},\ \bibinfo {pages} {1789} (\bibinfo {year}
  {2021})}\BibitemShut {NoStop}%
\bibitem [{\citenamefont {Ofek}\ \emph {et~al.}(2016)\citenamefont {Ofek},
  \citenamefont {Petrenko}, \citenamefont {Heeres}, \citenamefont {Reinhold},
  \citenamefont {Leghtas}, \citenamefont {Vlastakis}, \citenamefont {Liu},
  \citenamefont {Frunzio}, \citenamefont {Girvin}, \citenamefont {Jiang},
  \citenamefont {Mirrahimi}, \citenamefont {Devoret},\ and\ \citenamefont
  {Schoelkopf}}]{Ofek2016}%
  \BibitemOpen
  \bibfield  {author} {\bibinfo {author} {\bibfnamefont {N.}~\bibnamefont
  {Ofek}}, \bibinfo {author} {\bibfnamefont {A.}~\bibnamefont {Petrenko}},
  \bibinfo {author} {\bibfnamefont {R.}~\bibnamefont {Heeres}}, \bibinfo
  {author} {\bibfnamefont {P.}~\bibnamefont {Reinhold}}, \bibinfo {author}
  {\bibfnamefont {Z.}~\bibnamefont {Leghtas}}, \bibinfo {author} {\bibfnamefont
  {B.}~\bibnamefont {Vlastakis}}, \bibinfo {author} {\bibfnamefont
  {Y.}~\bibnamefont {Liu}}, \bibinfo {author} {\bibfnamefont {L.}~\bibnamefont
  {Frunzio}}, \bibinfo {author} {\bibfnamefont {S.~M.}\ \bibnamefont {Girvin}},
  \bibinfo {author} {\bibfnamefont {L.}~\bibnamefont {Jiang}}, \bibinfo
  {author} {\bibfnamefont {M.}~\bibnamefont {Mirrahimi}}, \bibinfo {author}
  {\bibfnamefont {M.~H.}\ \bibnamefont {Devoret}},\ and\ \bibinfo {author}
  {\bibfnamefont {R.~J.}\ \bibnamefont {Schoelkopf}},\ }\bibfield  {title}
  {\bibinfo {title} {Extending the {L}ifetime of a {Q}uantum {B}it with {E}rror
  {C}orrection in {S}uperconducting {C}ircuits},\ }\href
  {https://doi.org/10.1038/nature18949} {\bibfield  {journal} {\bibinfo
  {journal} {Nature}\ }\textbf {\bibinfo {volume} {536}},\ \bibinfo {pages}
  {441} (\bibinfo {year} {2016})}\BibitemShut {NoStop}%
\bibitem [{\citenamefont {Sivak}\ \emph {et~al.}(2023)\citenamefont {Sivak},
  \citenamefont {Eickbusch}, \citenamefont {Royer}, \citenamefont {Singh},
  \citenamefont {Tsioutsios}, \citenamefont {Ganjam}, \citenamefont {Miano},
  \citenamefont {Brock}, \citenamefont {Ding}, \citenamefont {Frunzio},
  \citenamefont {Girvin}, \citenamefont {Schoelkopf},\ and\ \citenamefont
  {Devoret}}]{sivak_real-time_2023}%
  \BibitemOpen
  \bibfield  {author} {\bibinfo {author} {\bibfnamefont {V.~V.}\ \bibnamefont
  {Sivak}}, \bibinfo {author} {\bibfnamefont {A.}~\bibnamefont {Eickbusch}},
  \bibinfo {author} {\bibfnamefont {B.}~\bibnamefont {Royer}}, \bibinfo
  {author} {\bibfnamefont {S.}~\bibnamefont {Singh}}, \bibinfo {author}
  {\bibfnamefont {I.}~\bibnamefont {Tsioutsios}}, \bibinfo {author}
  {\bibfnamefont {S.}~\bibnamefont {Ganjam}}, \bibinfo {author} {\bibfnamefont
  {A.}~\bibnamefont {Miano}}, \bibinfo {author} {\bibfnamefont {B.~L.}\
  \bibnamefont {Brock}}, \bibinfo {author} {\bibfnamefont {A.~Z.}\ \bibnamefont
  {Ding}}, \bibinfo {author} {\bibfnamefont {L.}~\bibnamefont {Frunzio}},
  \bibinfo {author} {\bibfnamefont {S.~M.}\ \bibnamefont {Girvin}}, \bibinfo
  {author} {\bibfnamefont {R.~J.}\ \bibnamefont {Schoelkopf}},\ and\ \bibinfo
  {author} {\bibfnamefont {M.~H.}\ \bibnamefont {Devoret}},\ }\bibfield
  {title} {\bibinfo {title} {Real-{T}ime {Q}uantum {E}rror {C}orrection
  {B}eyond {B}reak-{E}ven},\ }\href
  {https://doi.org/10.1038/s41586-023-05782-6} {\bibfield  {journal} {\bibinfo
  {journal} {Nature}\ }\textbf {\bibinfo {volume} {616}},\ \bibinfo {pages}
  {50} (\bibinfo {year} {2023})}\BibitemShut {NoStop}%
\bibitem [{\citenamefont {Ni}\ \emph {et~al.}(2023)\citenamefont {Ni},
  \citenamefont {Li}, \citenamefont {Deng}, \citenamefont {Cai}, \citenamefont
  {Zhang}, \citenamefont {Wang}, \citenamefont {Yang}, \citenamefont {Yu},
  \citenamefont {Yan}, \citenamefont {Liu}, \citenamefont {Zou}, \citenamefont
  {Sun}, \citenamefont {Zheng}, \citenamefont {Xu},\ and\ \citenamefont
  {Yu}}]{ni_beating_2023}%
  \BibitemOpen
  \bibfield  {author} {\bibinfo {author} {\bibfnamefont {Z.}~\bibnamefont
  {Ni}}, \bibinfo {author} {\bibfnamefont {S.}~\bibnamefont {Li}}, \bibinfo
  {author} {\bibfnamefont {X.}~\bibnamefont {Deng}}, \bibinfo {author}
  {\bibfnamefont {Y.}~\bibnamefont {Cai}}, \bibinfo {author} {\bibfnamefont
  {L.}~\bibnamefont {Zhang}}, \bibinfo {author} {\bibfnamefont
  {W.}~\bibnamefont {Wang}}, \bibinfo {author} {\bibfnamefont {Z.-B.}\
  \bibnamefont {Yang}}, \bibinfo {author} {\bibfnamefont {H.}~\bibnamefont
  {Yu}}, \bibinfo {author} {\bibfnamefont {F.}~\bibnamefont {Yan}}, \bibinfo
  {author} {\bibfnamefont {S.}~\bibnamefont {Liu}}, \bibinfo {author}
  {\bibfnamefont {C.-L.}\ \bibnamefont {Zou}}, \bibinfo {author} {\bibfnamefont
  {L.}~\bibnamefont {Sun}}, \bibinfo {author} {\bibfnamefont {S.-B.}\
  \bibnamefont {Zheng}}, \bibinfo {author} {\bibfnamefont {Y.}~\bibnamefont
  {Xu}},\ and\ \bibinfo {author} {\bibfnamefont {D.}~\bibnamefont {Yu}},\
  }\bibfield  {title} {\bibinfo {title} {Beating the {B}reak-{E}ven {P}oint
  with a {D}iscrete-{V}ariable-{E}ncoded {L}ogical {Q}ubit},\ }\href
  {https://doi.org/10.1038/s41586-023-05784-4} {\bibfield  {journal} {\bibinfo
  {journal} {Nature}\ }\textbf {\bibinfo {volume} {616}},\ \bibinfo {pages}
  {56} (\bibinfo {year} {2023})}\BibitemShut {NoStop}%
\bibitem [{\citenamefont {Hu}\ \emph {et~al.}(2019)\citenamefont {Hu},
  \citenamefont {Ma}, \citenamefont {Cai}, \citenamefont {Mu}, \citenamefont
  {Xu}, \citenamefont {Wang}, \citenamefont {Wu}, \citenamefont {Wang},
  \citenamefont {Song}, \citenamefont {Zou}, \citenamefont {Girvin},
  \citenamefont {Duan},\ and\ \citenamefont {Sun}}]{hu_qec_2019}%
  \BibitemOpen
  \bibfield  {author} {\bibinfo {author} {\bibfnamefont {L.}~\bibnamefont
  {Hu}}, \bibinfo {author} {\bibfnamefont {Y.}~\bibnamefont {Ma}}, \bibinfo
  {author} {\bibfnamefont {W.}~\bibnamefont {Cai}}, \bibinfo {author}
  {\bibfnamefont {X.}~\bibnamefont {Mu}}, \bibinfo {author} {\bibfnamefont
  {Y.}~\bibnamefont {Xu}}, \bibinfo {author} {\bibfnamefont {W.}~\bibnamefont
  {Wang}}, \bibinfo {author} {\bibfnamefont {Y.}~\bibnamefont {Wu}}, \bibinfo
  {author} {\bibfnamefont {H.}~\bibnamefont {Wang}}, \bibinfo {author}
  {\bibfnamefont {Y.~P.}\ \bibnamefont {Song}}, \bibinfo {author}
  {\bibfnamefont {C.-L.}\ \bibnamefont {Zou}}, \bibinfo {author} {\bibfnamefont
  {S.~M.}\ \bibnamefont {Girvin}}, \bibinfo {author} {\bibfnamefont {L.-M.}\
  \bibnamefont {Duan}},\ and\ \bibinfo {author} {\bibfnamefont
  {L.}~\bibnamefont {Sun}},\ }\bibfield  {title} {\bibinfo {title} {Quantum
  {E}rror {C}orrection and {U}niversal {G}ate {S}et {O}peration on a {B}inomial
  {B}osonic {L}ogical {Q}ubit},\ }\href
  {https://doi.org/10.1038/s41567-018-0414-3} {\bibfield  {journal} {\bibinfo
  {journal} {Nature Physics}\ }\textbf {\bibinfo {volume} {15}},\ \bibinfo
  {pages} {503} (\bibinfo {year} {2019})}\BibitemShut {NoStop}%
\bibitem [{\citenamefont {Aliferis}\ and\ \citenamefont
  {Preskill}(2008)}]{aliferis_biased_noise_2008}%
  \BibitemOpen
  \bibfield  {author} {\bibinfo {author} {\bibfnamefont {P.}~\bibnamefont
  {Aliferis}}\ and\ \bibinfo {author} {\bibfnamefont {J.}~\bibnamefont
  {Preskill}},\ }\bibfield  {title} {\bibinfo {title} {Fault-{T}olerant
  {Q}uantum {C}omputation {A}gainst {B}iased {N}oise},\ }\href
  {https://doi.org/10.1103/PhysRevA.78.052331} {\bibfield  {journal} {\bibinfo
  {journal} {Phys. Rev. A}\ }\textbf {\bibinfo {volume} {78}},\ \bibinfo
  {pages} {052331} (\bibinfo {year} {2008})}\BibitemShut {NoStop}%
\bibitem [{\citenamefont {Tuckett}\ \emph {et~al.}(2019)\citenamefont
  {Tuckett}, \citenamefont {Darmawan}, \citenamefont {Chubb}, \citenamefont
  {Bravyi}, \citenamefont {Bartlett},\ and\ \citenamefont
  {Flammia}}]{tuckett_tailoring_2019}%
  \BibitemOpen
  \bibfield  {author} {\bibinfo {author} {\bibfnamefont {D.~K.}\ \bibnamefont
  {Tuckett}}, \bibinfo {author} {\bibfnamefont {A.~S.}\ \bibnamefont
  {Darmawan}}, \bibinfo {author} {\bibfnamefont {C.~T.}\ \bibnamefont {Chubb}},
  \bibinfo {author} {\bibfnamefont {S.}~\bibnamefont {Bravyi}}, \bibinfo
  {author} {\bibfnamefont {S.~D.}\ \bibnamefont {Bartlett}},\ and\ \bibinfo
  {author} {\bibfnamefont {S.~T.}\ \bibnamefont {Flammia}},\ }\bibfield
  {title} {\bibinfo {title} {Tailoring {Surface} {Codes} for {Highly} {Biased}
  {Noise}},\ }\href {https://doi.org/10.1103/PhysRevX.9.041031} {\bibfield
  {journal} {\bibinfo  {journal} {Physical Review X}\ }\textbf {\bibinfo
  {volume} {9}},\ \bibinfo {pages} {041031} (\bibinfo {year}
  {2019})}\BibitemShut {NoStop}%
\bibitem [{\citenamefont {Guillaud}\ and\ \citenamefont
  {Mirrahimi}(2019)}]{guillaud_repetition_cat_qubits_2019}%
  \BibitemOpen
  \bibfield  {author} {\bibinfo {author} {\bibfnamefont {J.}~\bibnamefont
  {Guillaud}}\ and\ \bibinfo {author} {\bibfnamefont {M.}~\bibnamefont
  {Mirrahimi}},\ }\bibfield  {title} {\bibinfo {title} {Repetition {Cat}
  {Qubits} for {Fault}-{Tolerant} {Quantum} {Computation}},\ }\href
  {https://doi.org/10.1103/PhysRevX.9.041053} {\bibfield  {journal} {\bibinfo
  {journal} {Physical Review X}\ }\textbf {\bibinfo {volume} {9}},\ \bibinfo
  {pages} {041053} (\bibinfo {year} {2019})}\BibitemShut {NoStop}%
\bibitem [{\citenamefont {Darmawan}\ \emph {et~al.}(2021)\citenamefont
  {Darmawan}, \citenamefont {Brown}, \citenamefont {Grimsmo}, \citenamefont
  {Tuckett},\ and\ \citenamefont {Puri}}]{darmawan_practical_2021}%
  \BibitemOpen
  \bibfield  {author} {\bibinfo {author} {\bibfnamefont {A.~S.}\ \bibnamefont
  {Darmawan}}, \bibinfo {author} {\bibfnamefont {B.~J.}\ \bibnamefont {Brown}},
  \bibinfo {author} {\bibfnamefont {A.~L.}\ \bibnamefont {Grimsmo}}, \bibinfo
  {author} {\bibfnamefont {D.~K.}\ \bibnamefont {Tuckett}},\ and\ \bibinfo
  {author} {\bibfnamefont {S.}~\bibnamefont {Puri}},\ }\bibfield  {title}
  {\bibinfo {title} {Practical {Q}uantum {E}rror {C}orrection with the {XZZX}
  {C}ode and {K}err-{C}at {Q}ubits},\ }\href
  {https://doi.org/10.1103/PRXQuantum.2.030345} {\bibfield  {journal} {\bibinfo
   {journal} {PRX Quantum}\ }\textbf {\bibinfo {volume} {2}},\ \bibinfo {pages}
  {030345} (\bibinfo {year} {2021})}\BibitemShut {NoStop}%
\bibitem [{\citenamefont {Claes}\ \emph {et~al.}(2023)\citenamefont {Claes},
  \citenamefont {Bourassa},\ and\ \citenamefont {Puri}}]{claes_tailored_2023}%
  \BibitemOpen
  \bibfield  {author} {\bibinfo {author} {\bibfnamefont {J.}~\bibnamefont
  {Claes}}, \bibinfo {author} {\bibfnamefont {J.~E.}\ \bibnamefont
  {Bourassa}},\ and\ \bibinfo {author} {\bibfnamefont {S.}~\bibnamefont
  {Puri}},\ }\bibfield  {title} {\bibinfo {title} {Tailored {C}luster {S}tates
  with {H}igh {T}hreshold under {B}iased {N}oise},\ }\href
  {https://doi.org/10.1038/s41534-023-00677-w} {\bibfield  {journal} {\bibinfo
  {journal} {npj Quantum Information}\ }\textbf {\bibinfo {volume} {9}},\
  \bibinfo {pages} {1} (\bibinfo {year} {2023})}\BibitemShut {NoStop}%
\bibitem [{\citenamefont {Aliferis}\ \emph {et~al.}(2009)\citenamefont
  {Aliferis}, \citenamefont {Brito}, \citenamefont {DiVincenzo}, \citenamefont
  {Preskill}, \citenamefont {Steffen},\ and\ \citenamefont
  {Terhal}}]{aliferis_FT_biasednoise_scqubits_2009}%
  \BibitemOpen
  \bibfield  {author} {\bibinfo {author} {\bibfnamefont {P.}~\bibnamefont
  {Aliferis}}, \bibinfo {author} {\bibfnamefont {F.}~\bibnamefont {Brito}},
  \bibinfo {author} {\bibfnamefont {D.~P.}\ \bibnamefont {DiVincenzo}},
  \bibinfo {author} {\bibfnamefont {J.}~\bibnamefont {Preskill}}, \bibinfo
  {author} {\bibfnamefont {M.}~\bibnamefont {Steffen}},\ and\ \bibinfo {author}
  {\bibfnamefont {B.~M.}\ \bibnamefont {Terhal}},\ }\bibfield  {title}
  {\bibinfo {title} {Fault-{T}olerant {C}omputing with {B}iased-{N}oise
  {S}uperconducting {Q}ubits: a {C}ase {S}tudy},\ }\href
  {https://doi.org/10.1088/1367-2630/11/1/013061} {\bibfield  {journal}
  {\bibinfo  {journal} {New Journal of Physics}\ }\textbf {\bibinfo {volume}
  {11}},\ \bibinfo {pages} {013061} (\bibinfo {year} {2009})}\BibitemShut
  {NoStop}%
\bibitem [{\citenamefont {Grimm}\ \emph {et~al.}(2020)\citenamefont {Grimm},
  \citenamefont {Frattini}, \citenamefont {Puri}, \citenamefont {Mundhada},
  \citenamefont {Touzard}, \citenamefont {Mirrahimi}, \citenamefont {Girvin},
  \citenamefont {Shankar},\ and\ \citenamefont
  {Devoret}}]{grimm_kerr_cat_2020}%
  \BibitemOpen
  \bibfield  {author} {\bibinfo {author} {\bibfnamefont {A.}~\bibnamefont
  {Grimm}}, \bibinfo {author} {\bibfnamefont {N.~E.}\ \bibnamefont {Frattini}},
  \bibinfo {author} {\bibfnamefont {S.}~\bibnamefont {Puri}}, \bibinfo {author}
  {\bibfnamefont {S.~O.}\ \bibnamefont {Mundhada}}, \bibinfo {author}
  {\bibfnamefont {S.}~\bibnamefont {Touzard}}, \bibinfo {author} {\bibfnamefont
  {M.}~\bibnamefont {Mirrahimi}}, \bibinfo {author} {\bibfnamefont {S.~M.}\
  \bibnamefont {Girvin}}, \bibinfo {author} {\bibfnamefont {S.}~\bibnamefont
  {Shankar}},\ and\ \bibinfo {author} {\bibfnamefont {M.~H.}\ \bibnamefont
  {Devoret}},\ }\bibfield  {title} {\bibinfo {title} {Stabilization and
  {O}peration of a {Kerr}-{C}at {Q}ubit},\ }\href
  {https://doi.org/10.1038/s41586-020-2587-z} {\bibfield  {journal} {\bibinfo
  {journal} {Nature}\ }\textbf {\bibinfo {volume} {584}},\ \bibinfo {pages}
  {205} (\bibinfo {year} {2020})}\BibitemShut {NoStop}%
\bibitem [{\citenamefont {Puri}\ \emph {et~al.}(2020)\citenamefont {Puri},
  \citenamefont {St-Jean}, \citenamefont {Gross}, \citenamefont {Grimm},
  \citenamefont {Frattini}, \citenamefont {Iyer}, \citenamefont {Krishna},
  \citenamefont {Touzard}, \citenamefont {Jiang}, \citenamefont {Blais},
  \citenamefont {Flammia},\ and\ \citenamefont
  {Girvin}}]{puri_bias_preserving_2020}%
  \BibitemOpen
  \bibfield  {author} {\bibinfo {author} {\bibfnamefont {S.}~\bibnamefont
  {Puri}}, \bibinfo {author} {\bibfnamefont {L.}~\bibnamefont {St-Jean}},
  \bibinfo {author} {\bibfnamefont {J.~A.}\ \bibnamefont {Gross}}, \bibinfo
  {author} {\bibfnamefont {A.}~\bibnamefont {Grimm}}, \bibinfo {author}
  {\bibfnamefont {N.~E.}\ \bibnamefont {Frattini}}, \bibinfo {author}
  {\bibfnamefont {P.~S.}\ \bibnamefont {Iyer}}, \bibinfo {author}
  {\bibfnamefont {A.}~\bibnamefont {Krishna}}, \bibinfo {author} {\bibfnamefont
  {S.}~\bibnamefont {Touzard}}, \bibinfo {author} {\bibfnamefont
  {L.}~\bibnamefont {Jiang}}, \bibinfo {author} {\bibfnamefont
  {A.}~\bibnamefont {Blais}}, \bibinfo {author} {\bibfnamefont {S.~T.}\
  \bibnamefont {Flammia}},\ and\ \bibinfo {author} {\bibfnamefont {S.~M.}\
  \bibnamefont {Girvin}},\ }\bibfield  {title} {\bibinfo {title}
  {Bias-{P}reserving {G}ates with {S}tabilized {C}at {Q}ubits},\ }\href
  {https://doi.org/10.1126/sciadv.aay5901} {\bibfield  {journal} {\bibinfo
  {journal} {Science Advances}\ }\textbf {\bibinfo {volume} {6}},\ \bibinfo
  {pages} {eaay5901} (\bibinfo {year} {2020})}\BibitemShut {NoStop}%
\bibitem [{\citenamefont {Lescanne}\ \emph {et~al.}(2020)\citenamefont
  {Lescanne}, \citenamefont {Villiers}, \citenamefont {Peronnin}, \citenamefont
  {Sarlette}, \citenamefont {Delbecq}, \citenamefont {Huard}, \citenamefont
  {Kontos}, \citenamefont {Mirrahimi},\ and\ \citenamefont
  {Leghtas}}]{lescanne_suppression_bit_flips_2020}%
  \BibitemOpen
  \bibfield  {author} {\bibinfo {author} {\bibfnamefont {R.}~\bibnamefont
  {Lescanne}}, \bibinfo {author} {\bibfnamefont {M.}~\bibnamefont {Villiers}},
  \bibinfo {author} {\bibfnamefont {T.}~\bibnamefont {Peronnin}}, \bibinfo
  {author} {\bibfnamefont {A.}~\bibnamefont {Sarlette}}, \bibinfo {author}
  {\bibfnamefont {M.}~\bibnamefont {Delbecq}}, \bibinfo {author} {\bibfnamefont
  {B.}~\bibnamefont {Huard}}, \bibinfo {author} {\bibfnamefont
  {T.}~\bibnamefont {Kontos}}, \bibinfo {author} {\bibfnamefont
  {M.}~\bibnamefont {Mirrahimi}},\ and\ \bibinfo {author} {\bibfnamefont
  {Z.}~\bibnamefont {Leghtas}},\ }\bibfield  {title} {\bibinfo {title}
  {Exponential {S}uppression of {B}it-{F}lips in a {Q}ubit {E}ncoded in an
  {O}scillator},\ }\href {https://www.nature.com/articles/s41567-020-0824-x}
  {\bibfield  {journal} {\bibinfo  {journal} {Nature Physics}\ }\textbf
  {\bibinfo {volume} {16}} (\bibinfo {year} {2020})}\BibitemShut {NoStop}%
\bibitem [{\citenamefont {Berdou}\ \emph {et~al.}(2022)\citenamefont {Berdou},
  \citenamefont {Murani}, \citenamefont {Reglade}, \citenamefont {Smith},
  \citenamefont {Villiers}, \citenamefont {Palomo}, \citenamefont {Rosticher},
  \citenamefont {Denis}, \citenamefont {Morfin}, \citenamefont {Delbecq},
  \citenamefont {Kontos}, \citenamefont {Pankratova}, \citenamefont
  {Rautschke}, \citenamefont {Peronnin}, \citenamefont {Sellem}, \citenamefont
  {Rouchon}, \citenamefont {Sarlette}, \citenamefont {Mirrahimi}, \citenamefont
  {Campagne-Ibarcq}, \citenamefont {Jezouin}, \citenamefont {Lescanne},\ and\
  \citenamefont {Leghtas}}]{berdou_long_bit_flip_2022}%
  \BibitemOpen
  \bibfield  {author} {\bibinfo {author} {\bibfnamefont {C.}~\bibnamefont
  {Berdou}}, \bibinfo {author} {\bibfnamefont {A.}~\bibnamefont {Murani}},
  \bibinfo {author} {\bibfnamefont {U.}~\bibnamefont {Reglade}}, \bibinfo
  {author} {\bibfnamefont {W.~C.}\ \bibnamefont {Smith}}, \bibinfo {author}
  {\bibfnamefont {M.}~\bibnamefont {Villiers}}, \bibinfo {author}
  {\bibfnamefont {J.}~\bibnamefont {Palomo}}, \bibinfo {author} {\bibfnamefont
  {M.}~\bibnamefont {Rosticher}}, \bibinfo {author} {\bibfnamefont
  {A.}~\bibnamefont {Denis}}, \bibinfo {author} {\bibfnamefont
  {P.}~\bibnamefont {Morfin}}, \bibinfo {author} {\bibfnamefont
  {M.}~\bibnamefont {Delbecq}}, \bibinfo {author} {\bibfnamefont
  {T.}~\bibnamefont {Kontos}}, \bibinfo {author} {\bibfnamefont
  {N.}~\bibnamefont {Pankratova}}, \bibinfo {author} {\bibfnamefont
  {F.}~\bibnamefont {Rautschke}}, \bibinfo {author} {\bibfnamefont
  {T.}~\bibnamefont {Peronnin}}, \bibinfo {author} {\bibfnamefont {L.-A.}\
  \bibnamefont {Sellem}}, \bibinfo {author} {\bibfnamefont {P.}~\bibnamefont
  {Rouchon}}, \bibinfo {author} {\bibfnamefont {A.}~\bibnamefont {Sarlette}},
  \bibinfo {author} {\bibfnamefont {M.}~\bibnamefont {Mirrahimi}}, \bibinfo
  {author} {\bibfnamefont {P.}~\bibnamefont {Campagne-Ibarcq}}, \bibinfo
  {author} {\bibfnamefont {S.}~\bibnamefont {Jezouin}}, \bibinfo {author}
  {\bibfnamefont {R.}~\bibnamefont {Lescanne}},\ and\ \bibinfo {author}
  {\bibfnamefont {Z.}~\bibnamefont {Leghtas}},\ }\href
  {https://doi.org/10.48550/arXiv.2204.09128} {\bibinfo {title} {One {H}undred
  {S}econd {B}it-{F}lip {T}ime in a {T}wo-{P}hoton {D}issipative {O}scillator}}
  (\bibinfo {year} {2022}),\ \bibinfo {note} {arXiv:2204.09128}\BibitemShut
  {NoStop}%
\bibitem [{\citenamefont {Chao}\ and\ \citenamefont
  {Reichardt}(2020)}]{chao_flag_2020}%
  \BibitemOpen
  \bibfield  {author} {\bibinfo {author} {\bibfnamefont {R.}~\bibnamefont
  {Chao}}\ and\ \bibinfo {author} {\bibfnamefont {B.~W.}\ \bibnamefont
  {Reichardt}},\ }\bibfield  {title} {\bibinfo {title} {Flag {Fault}-{Tolerant}
  {Error} {Correction} for any {Stabilizer} {Code}},\ }\href
  {https://doi.org/10.1103/PRXQuantum.1.010302} {\bibfield  {journal} {\bibinfo
   {journal} {PRX Quantum}\ }\textbf {\bibinfo {volume} {1}},\ \bibinfo {pages}
  {010302} (\bibinfo {year} {2020})}\BibitemShut {NoStop}%
\bibitem [{\citenamefont {Chamberland}\ \emph {et~al.}(2020)\citenamefont
  {Chamberland}, \citenamefont {Zhu}, \citenamefont {Yoder}, \citenamefont
  {Hertzberg},\ and\ \citenamefont {Cross}}]{chamberland_topological_2020}%
  \BibitemOpen
  \bibfield  {author} {\bibinfo {author} {\bibfnamefont {C.}~\bibnamefont
  {Chamberland}}, \bibinfo {author} {\bibfnamefont {G.}~\bibnamefont {Zhu}},
  \bibinfo {author} {\bibfnamefont {T.~J.}\ \bibnamefont {Yoder}}, \bibinfo
  {author} {\bibfnamefont {J.~B.}\ \bibnamefont {Hertzberg}},\ and\ \bibinfo
  {author} {\bibfnamefont {A.~W.}\ \bibnamefont {Cross}},\ }\bibfield  {title}
  {\bibinfo {title} {Topological and {Subsystem} {Codes} on {Low}-{Degree}
  {Graphs} with {Flag} {Qubits}},\ }\href
  {https://link.aps.org/doi/10.1103/PhysRevX.10.011022} {\bibfield  {journal}
  {\bibinfo  {journal} {Physical Review X}\ }\textbf {\bibinfo {volume} {10}},\
  \bibinfo {pages} {011022} (\bibinfo {year} {2020})}\BibitemShut {NoStop}%
\bibitem [{\citenamefont {Ryan-Anderson}\ \emph {et~al.}(2021)\citenamefont
  {Ryan-Anderson}, \citenamefont {Bohnet}, \citenamefont {Lee}, \citenamefont
  {Gresh}, \citenamefont {Hankin}, \citenamefont {Gaebler}, \citenamefont
  {Francois}, \citenamefont {Chernoguzov}, \citenamefont {Lucchetti},
  \citenamefont {Brown}, \citenamefont {Gatterman}, \citenamefont {Halit},
  \citenamefont {Gilmore}, \citenamefont {Gerber}, \citenamefont {Neyenhuis},
  \citenamefont {Hayes},\ and\ \citenamefont
  {Stutz}}]{ryan_anderson_quantinuum_rt_ft_qec_2021}%
  \BibitemOpen
  \bibfield  {author} {\bibinfo {author} {\bibfnamefont {C.}~\bibnamefont
  {Ryan-Anderson}}, \bibinfo {author} {\bibfnamefont {J.}~\bibnamefont
  {Bohnet}}, \bibinfo {author} {\bibfnamefont {K.}~\bibnamefont {Lee}},
  \bibinfo {author} {\bibfnamefont {D.}~\bibnamefont {Gresh}}, \bibinfo
  {author} {\bibfnamefont {A.}~\bibnamefont {Hankin}}, \bibinfo {author}
  {\bibfnamefont {J.}~\bibnamefont {Gaebler}}, \bibinfo {author} {\bibfnamefont
  {D.}~\bibnamefont {Francois}}, \bibinfo {author} {\bibfnamefont
  {A.}~\bibnamefont {Chernoguzov}}, \bibinfo {author} {\bibfnamefont
  {D.}~\bibnamefont {Lucchetti}}, \bibinfo {author} {\bibfnamefont
  {N.}~\bibnamefont {Brown}}, \bibinfo {author} {\bibfnamefont
  {T.}~\bibnamefont {Gatterman}}, \bibinfo {author} {\bibfnamefont
  {S.}~\bibnamefont {Halit}}, \bibinfo {author} {\bibfnamefont
  {K.}~\bibnamefont {Gilmore}}, \bibinfo {author} {\bibfnamefont
  {J.}~\bibnamefont {Gerber}}, \bibinfo {author} {\bibfnamefont
  {B.}~\bibnamefont {Neyenhuis}}, \bibinfo {author} {\bibfnamefont
  {D.}~\bibnamefont {Hayes}},\ and\ \bibinfo {author} {\bibfnamefont
  {R.}~\bibnamefont {Stutz}},\ }\bibfield  {title} {\bibinfo {title}
  {Realization of {Real}-{Time} {Fault}-{Tolerant} {Quantum} {Error}
  {Correction}},\ }\href {https://doi.org/10.1103/PhysRevX.11.041058}
  {\bibfield  {journal} {\bibinfo  {journal} {Physical Review X}\ }\textbf
  {\bibinfo {volume} {11}},\ \bibinfo {pages} {041058} (\bibinfo {year}
  {2021})}\BibitemShut {NoStop}%
\bibitem [{\citenamefont {Ryan-Anderson}\ \emph {et~al.}(2022)\citenamefont
  {Ryan-Anderson}, \citenamefont {Brown}, \citenamefont {Allman}, \citenamefont
  {Arkin}, \citenamefont {Asa-Attuah}, \citenamefont {Baldwin}, \citenamefont
  {Berg}, \citenamefont {Bohnet}, \citenamefont {Braxton}, \citenamefont
  {Burdick}, \citenamefont {Campora}, \citenamefont {Chernoguzov},
  \citenamefont {Esposito}, \citenamefont {Evans}, \citenamefont {Francois},
  \citenamefont {Gaebler}, \citenamefont {Gatterman}, \citenamefont {Gerber},
  \citenamefont {Gilmore}, \citenamefont {Gresh}, \citenamefont {Hall},
  \citenamefont {Hankin}, \citenamefont {Hostetter}, \citenamefont {Lucchetti},
  \citenamefont {Mayer}, \citenamefont {Myers}, \citenamefont {Neyenhuis},
  \citenamefont {Santiago}, \citenamefont {Sedlacek}, \citenamefont {Skripka},
  \citenamefont {Slattery}, \citenamefont {Stutz}, \citenamefont {Tait},
  \citenamefont {Tobey}, \citenamefont {Vittorini}, \citenamefont {Walker},\
  and\ \citenamefont {Hayes}}]{ryan_anderson_implementing_2022}%
  \BibitemOpen
  \bibfield  {author} {\bibinfo {author} {\bibfnamefont {C.}~\bibnamefont
  {Ryan-Anderson}}, \bibinfo {author} {\bibfnamefont {N.~C.}\ \bibnamefont
  {Brown}}, \bibinfo {author} {\bibfnamefont {M.~S.}\ \bibnamefont {Allman}},
  \bibinfo {author} {\bibfnamefont {B.}~\bibnamefont {Arkin}}, \bibinfo
  {author} {\bibfnamefont {G.}~\bibnamefont {Asa-Attuah}}, \bibinfo {author}
  {\bibfnamefont {C.}~\bibnamefont {Baldwin}}, \bibinfo {author} {\bibfnamefont
  {J.}~\bibnamefont {Berg}}, \bibinfo {author} {\bibfnamefont {J.~G.}\
  \bibnamefont {Bohnet}}, \bibinfo {author} {\bibfnamefont {S.}~\bibnamefont
  {Braxton}}, \bibinfo {author} {\bibfnamefont {N.}~\bibnamefont {Burdick}},
  \bibinfo {author} {\bibfnamefont {J.~P.}\ \bibnamefont {Campora}}, \bibinfo
  {author} {\bibfnamefont {A.}~\bibnamefont {Chernoguzov}}, \bibinfo {author}
  {\bibfnamefont {J.}~\bibnamefont {Esposito}}, \bibinfo {author}
  {\bibfnamefont {B.}~\bibnamefont {Evans}}, \bibinfo {author} {\bibfnamefont
  {D.}~\bibnamefont {Francois}}, \bibinfo {author} {\bibfnamefont {J.~P.}\
  \bibnamefont {Gaebler}}, \bibinfo {author} {\bibfnamefont {T.~M.}\
  \bibnamefont {Gatterman}}, \bibinfo {author} {\bibfnamefont {J.}~\bibnamefont
  {Gerber}}, \bibinfo {author} {\bibfnamefont {K.}~\bibnamefont {Gilmore}},
  \bibinfo {author} {\bibfnamefont {D.}~\bibnamefont {Gresh}}, \bibinfo
  {author} {\bibfnamefont {A.}~\bibnamefont {Hall}}, \bibinfo {author}
  {\bibfnamefont {A.}~\bibnamefont {Hankin}}, \bibinfo {author} {\bibfnamefont
  {J.}~\bibnamefont {Hostetter}}, \bibinfo {author} {\bibfnamefont
  {D.}~\bibnamefont {Lucchetti}}, \bibinfo {author} {\bibfnamefont
  {K.}~\bibnamefont {Mayer}}, \bibinfo {author} {\bibfnamefont
  {J.}~\bibnamefont {Myers}}, \bibinfo {author} {\bibfnamefont
  {B.}~\bibnamefont {Neyenhuis}}, \bibinfo {author} {\bibfnamefont
  {J.}~\bibnamefont {Santiago}}, \bibinfo {author} {\bibfnamefont
  {J.}~\bibnamefont {Sedlacek}}, \bibinfo {author} {\bibfnamefont
  {T.}~\bibnamefont {Skripka}}, \bibinfo {author} {\bibfnamefont
  {A.}~\bibnamefont {Slattery}}, \bibinfo {author} {\bibfnamefont {R.~P.}\
  \bibnamefont {Stutz}}, \bibinfo {author} {\bibfnamefont {J.}~\bibnamefont
  {Tait}}, \bibinfo {author} {\bibfnamefont {R.}~\bibnamefont {Tobey}},
  \bibinfo {author} {\bibfnamefont {G.}~\bibnamefont {Vittorini}}, \bibinfo
  {author} {\bibfnamefont {J.}~\bibnamefont {Walker}},\ and\ \bibinfo {author}
  {\bibfnamefont {D.}~\bibnamefont {Hayes}},\ }\href
  {https://doi.org/10.48550/arXiv.2208.01863} {\bibinfo {title} {Implementing
  {Fault}-tolerant {Entangling} {Gates} on the {Five}-qubit {Code} and the
  {Color} {Code}}} (\bibinfo {year} {2022}),\ \bibinfo {note}
  {arXiv:2208.01863}\BibitemShut {NoStop}%
\bibitem [{\citenamefont {Chen}\ \emph {et~al.}(2022)\citenamefont {Chen},
  \citenamefont {Yoder}, \citenamefont {Kim}, \citenamefont {Sundaresan},
  \citenamefont {Srinivasan}, \citenamefont {Li}, \citenamefont {Córcoles},
  \citenamefont {Cross},\ and\ \citenamefont {Takita}}]{chen_calibrated_2022}%
  \BibitemOpen
  \bibfield  {author} {\bibinfo {author} {\bibfnamefont {E.~H.}\ \bibnamefont
  {Chen}}, \bibinfo {author} {\bibfnamefont {T.~J.}\ \bibnamefont {Yoder}},
  \bibinfo {author} {\bibfnamefont {Y.}~\bibnamefont {Kim}}, \bibinfo {author}
  {\bibfnamefont {N.}~\bibnamefont {Sundaresan}}, \bibinfo {author}
  {\bibfnamefont {S.}~\bibnamefont {Srinivasan}}, \bibinfo {author}
  {\bibfnamefont {M.}~\bibnamefont {Li}}, \bibinfo {author} {\bibfnamefont
  {A.~D.}\ \bibnamefont {Córcoles}}, \bibinfo {author} {\bibfnamefont {A.~W.}\
  \bibnamefont {Cross}},\ and\ \bibinfo {author} {\bibfnamefont
  {M.}~\bibnamefont {Takita}},\ }\href
  {https://doi.org/10.1103/PhysRevLett.128.110504} {\bibinfo {title}
  {Calibrated {D}ecoders for {E}xperimental {Q}uantum {E}rror {C}orrection}}
  (\bibinfo {year} {2022}),\ \bibinfo {note} {arXiv:2110.04285}\BibitemShut
  {NoStop}%
\bibitem [{\citenamefont {Kubica}\ \emph {et~al.}(2022)\citenamefont {Kubica},
  \citenamefont {Haim}, \citenamefont {Vaknin}, \citenamefont {Brandão},\ and\
  \citenamefont {Retzker}}]{kubica_erasure_2022}%
  \BibitemOpen
  \bibfield  {author} {\bibinfo {author} {\bibfnamefont {A.}~\bibnamefont
  {Kubica}}, \bibinfo {author} {\bibfnamefont {A.}~\bibnamefont {Haim}},
  \bibinfo {author} {\bibfnamefont {Y.}~\bibnamefont {Vaknin}}, \bibinfo
  {author} {\bibfnamefont {F.}~\bibnamefont {Brandão}},\ and\ \bibinfo
  {author} {\bibfnamefont {A.}~\bibnamefont {Retzker}},\ }\href
  {https://doi.org/10.48550/arXiv.2208.05461} {\bibinfo {title} {Erasure
  {Q}ubits: {Overcoming} the ${T}_{1}$ {L}imit in {S}uperconducting
  {C}ircuits}} (\bibinfo {year} {2022}),\ \bibinfo {note}
  {arXiv:2208.05461}\BibitemShut {NoStop}%
\bibitem [{\citenamefont {Kang}\ \emph {et~al.}(2023)\citenamefont {Kang},
  \citenamefont {Campbell},\ and\ \citenamefont
  {Brown}}]{kang_trapped_ions_erasure_2023}%
  \BibitemOpen
  \bibfield  {author} {\bibinfo {author} {\bibfnamefont {M.}~\bibnamefont
  {Kang}}, \bibinfo {author} {\bibfnamefont {W.~C.}\ \bibnamefont {Campbell}},\
  and\ \bibinfo {author} {\bibfnamefont {K.~R.}\ \bibnamefont {Brown}},\
  }\bibfield  {title} {\bibinfo {title} {Quantum {E}rror {C}orrection with
  {M}etastable {S}tates of {T}rapped {I}ons {U}sing {E}rasure {C}onversion},\
  }\href {https://doi.org/10.1103/PRXQuantum.4.020358} {\bibfield  {journal}
  {\bibinfo  {journal} {PRX Quantum}\ }\textbf {\bibinfo {volume} {4}},\
  \bibinfo {pages} {020358} (\bibinfo {year} {2023})}\BibitemShut {NoStop}%
\bibitem [{\citenamefont {Scholl}\ \emph {et~al.}(2023)\citenamefont {Scholl},
  \citenamefont {Shaw}, \citenamefont {Tsai}, \citenamefont {Finkelstein},
  \citenamefont {Choi},\ and\ \citenamefont {Endres}}]{scholl_erasure_2023}%
  \BibitemOpen
  \bibfield  {author} {\bibinfo {author} {\bibfnamefont {P.}~\bibnamefont
  {Scholl}}, \bibinfo {author} {\bibfnamefont {A.~L.}\ \bibnamefont {Shaw}},
  \bibinfo {author} {\bibfnamefont {R.~B.-S.}\ \bibnamefont {Tsai}}, \bibinfo
  {author} {\bibfnamefont {R.}~\bibnamefont {Finkelstein}}, \bibinfo {author}
  {\bibfnamefont {J.}~\bibnamefont {Choi}},\ and\ \bibinfo {author}
  {\bibfnamefont {M.}~\bibnamefont {Endres}},\ }\href@noop {} {\bibinfo {title}
  {Erasure conversion in a high-fidelity rydberg quantum simulator}} (\bibinfo
  {year} {2023}),\ \Eprint {https://arxiv.org/abs/2305.03406}
  {arXiv:2305.03406} \BibitemShut {NoStop}%
\bibitem [{\citenamefont {Ma}\ \emph {et~al.}(2023)\citenamefont {Ma},
  \citenamefont {Liu}, \citenamefont {Peng}, \citenamefont {Zhang},
  \citenamefont {Jandura}, \citenamefont {Claes}, \citenamefont {Burgers},
  \citenamefont {Pupillo}, \citenamefont {Puri},\ and\ \citenamefont
  {Thompson}}]{ma_high-fidelity_2023}%
  \BibitemOpen
  \bibfield  {author} {\bibinfo {author} {\bibfnamefont {S.}~\bibnamefont
  {Ma}}, \bibinfo {author} {\bibfnamefont {G.}~\bibnamefont {Liu}}, \bibinfo
  {author} {\bibfnamefont {P.}~\bibnamefont {Peng}}, \bibinfo {author}
  {\bibfnamefont {B.}~\bibnamefont {Zhang}}, \bibinfo {author} {\bibfnamefont
  {S.}~\bibnamefont {Jandura}}, \bibinfo {author} {\bibfnamefont
  {J.}~\bibnamefont {Claes}}, \bibinfo {author} {\bibfnamefont {A.~P.}\
  \bibnamefont {Burgers}}, \bibinfo {author} {\bibfnamefont {G.}~\bibnamefont
  {Pupillo}}, \bibinfo {author} {\bibfnamefont {S.}~\bibnamefont {Puri}},\ and\
  \bibinfo {author} {\bibfnamefont {J.~D.}\ \bibnamefont {Thompson}},\
  }\href@noop {} {\bibinfo {title} {High-fidelity gates with mid-circuit
  erasure conversion in a metastable neutral atom qubit}} (\bibinfo {year}
  {2023}),\ \Eprint {https://arxiv.org/abs/2305.05493} {arXiv:2305.05493}
  \BibitemShut {NoStop}%
\bibitem [{\citenamefont {Chuang}\ and\ \citenamefont
  {Yamamoto}(1995)}]{Chuang_1995_OG_dual_rail}%
  \BibitemOpen
  \bibfield  {author} {\bibinfo {author} {\bibfnamefont {I.~L.}\ \bibnamefont
  {Chuang}}\ and\ \bibinfo {author} {\bibfnamefont {Y.}~\bibnamefont
  {Yamamoto}},\ }\bibfield  {title} {\bibinfo {title} {Simple {Q}uantum
  {C}omputer},\ }\href {https://doi.org/10.1103/physreva.52.3489} {\bibfield
  {journal} {\bibinfo  {journal} {Physical Review A}\ }\textbf {\bibinfo
  {volume} {52}},\ \bibinfo {pages} {3489} (\bibinfo {year}
  {1995})}\BibitemShut {NoStop}%
\bibitem [{\citenamefont {Knill}\ \emph {et~al.}(2001)\citenamefont {Knill},
  \citenamefont {Laflamme},\ and\ \citenamefont
  {Milburn}}]{knill_KLM_linear_optics_2001}%
  \BibitemOpen
  \bibfield  {author} {\bibinfo {author} {\bibfnamefont {E.}~\bibnamefont
  {Knill}}, \bibinfo {author} {\bibfnamefont {R.}~\bibnamefont {Laflamme}},\
  and\ \bibinfo {author} {\bibfnamefont {G.~J.}\ \bibnamefont {Milburn}},\
  }\bibfield  {title} {\bibinfo {title} {A {S}cheme for {E}fficient {Q}uantum
  {C}omputation with {L}inear {O}ptics},\ }\href
  {https://doi.org/10.1038/35051009} {\bibfield  {journal} {\bibinfo  {journal}
  {Nature}\ }\textbf {\bibinfo {volume} {409}},\ \bibinfo {pages} {46}
  (\bibinfo {year} {2001})}\BibitemShut {NoStop}%
\bibitem [{\citenamefont {Kok}\ \emph {et~al.}(2007)\citenamefont {Kok},
  \citenamefont {Munro}, \citenamefont {Nemoto}, \citenamefont {Ralph},
  \citenamefont {Dowling},\ and\ \citenamefont {Milburn}}]{kok_linear_2007}%
  \BibitemOpen
  \bibfield  {author} {\bibinfo {author} {\bibfnamefont {P.}~\bibnamefont
  {Kok}}, \bibinfo {author} {\bibfnamefont {W.~J.}\ \bibnamefont {Munro}},
  \bibinfo {author} {\bibfnamefont {K.}~\bibnamefont {Nemoto}}, \bibinfo
  {author} {\bibfnamefont {T.~C.}\ \bibnamefont {Ralph}}, \bibinfo {author}
  {\bibfnamefont {J.~P.}\ \bibnamefont {Dowling}},\ and\ \bibinfo {author}
  {\bibfnamefont {G.~J.}\ \bibnamefont {Milburn}},\ }\bibfield  {title}
  {\bibinfo {title} {Linear {O}ptical {Q}uantum {C}omputing with {P}hotonic
  {Q}ubits},\ }\href {https://doi.org/10.1103/RevModPhys.79.135} {\bibfield
  {journal} {\bibinfo  {journal} {Reviews of Modern Physics}\ }\textbf
  {\bibinfo {volume} {79}},\ \bibinfo {pages} {135} (\bibinfo {year}
  {2007})}\BibitemShut {NoStop}%
\bibitem [{\citenamefont {Bartolucci}\ \emph {et~al.}(2023)\citenamefont
  {Bartolucci}, \citenamefont {Birchall}, \citenamefont {Bombín},
  \citenamefont {Cable}, \citenamefont {Dawson}, \citenamefont
  {Gimeno-Segovia}, \citenamefont {Johnston}, \citenamefont {Kieling},
  \citenamefont {Nickerson}, \citenamefont {Pant}, \citenamefont {Pastawski},
  \citenamefont {Rudolph},\ and\ \citenamefont
  {Sparrow}}]{bartolucci_fusion_based_2023}%
  \BibitemOpen
  \bibfield  {author} {\bibinfo {author} {\bibfnamefont {S.}~\bibnamefont
  {Bartolucci}}, \bibinfo {author} {\bibfnamefont {P.}~\bibnamefont
  {Birchall}}, \bibinfo {author} {\bibfnamefont {H.}~\bibnamefont {Bombín}},
  \bibinfo {author} {\bibfnamefont {H.}~\bibnamefont {Cable}}, \bibinfo
  {author} {\bibfnamefont {C.}~\bibnamefont {Dawson}}, \bibinfo {author}
  {\bibfnamefont {M.}~\bibnamefont {Gimeno-Segovia}}, \bibinfo {author}
  {\bibfnamefont {E.}~\bibnamefont {Johnston}}, \bibinfo {author}
  {\bibfnamefont {K.}~\bibnamefont {Kieling}}, \bibinfo {author} {\bibfnamefont
  {N.}~\bibnamefont {Nickerson}}, \bibinfo {author} {\bibfnamefont
  {M.}~\bibnamefont {Pant}}, \bibinfo {author} {\bibfnamefont {F.}~\bibnamefont
  {Pastawski}}, \bibinfo {author} {\bibfnamefont {T.}~\bibnamefont {Rudolph}},\
  and\ \bibinfo {author} {\bibfnamefont {C.}~\bibnamefont {Sparrow}},\
  }\bibfield  {title} {\bibinfo {title} {{F}usion-{B}ased {Q}uantum
  {C}omputation},\ }\href {https://doi.org/10.1038/s41467-023-36493-1}
  {\bibfield  {journal} {\bibinfo  {journal} {Nature Communications}\ }\textbf
  {\bibinfo {volume} {14}},\ \bibinfo {pages} {912} (\bibinfo {year}
  {2023})}\BibitemShut {NoStop}%
\bibitem [{\citenamefont {Zakka-Bajjani}\ \emph {et~al.}(2011)\citenamefont
  {Zakka-Bajjani}, \citenamefont {Nguyen}, \citenamefont {Lee}, \citenamefont
  {Vale}, \citenamefont {Simmonds},\ and\ \citenamefont
  {Aumentado}}]{zakka-bajjani_quantum_2011}%
  \BibitemOpen
  \bibfield  {author} {\bibinfo {author} {\bibfnamefont {E.}~\bibnamefont
  {Zakka-Bajjani}}, \bibinfo {author} {\bibfnamefont {F.}~\bibnamefont
  {Nguyen}}, \bibinfo {author} {\bibfnamefont {M.}~\bibnamefont {Lee}},
  \bibinfo {author} {\bibfnamefont {L.~R.}\ \bibnamefont {Vale}}, \bibinfo
  {author} {\bibfnamefont {R.~W.}\ \bibnamefont {Simmonds}},\ and\ \bibinfo
  {author} {\bibfnamefont {J.}~\bibnamefont {Aumentado}},\ }\bibfield  {title}
  {\bibinfo {title} {Quantum superposition of a single microwave photon in two
  different ’colour’ states},\ }\href {https://doi.org/10.1038/nphys2035}
  {\bibfield  {journal} {\bibinfo  {journal} {Nature Physics}\ }\textbf
  {\bibinfo {volume} {7}},\ \bibinfo {pages} {599} (\bibinfo {year} {2011})},\
  \bibinfo {note} {number: 8 Publisher: Nature Publishing Group}\BibitemShut
  {NoStop}%
\bibitem [{\citenamefont {Shim}\ and\ \citenamefont
  {Tahan}(2016)}]{shim_semiconductor-inspired_2016}%
  \BibitemOpen
  \bibfield  {author} {\bibinfo {author} {\bibfnamefont {Y.-P.}\ \bibnamefont
  {Shim}}\ and\ \bibinfo {author} {\bibfnamefont {C.}~\bibnamefont {Tahan}},\
  }\bibfield  {title} {\bibinfo {title} {Semiconductor-inspired design
  principles for superconducting quantum computing},\ }\href
  {https://doi.org/10.1038/ncomms11059} {\bibfield  {journal} {\bibinfo
  {journal} {Nature Communications}\ }\textbf {\bibinfo {volume} {7}},\
  \bibinfo {pages} {11059} (\bibinfo {year} {2016})},\ \bibinfo {note} {number:
  1 Publisher: Nature Publishing Group}\BibitemShut {NoStop}%
\bibitem [{\citenamefont {Campbell}\ \emph {et~al.}(2020)\citenamefont
  {Campbell}, \citenamefont {Shim}, \citenamefont {Kannan}, \citenamefont
  {Winik}, \citenamefont {Kim}, \citenamefont {Melville}, \citenamefont
  {Niedzielski}, \citenamefont {Yoder}, \citenamefont {Tahan}, \citenamefont
  {Gustavsson},\ and\ \citenamefont {Oliver}}]{campbell_small_gap_qubits_2020}%
  \BibitemOpen
  \bibfield  {author} {\bibinfo {author} {\bibfnamefont {D.~L.}\ \bibnamefont
  {Campbell}}, \bibinfo {author} {\bibfnamefont {Y.-P.}\ \bibnamefont {Shim}},
  \bibinfo {author} {\bibfnamefont {B.}~\bibnamefont {Kannan}}, \bibinfo
  {author} {\bibfnamefont {R.}~\bibnamefont {Winik}}, \bibinfo {author}
  {\bibfnamefont {D.~K.}\ \bibnamefont {Kim}}, \bibinfo {author} {\bibfnamefont
  {A.}~\bibnamefont {Melville}}, \bibinfo {author} {\bibfnamefont {B.~M.}\
  \bibnamefont {Niedzielski}}, \bibinfo {author} {\bibfnamefont {J.~L.}\
  \bibnamefont {Yoder}}, \bibinfo {author} {\bibfnamefont {C.}~\bibnamefont
  {Tahan}}, \bibinfo {author} {\bibfnamefont {S.}~\bibnamefont {Gustavsson}},\
  and\ \bibinfo {author} {\bibfnamefont {W.~D.}\ \bibnamefont {Oliver}},\
  }\bibfield  {title} {\bibinfo {title} {Universal nonadiabatic control of
  small-gap superconducting qubits},\ }\href
  {https://doi.org/10.1103/PhysRevX.10.041051} {\bibfield  {journal} {\bibinfo
  {journal} {Phys. Rev. X}\ }\textbf {\bibinfo {volume} {10}},\ \bibinfo
  {pages} {041051} (\bibinfo {year} {2020})}\BibitemShut {NoStop}%
\bibitem [{\citenamefont {Rosenblum}\ \emph {et~al.}(2018)\citenamefont
  {Rosenblum}, \citenamefont {Reinhold}, \citenamefont {Mirrahimi},
  \citenamefont {Jiang}, \citenamefont {Frunzio},\ and\ \citenamefont
  {Schoelkopf}}]{Rosenblum2018}%
  \BibitemOpen
  \bibfield  {author} {\bibinfo {author} {\bibfnamefont {S.}~\bibnamefont
  {Rosenblum}}, \bibinfo {author} {\bibfnamefont {P.}~\bibnamefont {Reinhold}},
  \bibinfo {author} {\bibfnamefont {M.}~\bibnamefont {Mirrahimi}}, \bibinfo
  {author} {\bibfnamefont {L.}~\bibnamefont {Jiang}}, \bibinfo {author}
  {\bibfnamefont {L.}~\bibnamefont {Frunzio}},\ and\ \bibinfo {author}
  {\bibfnamefont {R.~J.}\ \bibnamefont {Schoelkopf}},\ }\bibfield  {title}
  {\bibinfo {title} {Fault-{T}olerant {D}etection of a {Q}uantum {E}rror},\
  }\href {https://doi.org/10.1126/science.aat3996} {\bibfield  {journal}
  {\bibinfo  {journal} {Science}\ }\textbf {\bibinfo {volume} {361}},\ \bibinfo
  {pages} {266} (\bibinfo {year} {2018})}\BibitemShut {NoStop}%
\bibitem [{\citenamefont {Reinhold}\ \emph {et~al.}(2020)\citenamefont
  {Reinhold}, \citenamefont {Rosenblum}, \citenamefont {Ma}, \citenamefont
  {Frunzio}, \citenamefont {Jiang},\ and\ \citenamefont
  {Schoelkopf}}]{reinhold_error_corrected_gates_2020}%
  \BibitemOpen
  \bibfield  {author} {\bibinfo {author} {\bibfnamefont {P.}~\bibnamefont
  {Reinhold}}, \bibinfo {author} {\bibfnamefont {S.}~\bibnamefont {Rosenblum}},
  \bibinfo {author} {\bibfnamefont {W.-L.}\ \bibnamefont {Ma}}, \bibinfo
  {author} {\bibfnamefont {L.}~\bibnamefont {Frunzio}}, \bibinfo {author}
  {\bibfnamefont {L.}~\bibnamefont {Jiang}},\ and\ \bibinfo {author}
  {\bibfnamefont {R.~J.}\ \bibnamefont {Schoelkopf}},\ }\bibfield  {title}
  {\bibinfo {title} {{E}rror-{C}orrected {G}ates on an {E}ncoded {Q}ubit},\
  }\href {https://www.nature.com/articles/s41567-020-0931-8} {\bibfield
  {journal} {\bibinfo  {journal} {Nature Physics}\ }\textbf {\bibinfo {volume}
  {16}},\ \bibinfo {pages} {822} (\bibinfo {year} {2020})}\BibitemShut
  {NoStop}%
\bibitem [{\citenamefont {Ma}\ \emph {et~al.}(2020)\citenamefont {Ma},
  \citenamefont {Xu}, \citenamefont {Mu}, \citenamefont {Cai}, \citenamefont
  {Hu}, \citenamefont {Wang}, \citenamefont {Pan}, \citenamefont {Wang},
  \citenamefont {Song}, \citenamefont {Zou},\ and\ \citenamefont
  {Sun}}]{ma_error_transparent_2020}%
  \BibitemOpen
  \bibfield  {author} {\bibinfo {author} {\bibfnamefont {Y.}~\bibnamefont
  {Ma}}, \bibinfo {author} {\bibfnamefont {Y.}~\bibnamefont {Xu}}, \bibinfo
  {author} {\bibfnamefont {X.}~\bibnamefont {Mu}}, \bibinfo {author}
  {\bibfnamefont {W.}~\bibnamefont {Cai}}, \bibinfo {author} {\bibfnamefont
  {L.}~\bibnamefont {Hu}}, \bibinfo {author} {\bibfnamefont {W.}~\bibnamefont
  {Wang}}, \bibinfo {author} {\bibfnamefont {X.}~\bibnamefont {Pan}}, \bibinfo
  {author} {\bibfnamefont {H.}~\bibnamefont {Wang}}, \bibinfo {author}
  {\bibfnamefont {Y.~P.}\ \bibnamefont {Song}}, \bibinfo {author}
  {\bibfnamefont {C.-L.}\ \bibnamefont {Zou}},\ and\ \bibinfo {author}
  {\bibfnamefont {L.}~\bibnamefont {Sun}},\ }\bibfield  {title} {\bibinfo
  {title} {Error-{T}ransparent {O}perations on a {L}ogical {Q}ubit {P}rotected
  by {Q}uantum {E}rror {C}orrection},\ }\href
  {https://doi.org/10.1038/s41567-020-0893-x} {\bibfield  {journal} {\bibinfo
  {journal} {Nature Physics}\ }\textbf {\bibinfo {volume} {16}},\ \bibinfo
  {pages} {827} (\bibinfo {year} {2020})}\BibitemShut {NoStop}%
\bibitem [{\citenamefont {Chapman}\ \emph {et~al.}(2023)\citenamefont
  {Chapman}, \citenamefont {de~Graaf}, \citenamefont {Xue}, \citenamefont
  {Zhang}, \citenamefont {Teoh}, \citenamefont {Curtis}, \citenamefont
  {Tsunoda}, \citenamefont {Eickbusch}, \citenamefont {Read}, \citenamefont
  {Koottandavida}, \citenamefont {Mundhada}, \citenamefont {Frunzio},
  \citenamefont {Devoret}, \citenamefont {Girvin},\ and\ \citenamefont
  {Schoelkopf}}]{chapman_beamsplitter_2022}%
  \BibitemOpen
  \bibfield  {author} {\bibinfo {author} {\bibfnamefont {B.~J.}\ \bibnamefont
  {Chapman}}, \bibinfo {author} {\bibfnamefont {S.~J.}\ \bibnamefont
  {de~Graaf}}, \bibinfo {author} {\bibfnamefont {S.~H.}\ \bibnamefont {Xue}},
  \bibinfo {author} {\bibfnamefont {Y.}~\bibnamefont {Zhang}}, \bibinfo
  {author} {\bibfnamefont {J.}~\bibnamefont {Teoh}}, \bibinfo {author}
  {\bibfnamefont {J.~C.}\ \bibnamefont {Curtis}}, \bibinfo {author}
  {\bibfnamefont {T.}~\bibnamefont {Tsunoda}}, \bibinfo {author} {\bibfnamefont
  {A.}~\bibnamefont {Eickbusch}}, \bibinfo {author} {\bibfnamefont {A.~P.}\
  \bibnamefont {Read}}, \bibinfo {author} {\bibfnamefont {A.}~\bibnamefont
  {Koottandavida}}, \bibinfo {author} {\bibfnamefont {S.~O.}\ \bibnamefont
  {Mundhada}}, \bibinfo {author} {\bibfnamefont {L.}~\bibnamefont {Frunzio}},
  \bibinfo {author} {\bibfnamefont {M.}~\bibnamefont {Devoret}}, \bibinfo
  {author} {\bibfnamefont {S.}~\bibnamefont {Girvin}},\ and\ \bibinfo {author}
  {\bibfnamefont {R.}~\bibnamefont {Schoelkopf}},\ }\bibfield  {title}
  {\bibinfo {title} {High-on-off-ratio beam-splitter interaction for gates on
  bosonically encoded qubits},\ }\href
  {https://doi.org/10.1103/PRXQuantum.4.020355} {\bibfield  {journal} {\bibinfo
   {journal} {PRX Quantum}\ }\textbf {\bibinfo {volume} {4}},\ \bibinfo {pages}
  {020355} (\bibinfo {year} {2023})}\BibitemShut {NoStop}%
\bibitem [{\citenamefont {Lu}\ \emph {et~al.}(2023)\citenamefont {Lu},
  \citenamefont {Maiti}, \citenamefont {Garmon}, \citenamefont {Ganjam},
  \citenamefont {Zhang}, \citenamefont {Claes}, \citenamefont {Frunzio},
  \citenamefont {Girvin},\ and\ \citenamefont
  {Schoelkopf}}]{lu_beamsplitter_2023}%
  \BibitemOpen
  \bibfield  {author} {\bibinfo {author} {\bibfnamefont {Y.}~\bibnamefont
  {Lu}}, \bibinfo {author} {\bibfnamefont {A.}~\bibnamefont {Maiti}}, \bibinfo
  {author} {\bibfnamefont {J.~W.~O.}\ \bibnamefont {Garmon}}, \bibinfo {author}
  {\bibfnamefont {S.}~\bibnamefont {Ganjam}}, \bibinfo {author} {\bibfnamefont
  {Y.}~\bibnamefont {Zhang}}, \bibinfo {author} {\bibfnamefont
  {J.}~\bibnamefont {Claes}}, \bibinfo {author} {\bibfnamefont
  {L.}~\bibnamefont {Frunzio}}, \bibinfo {author} {\bibfnamefont {S.~M.}\
  \bibnamefont {Girvin}},\ and\ \bibinfo {author} {\bibfnamefont {R.~J.}\
  \bibnamefont {Schoelkopf}},\ }\href@noop {} {\bibinfo {title} {A
  {H}igh-{F}idelity {M}icrowave {B}eamsplitter with a {P}arity-{P}rotected
  {C}onverter}} (\bibinfo {year} {2023}),\ \Eprint
  {https://arxiv.org/abs/2303.00959} {arXiv:2303.00959} \BibitemShut {NoStop}%
\bibitem [{\citenamefont {de~Graaf}\ \emph {et~al.}()\citenamefont {de~Graaf}
  \emph {et~al.}}]{Stijn}%
  \BibitemOpen
  \bibfield  {author} {\bibinfo {author} {\bibfnamefont {S.~J.}\ \bibnamefont
  {de~Graaf}} \emph {et~al.},\ }\href@noop {} {\bibinfo {title} {In
  preparation}}\BibitemShut {NoStop}%
\bibitem [{\citenamefont {Frattini}\ \emph {et~al.}(2017)\citenamefont
  {Frattini}, \citenamefont {Vool}, \citenamefont {Shankar}, \citenamefont
  {Narla}, \citenamefont {Sliwa},\ and\ \citenamefont
  {Devoret}}]{frattini_SNAIL_2017}%
  \BibitemOpen
  \bibfield  {author} {\bibinfo {author} {\bibfnamefont {N.~E.}\ \bibnamefont
  {Frattini}}, \bibinfo {author} {\bibfnamefont {U.}~\bibnamefont {Vool}},
  \bibinfo {author} {\bibfnamefont {S.}~\bibnamefont {Shankar}}, \bibinfo
  {author} {\bibfnamefont {A.}~\bibnamefont {Narla}}, \bibinfo {author}
  {\bibfnamefont {K.~M.}\ \bibnamefont {Sliwa}},\ and\ \bibinfo {author}
  {\bibfnamefont {M.~H.}\ \bibnamefont {Devoret}},\ }\bibfield  {title}
  {\bibinfo {title} {3-{W}ave {M}ixing {Josephson} {D}ipole {E}lement},\ }\href
  {https://doi.org/10.1063/1.4984142} {\bibfield  {journal} {\bibinfo
  {journal} {Applied Physics Letters}\ }\textbf {\bibinfo {volume} {110}},\
  \bibinfo {pages} {222603} (\bibinfo {year} {2017})}\BibitemShut {NoStop}%
\bibitem [{\citenamefont {Verney}\ \emph {et~al.}(2019)\citenamefont {Verney},
  \citenamefont {Lescanne}, \citenamefont {Devoret}, \citenamefont {Leghtas},\
  and\ \citenamefont {Mirrahimi}}]{verney_driven_JJ_2019}%
  \BibitemOpen
  \bibfield  {author} {\bibinfo {author} {\bibfnamefont {L.}~\bibnamefont
  {Verney}}, \bibinfo {author} {\bibfnamefont {R.}~\bibnamefont {Lescanne}},
  \bibinfo {author} {\bibfnamefont {M.~H.}\ \bibnamefont {Devoret}}, \bibinfo
  {author} {\bibfnamefont {Z.}~\bibnamefont {Leghtas}},\ and\ \bibinfo {author}
  {\bibfnamefont {M.}~\bibnamefont {Mirrahimi}},\ }\bibfield  {title} {\bibinfo
  {title} {Structural {I}nstability of {D}riven {J}osephson {C}ircuits
  {P}revented by an {I}nductive {S}hunt},\ }\href
  {https://doi.org/10.1103/PhysRevApplied.11.024003} {\bibfield  {journal}
  {\bibinfo  {journal} {Phys. Rev. Appl.}\ }\textbf {\bibinfo {volume} {11}},\
  \bibinfo {pages} {024003} (\bibinfo {year} {2019})}\BibitemShut {NoStop}%
\bibitem [{\citenamefont {Gambetta}\ \emph {et~al.}(2007)\citenamefont
  {Gambetta}, \citenamefont {Braff}, \citenamefont {Wallraff}, \citenamefont
  {Girvin},\ and\ \citenamefont {Schoelkopf}}]{gambetta_optimal_readout_2007}%
  \BibitemOpen
  \bibfield  {author} {\bibinfo {author} {\bibfnamefont {J.}~\bibnamefont
  {Gambetta}}, \bibinfo {author} {\bibfnamefont {W.~A.}\ \bibnamefont {Braff}},
  \bibinfo {author} {\bibfnamefont {A.}~\bibnamefont {Wallraff}}, \bibinfo
  {author} {\bibfnamefont {S.~M.}\ \bibnamefont {Girvin}},\ and\ \bibinfo
  {author} {\bibfnamefont {R.~J.}\ \bibnamefont {Schoelkopf}},\ }\bibfield
  {title} {\bibinfo {title} {Protocols for {O}ptimal {R}eadout of {Q}ubits
  {U}sing a {C}ontinuous {Q}uantum {N}ondemolition {M}easurement},\ }\href
  {https://doi.org/10.1103/PhysRevA.76.012325} {\bibfield  {journal} {\bibinfo
  {journal} {Phys. Rev. A}\ }\textbf {\bibinfo {volume} {76}},\ \bibinfo
  {pages} {012325} (\bibinfo {year} {2007})}\BibitemShut {NoStop}%
\bibitem [{\citenamefont {Elder}\ \emph {et~al.}(2020)\citenamefont {Elder},
  \citenamefont {Wang}, \citenamefont {Reinhold}, \citenamefont {Hann},
  \citenamefont {Chou}, \citenamefont {Lester}, \citenamefont {Rosenblum},
  \citenamefont {Frunzio}, \citenamefont {Jiang},\ and\ \citenamefont
  {Schoelkopf}}]{elder_msmts_2020}%
  \BibitemOpen
  \bibfield  {author} {\bibinfo {author} {\bibfnamefont {S.~S.}\ \bibnamefont
  {Elder}}, \bibinfo {author} {\bibfnamefont {C.~S.}\ \bibnamefont {Wang}},
  \bibinfo {author} {\bibfnamefont {P.}~\bibnamefont {Reinhold}}, \bibinfo
  {author} {\bibfnamefont {C.~T.}\ \bibnamefont {Hann}}, \bibinfo {author}
  {\bibfnamefont {K.~S.}\ \bibnamefont {Chou}}, \bibinfo {author}
  {\bibfnamefont {B.~J.}\ \bibnamefont {Lester}}, \bibinfo {author}
  {\bibfnamefont {S.}~\bibnamefont {Rosenblum}}, \bibinfo {author}
  {\bibfnamefont {L.}~\bibnamefont {Frunzio}}, \bibinfo {author} {\bibfnamefont
  {L.}~\bibnamefont {Jiang}},\ and\ \bibinfo {author} {\bibfnamefont {R.~J.}\
  \bibnamefont {Schoelkopf}},\ }\bibfield  {title} {\bibinfo {title}
  {High-{F}idelity {M}easurement of {Q}ubits {E}ncoded in {M}ultilevel
  {S}uperconducting {C}ircuits},\ }\href
  {https://doi.org/10.1103/PhysRevX.10.011001} {\bibfield  {journal} {\bibinfo
  {journal} {Phys. Rev. X}\ }\textbf {\bibinfo {volume} {10}},\ \bibinfo
  {pages} {011001} (\bibinfo {year} {2020})}\BibitemShut {NoStop}%
\bibitem [{\citenamefont {Heeres}\ \emph {et~al.}(2017)\citenamefont {Heeres},
  \citenamefont {Reinhold}, \citenamefont {Ofek}, \citenamefont {Frunzio},
  \citenamefont {Jiang}, \citenamefont {Devoret},\ and\ \citenamefont
  {Schoelkopf}}]{Heeres_2017_OCP}%
  \BibitemOpen
  \bibfield  {author} {\bibinfo {author} {\bibfnamefont {R.~W.}\ \bibnamefont
  {Heeres}}, \bibinfo {author} {\bibfnamefont {P.}~\bibnamefont {Reinhold}},
  \bibinfo {author} {\bibfnamefont {N.}~\bibnamefont {Ofek}}, \bibinfo {author}
  {\bibfnamefont {L.}~\bibnamefont {Frunzio}}, \bibinfo {author} {\bibfnamefont
  {L.}~\bibnamefont {Jiang}}, \bibinfo {author} {\bibfnamefont {M.~H.}\
  \bibnamefont {Devoret}},\ and\ \bibinfo {author} {\bibfnamefont {R.~J.}\
  \bibnamefont {Schoelkopf}},\ }\bibfield  {title} {\bibinfo {title}
  {Implementing a universal gate set on a logical qubit encoded in an
  oscillator},\ }\href {https://doi.org/10.1038/s41467-017-00045-1} {\bibfield
  {journal} {\bibinfo  {journal} {Nature Communications}\ }\textbf {\bibinfo
  {volume} {8}},\ \bibinfo {pages} {94} (\bibinfo {year} {2017})}\BibitemShut
  {NoStop}%
\bibitem [{\citenamefont {Curtis}\ \emph {et~al.}(2021)\citenamefont {Curtis},
  \citenamefont {Hann}, \citenamefont {Elder}, \citenamefont {Wang},
  \citenamefont {Frunzio}, \citenamefont {Jiang},\ and\ \citenamefont
  {Schoelkopf}}]{curtis_single_shot_2021}%
  \BibitemOpen
  \bibfield  {author} {\bibinfo {author} {\bibfnamefont {J.~C.}\ \bibnamefont
  {Curtis}}, \bibinfo {author} {\bibfnamefont {C.~T.}\ \bibnamefont {Hann}},
  \bibinfo {author} {\bibfnamefont {S.~S.}\ \bibnamefont {Elder}}, \bibinfo
  {author} {\bibfnamefont {C.~S.}\ \bibnamefont {Wang}}, \bibinfo {author}
  {\bibfnamefont {L.}~\bibnamefont {Frunzio}}, \bibinfo {author} {\bibfnamefont
  {L.}~\bibnamefont {Jiang}},\ and\ \bibinfo {author} {\bibfnamefont {R.~J.}\
  \bibnamefont {Schoelkopf}},\ }\bibfield  {title} {\bibinfo {title}
  {Single-{S}hot {N}umber-{R}esolved {D}etection of {M}icrowave {P}hotons with
  {E}rror {M}itigation},\ }\href {https://doi.org/10.1103/PhysRevA.103.023705}
  {\bibfield  {journal} {\bibinfo  {journal} {Physical Review A}\ }\textbf
  {\bibinfo {volume} {103}},\ \bibinfo {pages} {023705} (\bibinfo {year}
  {2021})}\BibitemShut {NoStop}%
\bibitem [{\citenamefont {Fowler}(2013)}]{fowler_qubit_leakage_2013}%
  \BibitemOpen
  \bibfield  {author} {\bibinfo {author} {\bibfnamefont {A.~G.}\ \bibnamefont
  {Fowler}},\ }\bibfield  {title} {\bibinfo {title} {Coping with {Q}ubit
  {L}eakage in {T}opological {C}odes},\ }\href
  {https://doi.org/10.1103/PhysRevA.88.042308} {\bibfield  {journal} {\bibinfo
  {journal} {Phys. Rev. A}\ }\textbf {\bibinfo {volume} {88}},\ \bibinfo
  {pages} {042308} (\bibinfo {year} {2013})}\BibitemShut {NoStop}%
\bibitem [{\citenamefont {Ghosh}\ \emph {et~al.}(2013)\citenamefont {Ghosh},
  \citenamefont {Fowler}, \citenamefont {Martinis},\ and\ \citenamefont
  {Geller}}]{ghosh_leakage_2013}%
  \BibitemOpen
  \bibfield  {author} {\bibinfo {author} {\bibfnamefont {J.}~\bibnamefont
  {Ghosh}}, \bibinfo {author} {\bibfnamefont {A.~G.}\ \bibnamefont {Fowler}},
  \bibinfo {author} {\bibfnamefont {J.~M.}\ \bibnamefont {Martinis}},\ and\
  \bibinfo {author} {\bibfnamefont {M.~R.}\ \bibnamefont {Geller}},\ }\bibfield
   {title} {\bibinfo {title} {Understanding the {E}ffects of {L}eakage in
  {S}uperconducting {Q}uantum-{E}rror-{D}etection {C}ircuits},\ }\href
  {https://doi.org/10.1103/PhysRevA.88.062329} {\bibfield  {journal} {\bibinfo
  {journal} {Phys. Rev. A}\ }\textbf {\bibinfo {volume} {88}},\ \bibinfo
  {pages} {062329} (\bibinfo {year} {2013})}\BibitemShut {NoStop}%
\bibitem [{\citenamefont {Bultink}\ \emph {et~al.}(2020)\citenamefont
  {Bultink}, \citenamefont {O’Brien}, \citenamefont {Vollmer}, \citenamefont
  {Muthusubramanian}, \citenamefont {Beekman}, \citenamefont {Rol},
  \citenamefont {Fu}, \citenamefont {Tarasinski}, \citenamefont {Ostroukh},
  \citenamefont {Varbanov}, \citenamefont {Bruno},\ and\ \citenamefont
  {DiCarlo}}]{bultink_protecting_2020}%
  \BibitemOpen
  \bibfield  {author} {\bibinfo {author} {\bibfnamefont {C.~C.}\ \bibnamefont
  {Bultink}}, \bibinfo {author} {\bibfnamefont {T.~E.}\ \bibnamefont
  {O’Brien}}, \bibinfo {author} {\bibfnamefont {R.}~\bibnamefont {Vollmer}},
  \bibinfo {author} {\bibfnamefont {N.}~\bibnamefont {Muthusubramanian}},
  \bibinfo {author} {\bibfnamefont {M.~W.}\ \bibnamefont {Beekman}}, \bibinfo
  {author} {\bibfnamefont {M.~A.}\ \bibnamefont {Rol}}, \bibinfo {author}
  {\bibfnamefont {X.}~\bibnamefont {Fu}}, \bibinfo {author} {\bibfnamefont
  {B.}~\bibnamefont {Tarasinski}}, \bibinfo {author} {\bibfnamefont
  {V.}~\bibnamefont {Ostroukh}}, \bibinfo {author} {\bibfnamefont
  {B.}~\bibnamefont {Varbanov}}, \bibinfo {author} {\bibfnamefont
  {A.}~\bibnamefont {Bruno}},\ and\ \bibinfo {author} {\bibfnamefont
  {L.}~\bibnamefont {DiCarlo}},\ }\bibfield  {title} {\bibinfo {title}
  {Protecting quantum entanglement from leakage and qubit errors via repetitive
  parity measurements},\ }\href@noop {} {\bibfield  {journal} {\bibinfo
  {journal} {Science Advances}\ }\textbf {\bibinfo {volume} {6}},\ \bibinfo
  {pages} {eaay3050} (\bibinfo {year} {2020})}\BibitemShut {NoStop}%
\bibitem [{\citenamefont {McEwen}\ \emph {et~al.}(2021)\citenamefont {McEwen},
  \citenamefont {Kafri}, \citenamefont {Chen}, \citenamefont {Atalaya},
  \citenamefont {Satzinger}, \citenamefont {Quintana}, \citenamefont {Klimov},
  \citenamefont {Sank}, \citenamefont {Gidney}, \citenamefont {Fowler},
  \citenamefont {Arute}, \citenamefont {Arya}, \citenamefont {Buckley},
  \citenamefont {Burkett}, \citenamefont {Bushnell}, \citenamefont {Chiaro},
  \citenamefont {Collins}, \citenamefont {Demura}, \citenamefont {Dunsworth},
  \citenamefont {Erickson}, \citenamefont {Foxen}, \citenamefont {Giustina},
  \citenamefont {Huang}, \citenamefont {Hong}, \citenamefont {Jeffrey},
  \citenamefont {Kim}, \citenamefont {Kechedzhi}, \citenamefont {Kostritsa},
  \citenamefont {Laptev}, \citenamefont {Megrant}, \citenamefont {Mi},
  \citenamefont {Mutus}, \citenamefont {Naaman}, \citenamefont {Neeley},
  \citenamefont {Neill}, \citenamefont {Niu}, \citenamefont {Paler},
  \citenamefont {Redd}, \citenamefont {Roushan}, \citenamefont {White},
  \citenamefont {Yao}, \citenamefont {Yeh}, \citenamefont {Zalcman},
  \citenamefont {Chen}, \citenamefont {Smelyanskiy}, \citenamefont {Martinis},
  \citenamefont {Neven}, \citenamefont {Kelly}, \citenamefont {Korotkov},
  \citenamefont {Petukhov},\ and\ \citenamefont
  {Barends}}]{mcewen_removing_2021}%
  \BibitemOpen
  \bibfield  {author} {\bibinfo {author} {\bibfnamefont {M.}~\bibnamefont
  {McEwen}}, \bibinfo {author} {\bibfnamefont {D.}~\bibnamefont {Kafri}},
  \bibinfo {author} {\bibfnamefont {Z.}~\bibnamefont {Chen}}, \bibinfo {author}
  {\bibfnamefont {J.}~\bibnamefont {Atalaya}}, \bibinfo {author} {\bibfnamefont
  {K.~J.}\ \bibnamefont {Satzinger}}, \bibinfo {author} {\bibfnamefont
  {C.}~\bibnamefont {Quintana}}, \bibinfo {author} {\bibfnamefont {P.~V.}\
  \bibnamefont {Klimov}}, \bibinfo {author} {\bibfnamefont {D.}~\bibnamefont
  {Sank}}, \bibinfo {author} {\bibfnamefont {C.}~\bibnamefont {Gidney}},
  \bibinfo {author} {\bibfnamefont {A.~G.}\ \bibnamefont {Fowler}}, \bibinfo
  {author} {\bibfnamefont {F.}~\bibnamefont {Arute}}, \bibinfo {author}
  {\bibfnamefont {K.}~\bibnamefont {Arya}}, \bibinfo {author} {\bibfnamefont
  {B.}~\bibnamefont {Buckley}}, \bibinfo {author} {\bibfnamefont
  {B.}~\bibnamefont {Burkett}}, \bibinfo {author} {\bibfnamefont
  {N.}~\bibnamefont {Bushnell}}, \bibinfo {author} {\bibfnamefont
  {B.}~\bibnamefont {Chiaro}}, \bibinfo {author} {\bibfnamefont
  {R.}~\bibnamefont {Collins}}, \bibinfo {author} {\bibfnamefont
  {S.}~\bibnamefont {Demura}}, \bibinfo {author} {\bibfnamefont
  {A.}~\bibnamefont {Dunsworth}}, \bibinfo {author} {\bibfnamefont
  {C.}~\bibnamefont {Erickson}}, \bibinfo {author} {\bibfnamefont
  {B.}~\bibnamefont {Foxen}}, \bibinfo {author} {\bibfnamefont
  {M.}~\bibnamefont {Giustina}}, \bibinfo {author} {\bibfnamefont
  {T.}~\bibnamefont {Huang}}, \bibinfo {author} {\bibfnamefont
  {S.}~\bibnamefont {Hong}}, \bibinfo {author} {\bibfnamefont {E.}~\bibnamefont
  {Jeffrey}}, \bibinfo {author} {\bibfnamefont {S.}~\bibnamefont {Kim}},
  \bibinfo {author} {\bibfnamefont {K.}~\bibnamefont {Kechedzhi}}, \bibinfo
  {author} {\bibfnamefont {F.}~\bibnamefont {Kostritsa}}, \bibinfo {author}
  {\bibfnamefont {P.}~\bibnamefont {Laptev}}, \bibinfo {author} {\bibfnamefont
  {A.}~\bibnamefont {Megrant}}, \bibinfo {author} {\bibfnamefont
  {X.}~\bibnamefont {Mi}}, \bibinfo {author} {\bibfnamefont {J.}~\bibnamefont
  {Mutus}}, \bibinfo {author} {\bibfnamefont {O.}~\bibnamefont {Naaman}},
  \bibinfo {author} {\bibfnamefont {M.}~\bibnamefont {Neeley}}, \bibinfo
  {author} {\bibfnamefont {C.}~\bibnamefont {Neill}}, \bibinfo {author}
  {\bibfnamefont {M.}~\bibnamefont {Niu}}, \bibinfo {author} {\bibfnamefont
  {A.}~\bibnamefont {Paler}}, \bibinfo {author} {\bibfnamefont
  {N.}~\bibnamefont {Redd}}, \bibinfo {author} {\bibfnamefont {P.}~\bibnamefont
  {Roushan}}, \bibinfo {author} {\bibfnamefont {T.~C.}\ \bibnamefont {White}},
  \bibinfo {author} {\bibfnamefont {J.}~\bibnamefont {Yao}}, \bibinfo {author}
  {\bibfnamefont {P.}~\bibnamefont {Yeh}}, \bibinfo {author} {\bibfnamefont
  {A.}~\bibnamefont {Zalcman}}, \bibinfo {author} {\bibfnamefont
  {Y.}~\bibnamefont {Chen}}, \bibinfo {author} {\bibfnamefont {V.~N.}\
  \bibnamefont {Smelyanskiy}}, \bibinfo {author} {\bibfnamefont {J.~M.}\
  \bibnamefont {Martinis}}, \bibinfo {author} {\bibfnamefont {H.}~\bibnamefont
  {Neven}}, \bibinfo {author} {\bibfnamefont {J.}~\bibnamefont {Kelly}},
  \bibinfo {author} {\bibfnamefont {A.~N.}\ \bibnamefont {Korotkov}}, \bibinfo
  {author} {\bibfnamefont {A.~G.}\ \bibnamefont {Petukhov}},\ and\ \bibinfo
  {author} {\bibfnamefont {R.}~\bibnamefont {Barends}},\ }\bibfield  {title}
  {\bibinfo {title} {Removing {L}eakage-{I}nduced {C}orrelated {E}rrors in
  {S}uperconducting {Q}uantum {E}rror {C}orrection},\ }\href
  {https://doi.org/10.1038/s41467-021-21982-y} {\bibfield  {journal} {\bibinfo
  {journal} {Nature Communications}\ }\textbf {\bibinfo {volume} {12}},\
  \bibinfo {pages} {1761} (\bibinfo {year} {2021})}\BibitemShut {NoStop}%
\bibitem [{\citenamefont {Tsunoda}\ \emph {et~al.}(2023)\citenamefont
  {Tsunoda}, \citenamefont {Teoh}, \citenamefont {Kalfus}, \citenamefont
  {de~Graaf}, \citenamefont {Chapman}, \citenamefont {Curtis}, \citenamefont
  {Thakur}, \citenamefont {Girvin},\ and\ \citenamefont
  {Schoelkopf}}]{tsunoda_error-detectable_2022}%
  \BibitemOpen
  \bibfield  {author} {\bibinfo {author} {\bibfnamefont {T.}~\bibnamefont
  {Tsunoda}}, \bibinfo {author} {\bibfnamefont {J.~D.}\ \bibnamefont {Teoh}},
  \bibinfo {author} {\bibfnamefont {W.~D.}\ \bibnamefont {Kalfus}}, \bibinfo
  {author} {\bibfnamefont {S.~J.}\ \bibnamefont {de~Graaf}}, \bibinfo {author}
  {\bibfnamefont {B.~J.}\ \bibnamefont {Chapman}}, \bibinfo {author}
  {\bibfnamefont {J.~C.}\ \bibnamefont {Curtis}}, \bibinfo {author}
  {\bibfnamefont {N.}~\bibnamefont {Thakur}}, \bibinfo {author} {\bibfnamefont
  {S.~M.}\ \bibnamefont {Girvin}},\ and\ \bibinfo {author} {\bibfnamefont
  {R.~J.}\ \bibnamefont {Schoelkopf}},\ }\bibfield  {title} {\bibinfo {title}
  {Error-{D}etectable {B}osonic {E}ntangling {G}ates with a {N}oisy
  {A}ncilla},\ }\href {https://doi.org/10.1103/PRXQuantum.4.020354} {\bibfield
  {journal} {\bibinfo  {journal} {PRX Quantum}\ }\textbf {\bibinfo {volume}
  {4}},\ \bibinfo {pages} {020354} (\bibinfo {year} {2023})}\BibitemShut
  {NoStop}%
\bibitem [{\citenamefont {Winkel}\ \emph {et~al.}()\citenamefont {Winkel} \emph
  {et~al.}}]{Winkel}%
  \BibitemOpen
  \bibfield  {author} {\bibinfo {author} {\bibfnamefont {P.}~\bibnamefont
  {Winkel}} \emph {et~al.},\ }\href@noop {} {\bibinfo {title} {In
  preparation}}\BibitemShut {NoStop}%
\bibitem [{\citenamefont {Levine}\ \emph {et~al.}(2023)\citenamefont {Levine},
  \citenamefont {Haim}, \citenamefont {Hung}, \citenamefont {Alidoust},
  \citenamefont {Kalaee}, \citenamefont {DeLorenzo}, \citenamefont {Wollack},
  \citenamefont {Arriola}, \citenamefont {Khalajhedayati}, \citenamefont
  {Sanil}, \citenamefont {Vaknin}, \citenamefont {Kubica}, \citenamefont
  {Clerk}, \citenamefont {Hover}, \citenamefont {Brandão}, \citenamefont
  {Retzker},\ and\ \citenamefont {Painter}}]{levine_dual_rail_transmons_2023}%
  \BibitemOpen
  \bibfield  {author} {\bibinfo {author} {\bibfnamefont {H.}~\bibnamefont
  {Levine}}, \bibinfo {author} {\bibfnamefont {A.}~\bibnamefont {Haim}},
  \bibinfo {author} {\bibfnamefont {J.~S.~C.}\ \bibnamefont {Hung}}, \bibinfo
  {author} {\bibfnamefont {N.}~\bibnamefont {Alidoust}}, \bibinfo {author}
  {\bibfnamefont {M.}~\bibnamefont {Kalaee}}, \bibinfo {author} {\bibfnamefont
  {L.}~\bibnamefont {DeLorenzo}}, \bibinfo {author} {\bibfnamefont {E.~A.}\
  \bibnamefont {Wollack}}, \bibinfo {author} {\bibfnamefont {P.~A.}\
  \bibnamefont {Arriola}}, \bibinfo {author} {\bibfnamefont {A.}~\bibnamefont
  {Khalajhedayati}}, \bibinfo {author} {\bibfnamefont {R.}~\bibnamefont
  {Sanil}}, \bibinfo {author} {\bibfnamefont {Y.}~\bibnamefont {Vaknin}},
  \bibinfo {author} {\bibfnamefont {A.}~\bibnamefont {Kubica}}, \bibinfo
  {author} {\bibfnamefont {A.~A.}\ \bibnamefont {Clerk}}, \bibinfo {author}
  {\bibfnamefont {D.}~\bibnamefont {Hover}}, \bibinfo {author} {\bibfnamefont
  {F.}~\bibnamefont {Brandão}}, \bibinfo {author} {\bibfnamefont
  {A.}~\bibnamefont {Retzker}},\ and\ \bibinfo {author} {\bibfnamefont
  {O.}~\bibnamefont {Painter}},\ }\href@noop {} {\bibinfo {title}
  {Demonstrating a {L}ong-{C}oherence {D}ual-{R}ail {E}rasure {Q}ubit {U}sing
  {T}unable {T}ransmons}} (\bibinfo {year} {2023}),\ \Eprint
  {https://arxiv.org/abs/2307.08737} {arXiv:2307.08737 [quant-ph]} \BibitemShut
  {NoStop}%
\bibitem [{\citenamefont {Thorbeck}\ \emph {et~al.}(2023)\citenamefont
  {Thorbeck}, \citenamefont {Xiao}, \citenamefont {Kamal},\ and\ \citenamefont
  {Govia}}]{thorbeck2023readoutinduced}%
  \BibitemOpen
  \bibfield  {author} {\bibinfo {author} {\bibfnamefont {T.}~\bibnamefont
  {Thorbeck}}, \bibinfo {author} {\bibfnamefont {Z.}~\bibnamefont {Xiao}},
  \bibinfo {author} {\bibfnamefont {A.}~\bibnamefont {Kamal}},\ and\ \bibinfo
  {author} {\bibfnamefont {L.~C.~G.}\ \bibnamefont {Govia}},\ }\href@noop {}
  {\bibinfo {title} {Readout-{I}nduced {S}uppression and {E}nhancement of
  {S}uperconducting {Q}ubit {L}ifetimes}} (\bibinfo {year} {2023}),\ \Eprint
  {https://arxiv.org/abs/2305.10508} {arXiv:2305.10508} \BibitemShut {NoStop}%
\bibitem [{\citenamefont {Magesan}\ \emph {et~al.}(2011)\citenamefont
  {Magesan}, \citenamefont {Gambetta},\ and\ \citenamefont
  {Emerson}}]{Magesan_2011}%
  \BibitemOpen
  \bibfield  {author} {\bibinfo {author} {\bibfnamefont {E.}~\bibnamefont
  {Magesan}}, \bibinfo {author} {\bibfnamefont {J.~M.}\ \bibnamefont
  {Gambetta}},\ and\ \bibinfo {author} {\bibfnamefont {J.}~\bibnamefont
  {Emerson}},\ }\bibfield  {title} {\bibinfo {title} {Scalable and {R}obust
  {R}andomized {B}enchmarking of {Q}uantum {P}rocesses},\ }\href
  {https://doi.org/10.1103/PhysRevLett.106.180504} {\bibfield  {journal}
  {\bibinfo  {journal} {Phys. Rev. Lett.}\ }\textbf {\bibinfo {volume} {106}},\
  \bibinfo {pages} {180504} (\bibinfo {year} {2011})}\BibitemShut {NoStop}%
\bibitem [{\citenamefont {{Qiskit contributors}}(2023)}]{Qiskit}%
  \BibitemOpen
  \bibfield  {author} {\bibinfo {author} {\bibnamefont {{Qiskit
  contributors}}},\ }\href {https://doi.org/10.5281/zenodo.2573505} {\bibinfo
  {title} {Qiskit: An {O}pen-{S}ource {F}ramework for {Q}uantum {C}omputing}}
  (\bibinfo {year} {2023})\BibitemShut {NoStop}%
\bibitem [{\citenamefont {Blais}\ \emph {et~al.}(2021)\citenamefont {Blais},
  \citenamefont {Grimsmo}, \citenamefont {Girvin},\ and\ \citenamefont
  {Wallraff}}]{Blais_2021}%
  \BibitemOpen
  \bibfield  {author} {\bibinfo {author} {\bibfnamefont {A.}~\bibnamefont
  {Blais}}, \bibinfo {author} {\bibfnamefont {A.~L.}\ \bibnamefont {Grimsmo}},
  \bibinfo {author} {\bibfnamefont {S.~M.}\ \bibnamefont {Girvin}},\ and\
  \bibinfo {author} {\bibfnamefont {A.}~\bibnamefont {Wallraff}},\ }\bibfield
  {title} {\bibinfo {title} {Circuit quantum electrodynamics},\ }\href
  {https://doi.org/10.1103/RevModPhys.93.025005} {\bibfield  {journal}
  {\bibinfo  {journal} {Rev. Mod. Phys.}\ }\textbf {\bibinfo {volume} {93}},\
  \bibinfo {pages} {025005} (\bibinfo {year} {2021})}\BibitemShut {NoStop}%
\bibitem [{\citenamefont {Blais}\ \emph {et~al.}(2004)\citenamefont {Blais},
  \citenamefont {Huang}, \citenamefont {Wallraff}, \citenamefont {Girvin},\
  and\ \citenamefont {Schoelkopf}}]{Blais_2004}%
  \BibitemOpen
  \bibfield  {author} {\bibinfo {author} {\bibfnamefont {A.}~\bibnamefont
  {Blais}}, \bibinfo {author} {\bibfnamefont {R.-S.}\ \bibnamefont {Huang}},
  \bibinfo {author} {\bibfnamefont {A.}~\bibnamefont {Wallraff}}, \bibinfo
  {author} {\bibfnamefont {S.~M.}\ \bibnamefont {Girvin}},\ and\ \bibinfo
  {author} {\bibfnamefont {R.~J.}\ \bibnamefont {Schoelkopf}},\ }\bibfield
  {title} {\bibinfo {title} {Cavity quantum electrodynamics for superconducting
  electrical circuits: An architecture for quantum computation},\ }\href
  {https://doi.org/10.1103/PhysRevA.69.062320} {\bibfield  {journal} {\bibinfo
  {journal} {Phys. Rev. A}\ }\textbf {\bibinfo {volume} {69}},\ \bibinfo
  {pages} {062320} (\bibinfo {year} {2004})}\BibitemShut {NoStop}%
\bibitem [{\citenamefont {Johansson}\ \emph {et~al.}(2012)\citenamefont
  {Johansson}, \citenamefont {Nation},\ and\ \citenamefont {Nori}}]{QuTiP_1}%
  \BibitemOpen
  \bibfield  {author} {\bibinfo {author} {\bibfnamefont {J.}~\bibnamefont
  {Johansson}}, \bibinfo {author} {\bibfnamefont {P.}~\bibnamefont {Nation}},\
  and\ \bibinfo {author} {\bibfnamefont {F.}~\bibnamefont {Nori}},\ }\bibfield
  {title} {\bibinfo {title} {Qutip: An {O}pen-{S}ource {P}ython {F}ramework for
  the {D}ynamics of {O}pen {Q}uantum {S}ystems},\ }\href
  {https://doi.org/https://doi.org/10.1016/j.cpc.2012.02.021} {\bibfield
  {journal} {\bibinfo  {journal} {Computer Physics Communications}\ }\textbf
  {\bibinfo {volume} {183}},\ \bibinfo {pages} {1760} (\bibinfo {year}
  {2012})}\BibitemShut {NoStop}%
\bibitem [{\citenamefont {Johansson}\ \emph {et~al.}(2013)\citenamefont
  {Johansson}, \citenamefont {Nation},\ and\ \citenamefont {Nori}}]{QuTiP_2}%
  \BibitemOpen
  \bibfield  {author} {\bibinfo {author} {\bibfnamefont {J.}~\bibnamefont
  {Johansson}}, \bibinfo {author} {\bibfnamefont {P.}~\bibnamefont {Nation}},\
  and\ \bibinfo {author} {\bibfnamefont {F.}~\bibnamefont {Nori}},\ }\bibfield
  {title} {\bibinfo {title} {Qutip 2: A {P}ython {F}ramework for the {D}ynamics
  of {O}pen {Q}uantum {S}ystems},\ }\href
  {https://doi.org/https://doi.org/10.1016/j.cpc.2012.11.019} {\bibfield
  {journal} {\bibinfo  {journal} {Computer Physics Communications}\ }\textbf
  {\bibinfo {volume} {184}},\ \bibinfo {pages} {1234} (\bibinfo {year}
  {2013})}\BibitemShut {NoStop}%
\bibitem [{\citenamefont {Siddhu}\ \emph {et~al.}(2021)\citenamefont {Siddhu},
  \citenamefont {Chatterjee}, \citenamefont {Jagannathan}, \citenamefont
  {Mandayam},\ and\ \citenamefont {Tayur}}]{siddhu2021queuechannel}%
  \BibitemOpen
  \bibfield  {author} {\bibinfo {author} {\bibfnamefont {V.}~\bibnamefont
  {Siddhu}}, \bibinfo {author} {\bibfnamefont {A.}~\bibnamefont {Chatterjee}},
  \bibinfo {author} {\bibfnamefont {K.}~\bibnamefont {Jagannathan}}, \bibinfo
  {author} {\bibfnamefont {P.}~\bibnamefont {Mandayam}},\ and\ \bibinfo
  {author} {\bibfnamefont {S.}~\bibnamefont {Tayur}},\ }\href@noop {} {\bibinfo
  {title} {Queue-{C}hannel {C}apacities with {G}eneralized {A}mplitude
  {D}amping}} (\bibinfo {year} {2021}),\ \Eprint
  {https://arxiv.org/abs/2107.13486} {arXiv:2107.13486} \BibitemShut {NoStop}%
\bibitem [{\citenamefont {Nielsen}\ and\ \citenamefont
  {Chuang}(2010)}]{nielsen_chuang_2010}%
  \BibitemOpen
  \bibfield  {author} {\bibinfo {author} {\bibfnamefont {M.~A.}\ \bibnamefont
  {Nielsen}}\ and\ \bibinfo {author} {\bibfnamefont {I.~L.}\ \bibnamefont
  {Chuang}},\ }\href {https://doi.org/10.1017/CBO9780511976667} {\emph
  {\bibinfo {title} {Quantum {C}omputation and {Q}uantum {I}nformation: 10th
  {A}nniversary {E}dition}}}\ (\bibinfo  {publisher} {Cambridge University
  Press},\ \bibinfo {year} {2010})\BibitemShut {NoStop}%
\bibitem [{\citenamefont {Horn}(2022)}]{sequencing}%
  \BibitemOpen
  \bibfield  {author} {\bibinfo {author} {\bibfnamefont {L.~B.-V.}\
  \bibnamefont {Horn}},\ }\bibfield  {title} {\bibinfo {title}
  {sequencing-dev/sequencing: v1.2.0}\ }\href
  {https://doi.org/10.5281/zenodo.7036085} {10.5281/zenodo.7036085} (\bibinfo
  {year} {2022})\BibitemShut {NoStop}%
\end{thebibliography}%
